\documentclass[acmsmall]{acmart}
\AtBeginDocument{%
  }

\setcopyright{acmlicensed}
\copyrightyear{2025}
\acmYear{2025}
\acmDOI{XXXXXXX.XXXXXXX}

\acmJournal{JACM}
\acmVolume{37}
\acmNumber{4}
\acmArticle{111}
\acmMonth{8}

\usepackage{array}
\usepackage{float}
\usepackage{booktabs}
\usepackage{multirow}
\usepackage{makecell}
\usepackage{dashrule}
\usepackage{amsmath}
\DeclareMathOperator*{\med}{med}
\usepackage{enumitem}
\usepackage{longtable}
\usepackage{soul}
\usepackage{arydshln}
\usepackage[ruled, lined, linesnumbered, commentsnumbered, noend]{algorithm2e}

\begin{document}

\title{Engineering Serendipity through Recommendations of Items with Atypical Aspects}

\author{Ramit Aditya}
\authornote{Both authors contributed equally to this research.}
\email{raditya@charlotte.edu}
\orcid{0009-0000-2405-989X}
\author{Razvan Bunescu}
\authornotemark[1]
\email{rbunescu@charlotte.edu}
\orcid{0000-0003-2919-3566}
\author{Smita Nannaware}
\email{snannawa@charlotte.edu}
\orcid{0009-0002-7185-7186}
\author{Erfan Al-Hossami}
\email{ealhossa@charlotte.edu}
\orcid{0000-0002-8436-8974}

\affiliation{%
  \institution{\\University of North Carolina at Charlotte}
  \city{Charlotte}
  \state{North Carolina}
  \country{USA}
}

\renewcommand{\shortauthors}{Aditya et al.}

\begin{abstract}
A restaurant dinner or a hotel stay may lead to memorable experiences when guests encounter unexpected aspects that also match their interests. For example, an origami-making station in the waiting area of a restaurant may be both surprising and enjoyable for a customer who is passionate about paper crafts. Similarly, an exhibit of 18th century harpsichords would be atypical for a hotel lobby and likely pique the interest of a guest who has a passion for Baroque music. 
Motivated by this insight, in this paper we introduce the new task of engineering serendipity through recommendations of items with atypical aspects. We describe an LLM-based system pipeline that extracts atypical aspects from item reviews, then estimates and aggregates their user-specific utility in a measure of serendipity potential that is used to rerank a list of items recommended to the user. To facilitate system development and evaluation, we introduce a dataset of Yelp reviews that are manually annotated with atypical aspects and a dataset of artificially generated user profiles, together with crowdsourced annotations of user-aspect utility values. Furthermore, we introduce a custom procedure for dynamic selection of in-context learning examples, which is shown to improve LLM-based judgments of atypicality and utility. Experimental evaluations show that serendipity-based rankings generated by the system are highly correlated with ground truth rankings for which serendipity scores are computed from manual annotations of atypical aspects and their user-dependent utility. Overall, we hope that the new recommendation task and the associated system presented in this paper catalyze further research into recommendation approaches that go beyond accuracy in their pursuit of enhanced user satisfaction.

The datasets and the code are made publicly available at \href{https://github.com/ramituncc49er/ATARS}{https://github.com/ramituncc49er/ATARS}.




\end{abstract}



\received{20 February 2007}
\received[revised]{12 March 2009}
\received[accepted]{5 June 2009}

\maketitle

\section{Introduction and Motivation}
\label{sec:introduction}

When searching for services such as hotels or restaurants, people are often presented with an overwhelming number of high-quality options that match their search criteria, especially in large metropolitan areas. This may lead to choice overload, or overchoice \cite{doi:10.1177/019263657005434912,CHERNEV2015333}, where an individual has to choose from a large number of items that are too difficult to compare, particularly under time constraints \cite{inbar_decision_2011}. Making a decision in the presence of overchoice becomes mentally exhausting and can trigger impaired self-regulation \cite{article}, decision paralysis, or anxiety \cite{article2}. The level of satisfaction that people experience when faced with an increasing number of choices has been observed to follow the well-known Wundt curve \cite{berlyne1973aesthetics}, an inverted U-shape curve originally relating stimulus intensity with its pleasantness. According to this functional dependency, as the number of choices goes up, satisfaction initially increases and then decreases \cite{Shah2007BuyingBA,kaiman:mde18}. Choice overload can be mitigated by reducing the number of consumer choices \cite{schwartz2004paradox} or by making one item stand out and appear better than the others \cite{scheibehenne_can_2010}. In this paper, we follow the latter approach and introduce a recommender system pipeline that promotes items within the search category that possess aspects with the potential to generate serendipitous experiences for the user. Our serendipity-focused approach is justified by empirical studies showing that serendipitous encounters contribute to significant increases in satisfaction, enjoyment, and meaningfulness \cite{kim_serendipity_2021}.

\begin{figure}[t]
    \centering
    \includegraphics[width=\linewidth]{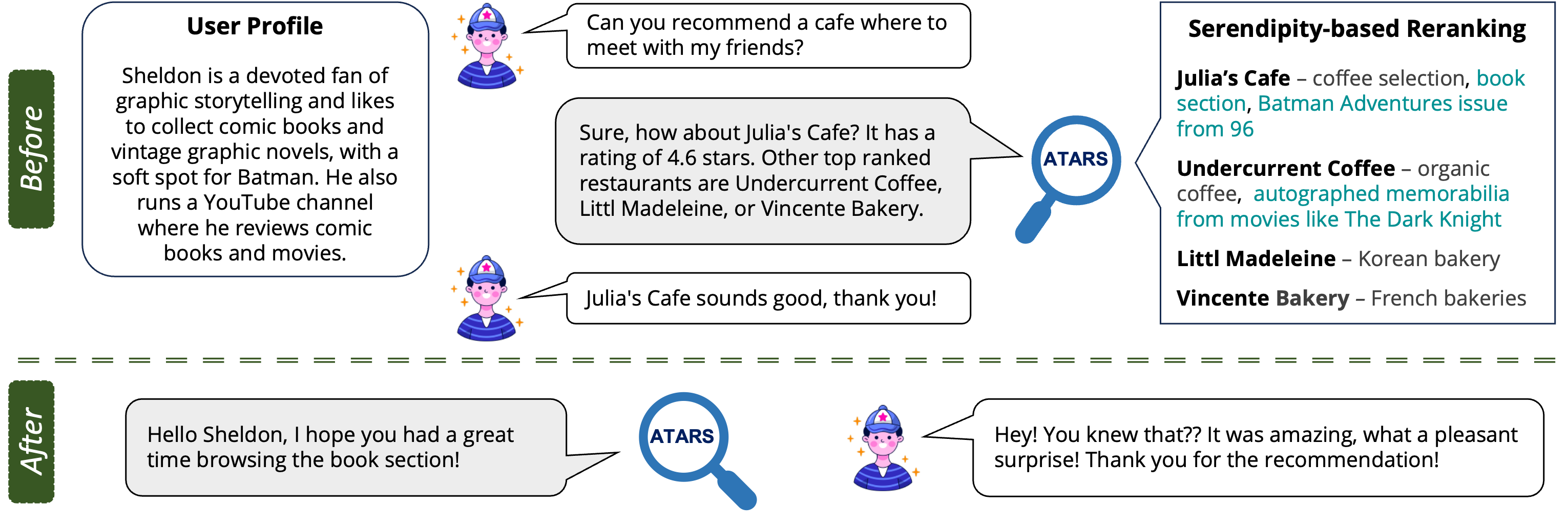}
    \caption{Example interaction with the proposed recommender system, where items with \textcolor{teal}{atypical aspects relevant to the user} are promoted at the top of the ranking ({\it before}), in order to engineer a serendipitous experience ({\it after}).}
    \label{fig:atars-interaction}
\end{figure}

Figure~\ref{fig:atars-interaction} illustrates a prototypical interaction between a user and the proposed Atypical Aspect-Based Recommender System (ATARS). The user, named Sheldon, is searching for a cafe where to meet with his friends. Correspondingly, the system compiles a ranked list of restaurants where top ranked items have atypical aspects that match the user's interests as described in his profile, in this case graphic storytelling, comic books, and Batman. The top ranked restaurant is then recommended to the user, who also has the option of choosing other top items. Importantly, the system does not disclose the extracted atypical aspects, such that, upon visiting the restaurant, its atypical aspects both surprise the user and engage his interest.

\begin{figure}[t]
    \centering
    \includegraphics[width=\linewidth]{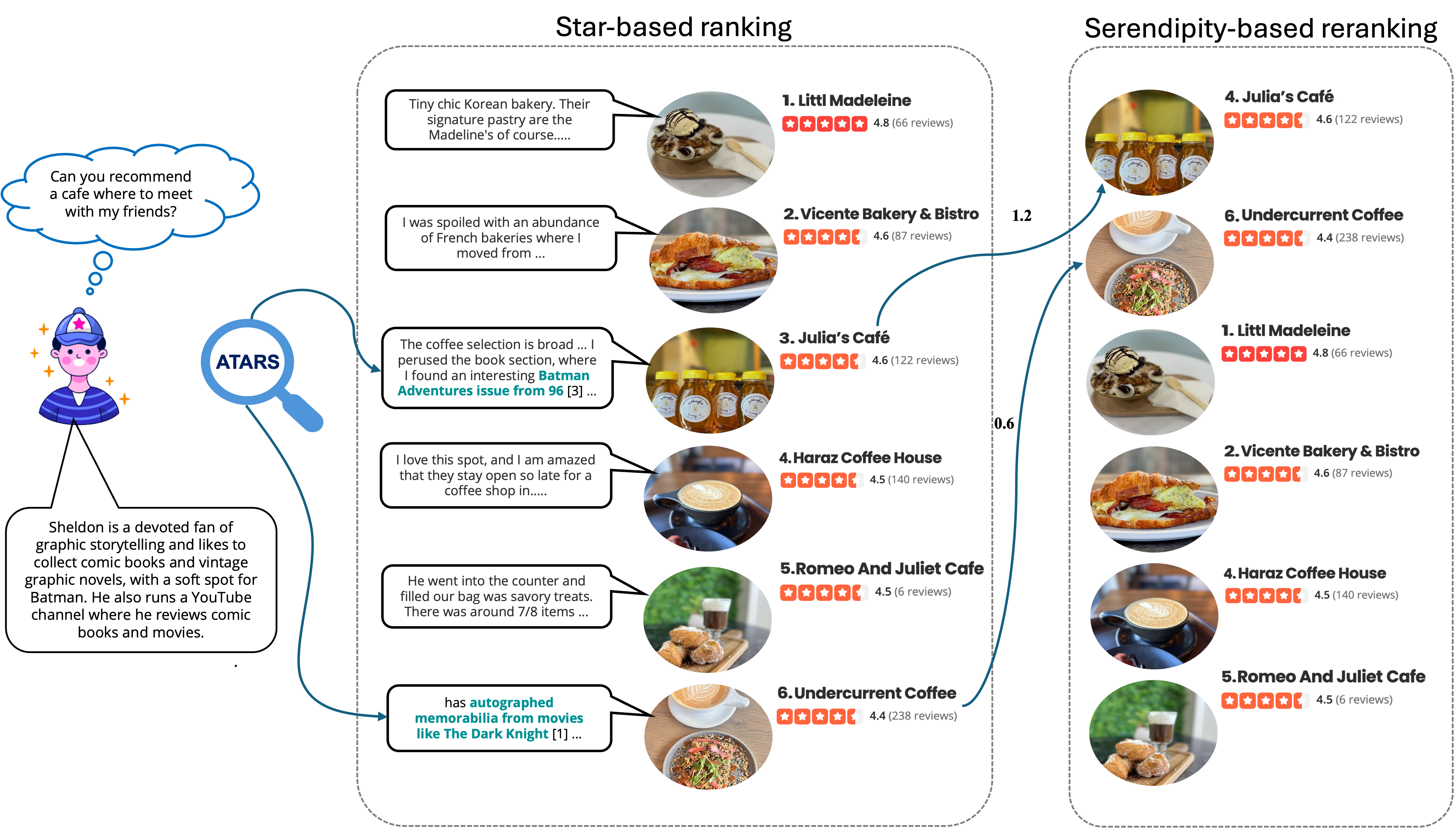}
    \caption{Example of the 3 steps of the Atypical Aspect-Based Recommender System (ATARS), where the left pane shows the user query and the user profile, given as input to the system. Middle pane shows the original star-based ranking of a sample of cafes, together with a sample of reviews showing \textcolor{teal}{atypical aspects} extracted in Step 1 and their user-dependent utilities extracted at Step 2 (between brackets). Right pane shows the re-ranking done based on the serendipity scores computed at Step 3 (shown on the connecting arrows).}
    \label{fig:atars-example}
\end{figure}

Overall, the proposed recommender system is structured as a pipeline of 3 main components:
\begin{enumerate}
    \item {\bf Atypical Aspect Extraction}: For each item that satisfies the user query, process its reviews in order to identify aspects that it possesses that are atypical for its category and hence have the potential to surprise (Section~\ref{sec:extrata}).
    \item {\bf Aspect Utility Estimation}: For each atypical aspect identified above, estimate its utility to the user based on information from a user profile (Section~\ref{sec:utility}); also referred to as Utility Classification.
    \item {\bf Serendipity-Based Re-ranking}: For each item, estimate its potential for serendipity by computing a serendipity score that aggregates utility values across all its atypical aspects. Use the serendipity score to re-rank the items from an original star-based ranking (Section~\ref{sec:reranking}).
\end{enumerate}
Figure~\ref{fig:atars-example} illustrates the outputs of the 3 modules, using the example introduced in Figure~\ref{fig:atars-interaction}. The system first identifies all the cafes in the user's area in its database and processes their reviews in order to identify atypical aspects. The middle pane shows a sample of cafes ranked by their star rating, as well as a sample of reviews. In these reviews, the module at Step 1 identifies phrases such as "Batman Adventures issue from 96" and "autographed memorabilia from movies like the Dark Knight" to be atypical for restaurants. The module at Step 2 assigns the first aspect a higher utility of 3, due to the user's interest in collecting comic books and vintage graphic novels (especially Batman), whereas the second aspect is assigned a relatively lower utility value of 1, due to the less direct relevance to the user's interests (through the movies category). The module at Step 3 then aggregates the utility values into a user-dependent serendipity score that is used to re-rank the items. As can be seen in the pane on the right, the items that have atypical aspects are now promoted to the top of the ranking in order of their overall utility for the user, as reflected in their serendipity scores. Importantly, \textit{the system does not disclose to the user} how the ranking was computed, in particular the roles of surprise (through atypical aspects) and utility (through relevance to the user profile). Therefore, if the user decides to go to one of the top two cafes, he is likely to experience serendipity by virtue of being surprised (due to the atypical aspects) in a positive way (due to atypical aspects matching user interests).

Surprise is an emotion caused by stimuli that confound or deviate from an individual's expectations \cite{scherer2005emotions}. Atypical aspects have a strong potential for triggering surprise in the user. For example, a user seeking restaurant recommendations is likely to be surprised by a restaurant that has an {\it origami making station} in the waiting area, because this is atypical for a restaurant. The surprise-evoking potential of atypical aspects is due to their {\it schema-discrepancy} \cite{reisenzein_cognitive-evolutionary_2019}, in that atypical aspects are unexpected relative to the schema associated with the corresponding domain, e.g., a comic book section relative to the restaurant schema, or a harpsichord exhibition relative to the hotel schema. In general, surprising events have the potential to increase user satisfaction because they can easily capture attention, leading to increased engagement and long dwell times \cite{horstmann_surpriseattention_2015}. However, serendipity requires not only surprise, but also relevance or utility for the user, hence the importance of utility estimation. For example, users interested in paper crafts are likely to find the origami making station valuable (high utility) , whereas users with different interests may choose to ignore it (low utility). When combined with atypical aspect extraction, utility estimation will enable determining that the user who likes paper crafts is likely to experience serendipity upon visiting the restaurant, whereas other users may only be surprised by the sight of the origami making station.

The serendipity score assigned to an item aggregates the utility values of all atypical aspects across the item's reviews, and then is used for re-ranking items in the original star-based ranking (Section~\ref{sec:reranking}). However, if done repeatedly across multiple recommendation sessions, this deterministic approach can lead to diminished returns in terms of surprise, as the user can learn that the recommender system always promotes items with atypical aspects relevant to their interests. This phenomenon, where a hardwired serendipity strategy makes users start to "expect the unexpected", has been documented in HCI literature, as reviewed in \cite{makri_making_2014}. Possible solutions involve inserting randomness in the serendipity engineering aspects of the recommendation process. For example, at times that are chosen in a nondeterministic fashion, the system can promote items with atypical aspects that are not necessarily known to be relevant to the user, effectively achieving a surprise-based re-ranking (Section~\ref{sec:reranking}). Since the space of atypical aspects is virtually unbounded, this approach ensures that users continue encountering surprise-generating atypical aspects indefinitely. Alternatively or complementary, the system can ignore atypical aspects altogether and use different sources of surprise, or even fall back on the traditional star-based ranking. All of these techniques imply a trade-off between surprise potential and expected utility, and require maintaining a user-level history of items and their atypical aspects. These and other possible approaches for cross-session maintenance of serendipity potential are left for future work.



The rest of the paper proceeds as follows. In Section~\ref{sec:related}, we summarize related work and recent progress in serendipity recommendations. In Section~\ref{sec:architecture}, we introduce the overall pipeline architecture of the proposed Atypical Aspect-Based Recommender System (ATARS) and formally describe the input and output of each of the 3 processing components. In Section~\ref{sec:extrata}, we describe and evaluate LLM-based approaches that identify atypical aspects in customer reviews in 3 target domains: restaurants, hotels, and hair salons. In Section~\ref{sec:utility}, we introduce a dataset of user profiles (Section~\ref{sec:profiles-dataset}), a dataset of aspect-profile pairs that are manually annotated with utility values (Section~\ref{sec:utility-dataset}), as well as the LLM-based approaches for utility classification (Section~\ref{sec:utility-llm}). Atypical aspect extraction and utility classification are brought together in Section~\ref{sec:reranking}, where we present a method for computing the utility-based serendipity score, which is then used for the serendipity-based re-ranking of items. The paper ends with concluding remarks and ideas for future work in Section~\ref{sec:conclusion}.


\section{Related Work}
\label{sec:related}

Choice overload or overchoice in recommender systems has previously been addressed by recommending a smaller set of items\cite{bollen_overchoice_2010} that are attractive to users, based on their preferences~\cite{jakob_reviewrec_2009, zhang_reviewrec_2014}, ratings, item attributes, or reviews~\cite{musto_absarec_2017, selmene_sentanalysis_2020, li_latentfeat_2022}. Prior studies on mitigating choice overload have focused on item assortment strategies such as categorization \cite{huber_adapting_cutoffs_1991}. \citet{diehl_smart_agents_2003} discovered that customers who had their available options ranked chose better products at lower prices than customers who had their available options in random order.

Serendipity has been recognized as an important factor for increasing user satisfaction in recommender systems, as documented in various surveys \cite{xi_2023_serendipity,ziarani_serendipity_rs_survey_2021, kotkov_serendipity_survey_2016}. There is a wide variety of approaches that aim to generate serendipitous recommendations. For example, \citet{li_unexrec_2020} recommended unexpected items by modeling user interests as clusters of historical data in a latent space and calculating the weighted distance between a new item and the clusters of interests. \citet{kotkov_serDataset_2018} crowdsourced serendipity labels for a movie dataset using multiple definitions of serendipity. \citet{niu_serendipity_2018} and \citet{niu:acm18} introduced computational approaches for serendipity and evaluated them on recommendations for online health news. \citet{tonmoy_bunescu_serendipity_2023} defined serendipity as the product between the user's rating of an item and Bayesian surprise, calculated as the KL divergence between the user's prior preference distribution (before consuming an item) and posterior distribution (after consuming an item) over item topics. Recommendations were then made by using collaborative filtering to find users with similar past preferences who encountered positively rated items that were surprising to them.

Serendipity-based recommendation approaches are typically organized into three stages: pre-processing, in-processing, and post-processing. The pre-processing typically involves extracting or engineering serendipity features. For example, \citet{li_elastic_serendipity_2019} constructs an asymmetric relevance network to discover weak ties between users and movies, allowing them to recommend films just beyond a user's usual choices yet still meaningfully connected. Another pre-processing example is the task of atypical aspect extraction (ExtrAtA) from customer reviews, which was originally introduced in our work from \cite{nannaware2023extraction}, together with LM-based approaches.
In-processing approaches incorporate serendipity during model training to either learn a serendipity representation or predict a serendipity score, as in \cite{xi_2023_serendipity,li_desr_2020,zhang_snpr_2021,wang_serendipity_2023}). Post-processing methods re-rank items to generate serendipity oriented recommendation lists, as in \cite{liu_neural_reranking_rs_2023}.

LLMs have been leveraged for serendipity-based recommendations in various studies. \citet{niu_serenprompt_2024} introduced SerenPrompt, a prompt-based framework to generate serendipitous recommendations. They proposed 3 types of prompts, discrete, continuous, and hybrid, and 2 prompting styles, direct and indirect, to investigate different strategies for creating serendipity. \citet{tokutake_zsserendipity_2024} use LLMs to evaluate serendipity of items in a zero-shot setting. \citet{xi_allm_2025} demonstrates using LLM to compress user information and long behavior sequences to generate profiles and train an LLM-based recommender, SerenGPT, to predict serendipitous items and align its serendipity judgments with human assessments.

The serendipitous recommender system introduced in this paper leverages surprise that is triggered by encounters of atypical aspects, which are inherently unexpected. To the best of our knowledge, no prior serendipity recommendation frameworks have leveraged atypical aspects as a source of surprise.


\section{ATARS TASK DEFINITION AND SYSTEM ARCHITECTURE}
\label{sec:architecture}

The primary objective of this work is to develop an Atypical Aspect-Based Recommender System (ATARS) that takes as input information about a user's general interests together with a search query, and produces as output a list of items whose ranking is determined by two criteria:
\begin{enumerate}
    \item How well the item satisfies the user query, including ratings of item quality.
    \item How much serendipity the user is likely to experience upon consuming the item.
\end{enumerate}
In this work, we consider items from three domain categories: {\sc Restaurants}, {\sc Hotels}, and {\sc Hair Salons}. To illustrate the various aspects of the task, throughout most of the narrative in this paper we will primarily use examples drawn from the restaurant category.

\begin{figure}[t]
    \centering
    \includegraphics[width=\linewidth]{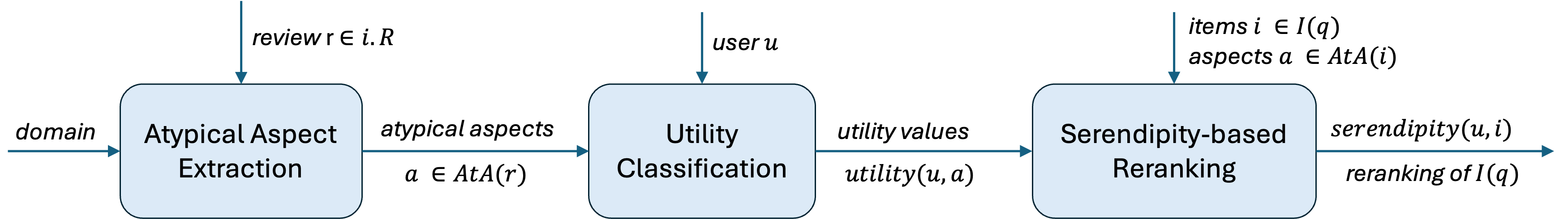}
    \caption{ATARS system architecture.}
    \label{fig:pipeline}
\end{figure}

For each domain, the input to the ATARS system will include:
\begin{itemize}
    \item A {\it user profile} $u$ that contains relevant information about the user, especially in terms of their interests, hobbies, passions and lifestyle.
    \item A {\it search query} $q$, be it a standalone request or a turn in a longer conversation.
    \item A {\it set of items} $i \in \mathcal{I}$, each of which is associated with a {\it star} rating $i.star \in [0, 5]$ and a number of {\it reviews} $r \in i.R$ that may mention one or more aspects of the item that are atypical for the domain.
\end{itemize}
Given the input above, the ATARS system will output a {\it re-ranked list of items}, where items that match the user's search query are ranked by considering both the original star rating as well as how much serendipity the atypical aspects of the restaurant is expected to generate for the user. More formally, the recommender system first identifies a subset of items $\mathcal{I}(q) \subseteq \mathcal{I}$ that are relevant to the user query $q$. Then the items in $I(q)$, their reviews, and the user profile will be processed by the 3 component pipeline as follows:
\begin{enumerate}
    \item For each item $i \in \mathcal{I}(q)$, for each review $r \in i.R$, extract all aspects $AtA(r)$ that are mentioned in the review and that are atypical for the target domain.
    \begin{itemize}
        \item Collect atypical aspects across that item's reviews in $AtA(i) =  \displaystyle\bigcup_{r \in i.R} AtA(r)$.
    \end{itemize}
    \item Given a user profile $u$ and an atypical aspect $a$, estimate a $utility(u, a)$ measure that captures how well the atypical aspect matches the user profile in terms of being {\it interesting}, {\it appealing}, or {\it useful} (henceforth referred to as {\it useful}).
    \item For each item $i \in \mathcal{I}(q)$, compute the user-dependent $serendipity(u, i)$ by aggregating the $utility(u, a)$ over all the atypical aspects $a \in AtA(i)$ of that item. 
    \begin{itemize}
        \item Rank all items based on their serendipity scores and star ratings.
    \end{itemize}
\end{enumerate}
The 3 processing steps and their associated notation are illustrated in Figure~\ref{fig:pipeline}.


\section{EXTRACTION OF ATYPICAL ASPECTS FROM REVIEWS}
\label{sec:extrata}

\begin{figure}[t]
    \centering
    \includegraphics[width=\linewidth]{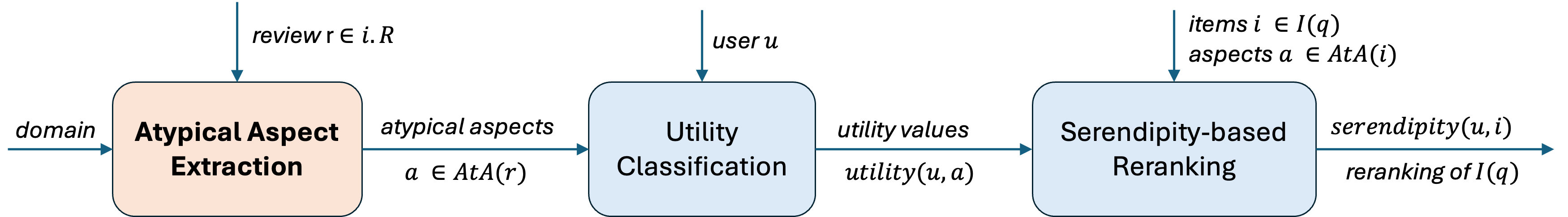}
    \caption{ATARS system architecture: Atypical Aspect Extraction (ExtrAtA).}
    \label{fig:pipeline1}
\end{figure}

The first step in the system architecture pipeline, as highlighted in Figure~\ref{fig:pipeline1}, is that of extracting atypical aspects from item reviews. Given a domain category, e.g. {\sc restaurants}, and a customer review of a particular item in that category, e.g. {\it a restaurant}, the task is to extract aspects of that particular item that are {\it atypical of items in its category}. Throughout most of this paper we will use the category of restaurants as an example. All aspects that are related to the core business of a restaurant, including but not limited to food, service, price, opening hours, parking, are considered typical aspects and are not annotated. Conversely, we define and annotate an aspect as atypical if it is not related to the core business of {\sc restaurants}, yet it belongs to or is a feature of {\it the restaurant} that is reviewed ({\sc restaurants} refers generically \cite{carlson_generic_1995} to the restaurant category, whereas {\it the restaurant} refers to a specific restaurant). Table~\ref{tab:examples_restaurant} shows samples taken from 4 reviews, illustrating two types of manual annotations, {\it extractive} and {\it abstractive}, analogous to the extractive \cite{hongyan:sigir99,knight_summarization_2002} and abstractive \cite{jing-2002-using,rush-etal-2015-neural} annotation schemes widely used in summarization datasets\footnote{\url{https://duc.nist.gov}, \url{https://tac.nist.gov}}. A special case is made of aspects related to the ambience or atmosphere of a restaurant: while ambience might be considered as an important part of, and thus subordinated to, the core business of restaurants, there are cases where ambience aspects stand out and become an attraction on their own. When that happens, we annotate them on a secondary, optional layer.

\begin{table*}[t]
    \small
    \centering
    \begin{tabular}{p{\textwidth}}
    \toprule
    \noindent $\blacktriangleright$ A group of work friends and I stumbled upon Upper Deck a little over a year ago and everyone from our office has turned Upper Deck into our local watering hole ever since. Their happy hour special is unbeatable, they have a good selection of draft beers, and the food is out of this world good. The stand out feature of Upper Deck is the offering of {\bf life size beer pong} at their outside patio. This takes traditional beer pong and substitutes solo cups with garbage cans (painted to look like solo cups) and Ping Pong balls with dodgeballs They also have a {\bf pool table} and recently added {\bf arcade games} ({\bf nfl blitz 99} beats {\bf madden 15} all day). Get some friends and bring your appetites and some quarters, you won't be disappointed.  \\
    \hdashline[1pt/2.5pt]
    {\it \textbf{The restaurant offers life size beer pong at their outside patio. They have a pool table. They recently added arcade games, such as nfl blitz 99 and madden 15.}} \\
    \midrule
    \noindent $\blacktriangleright$ The big draw of this place is the excellent pizza, which you can have with beer on an outside deck with a view of the {\bf parklands}. It's a nice place to hang out on a sunny afternoon. You can even go for a walk in the {\bf Goose Creek Park} behind the restaurant afterwards to burn off the calories you just consumed. The big minus is that if you go for lunch or in early afternoon, the menu is really limited. This place also seems to attract a goodly number of families with kids at lunch times, probably because it serves pizza and there's a {\bf playground} in the adjoining park, \\
    \hdashline[1pt/2.5pt]
    {\it \textbf{The restaurant has an outside deck with a view of the parklands. Customers can go for a walk in the Goose Creek Park behind the restaurant. There's a playground in the adjoining park.}} \\
    \midrule
    \noindent $\blacktriangleright$ Classic West Philly spot where you can see {\bf local wildlife}. Everyone from {\bf moms} to {\bf anarchists} to {\bf hackers} to {\bf organic gardeners} to {\bf activists} hangs out there. The coffee is excellent, the baked goods are great as well, and if you're working on something, you might run into a possible collaborator there. If you're thinking of moving to West Philly, definitely check out Satellite and the farmer's market. They've replaced the cracked and chipped cups with awesome new cups, which are awesome. \\
    \hdashline[1pt/2.5pt]
    {\it \textbf{In this restaurant you can see local wildlife. Everyone from moms to anarchists to hackers to organic gardeners to activists hangs out there.}}\\
    \midrule
    \noindent $\blacktriangleright$ This is such a cool place! Three words was all it took to add this gem to my list of places to visit while in St. Louis - "{\bf Good Burger Car}!!" YES! They have the car from the movie Good Burger! A movie I was obsessed with as a child \& have since gotten my kids to love just as much! The place is made out of \textcolor{teal}{\textbf{cool, colorful shipping containers}} with many neat decorations, what looks like an \textcolor{teal}{\textbf{alien spaceship}} from Toy Story with a laser on it adorned the top of the place along with a \textcolor{teal}{\textbf{cow}}. Now, on to the food. They have many different options to choose from including create-your-own burgers \& many of their own creations, sandwiches, salads, sides, kid's meals, \& shakes \& floats ... Such a unique place \& worth a visit! They also sell {\bf souvenir T-shirts} \& {\bf hats}, \& my fiance had to get himself a {\bf "HI AF" shirt}. The shirts were heavily influenced by the {\bf movie Good Burger} \& there was one in particular I had my heart set on, unfortunately they did not have it in my size.\\
    \hdashline[1pt/2.5pt]
    {\it \textbf{The restaurant has the car from the movie Good Burger. They sell souvenir T-shirts and hats, and a customer got a "HI AF" shirt. The shirts are heavily influenced by the movie Good Burger.}}\\
    \textcolor{teal}{\textbf{The place is made out of cool, colorful shipping containers, with many neat decorations, what looks like an alien spaceship from Toy Story with a laser on it adorned the top of the place along with a cow.}}\\
    \bottomrule
    \end{tabular}
    \caption{Examples of {\bf extractive} and {\it \textbf{abstractive}} annotations of customer reviews in the restaurants domain. Secondary or optional annotations are shown in \textcolor{teal}{\textbf{blue}}.}
    \label{tab:examples_restaurant}
\end{table*}

In the extractive annotation, only base noun phrases referring to atypical aspects are annotated. If multiple phrases refer to the same atypical aspects, we only annotate the grounding instance of the coreference chain. For example, the noun phrase "adjoining park" in the second review is not annotated, as it refers to the "Goose Creek Park" which is already annotated. However, if a review mentions an atypical category, such as "arcade games" in the first review or "local wildlife" in the third review, any category instance that is mentioned will also be annotated as atypical, such as "nfl blitz 99" in the first review or "anarchists" in the third review, respectively.
If the atypical aspect is a more complex noun phrase, we only annotate the base noun phrase that expresses the semantic core (often the syntactic head), as in "the {\it Goose Creek Park} behind the restaurant". The extractive annotation is meant to be used together with the original review in downstream applications, which makes it acceptable to annotate only the most important part of the phrase.

In the abstractive annotation, one or more sentences are generated that enumerate the atypical aspects mentioned in the review. The formulation is kept as close as possible to the original text while maintaining naturalness. The generated sentences are intended to be concise, usually maintaining details that are expressed in the same sentence in the review, however keeping out unimportant information about the atypical aspect that is mentioned in other sentences, or details that are vague or uncertain. Sentiment words are maintained only if it helps keep the text natural and faithful to the original. The abstractive annotation is meant to be standalone and used without the original review in downstream tasks, as such it may require some minimal rewriting of the original review formulation, e.g. adding the phrase "the restaurant", or removing opinion words such as "beats" in the first example. Sometimes reviewers use metaphors to refer to atypical aspects, in which case it is important that the abstractive version preserves the metaphorical meaning. This is the case for "local wildlife" in the third review, which refers metaphorically to types of customers that are seen relatively less often in that context.


In many aspect-based sentiment analysis approaches \cite{Semeval-2014}, identifying {\it typical} aspects that are mentioned in a review is done explicitly as {\it aspect term extraction}. However, the task of extracting {\it atypical} aspects, as introduced above, cannot be solved simply by first (a) identifying all typical aspects of restaurants that are mentioned in a review, followed by (b) extracting all other noun phrases, i.e. phrases that do not refer to a typical aspect of a restaurant. In their reviews, people often mention entities or events that are not associated with the restaurant, such as "our office" in the first review, or "the farmer's market" in the third review, and these phrases should not be extracted either. Thus, it is important that the noun phrase refers to an aspect that is {\it associated with the reviewed restaurant}, and that at the same time is atypical of restaurants (or unexpected for a restaurant). Finally, while Table~\ref{tab:examples_restaurant} may induce the perception that atypical aspects are common, the opposite is actually true. As will be detailed in Section~\ref{sec:extrata-dataset} below, it takes going through at least 50 reviews in order to find one review that mentions an atypical aspect. The difficulty of manually finding this "needle in a haystack" further motivates the development of automated approaches for atypical aspect extraction. In Appendix~\ref{app:annotatedata-examples}, we provide further examples of atypical aspect annotations drawn from reviews of hotels (Table~\ref{tab:examples_hotels-lt}) and hair salons (Table~\ref{tab:examples_hairsalons-lt}).

\subsection{The ExtrAtA Dataset of Atypical Aspect Annotations}
\label{sec:extrata-dataset}

We used the Yelp dataset \cite{yelp} as a source of reviews for the 3 target categories: Restaurant ($\sim$5M reviews), Hotel ($\sim$190K reviews), and Hair Salon ($\sim$115K reviews). Because most aspects are expressed as nouns and less frequently as verbs, we use \citet{spacy} to collect lemmas of all nouns and verbs and compute their frequencies for each domain. We rank words in ascending order based on their counts and filter out words that appear with very low frequency, e.g., less than 10 times for the Restaurant domain, as these tend to be spelling mistakes or interjections that are purposely misspelled for extra emphasis, e.g., "amaazzing". We then consider the remaining rare words in ascending order of their frequency as candidate atypical words, extract the reviews that mention them, and read these reviews to determine which occurrences truly refer to an atypical aspect. When reading a review, all atypical aspects are annotated, not only the ones corresponding to the search word. Notwithstanding the heuristic selection of reviews based on the occurrence of rare words, overall this was still a very time-consuming process, because rare words very often appear in a review without necessarily referring to atypical aspects. For example, out of the 43 restaurant reviews that contain the lemma "poncho", in only 1 review the word "ponchos" was deemed to refer to an atypical aspect (the restaurant was selling them). The other 42 reviews contained references to ponchos that were not associated with the restaurants itself, e.g. staff helping customers put their ponchos on a rainy day, or customers describing their arrival at the restaurant on a rainy day. As we went down the list of rare words, their frequency increased, resulting in a larger number of reviews to skim through for each rare word. Overall, for the Restaurant dataset, we used as search words the rare words that appeared with a frequency of up to 187. Upon semi-automatically sifting through the $\sim$97K reviews found to contain these words, we were able to collect 114 reviews that contained atypical aspects. On average, one hour of following this process led to finding between 2 and 3 reviews containing atypical aspects for the restaurant category, whereas for the hair salon category it took on average two hours to find 1 atypical reviews. Henceforth the term {\it atypical review} will be used to refer to a review that contains one or more atypical aspects; analogously, the term {\it typical review} will be used to refer to reviews that do not contain any atypical aspect.

As illustrated in the examples from Table~\ref{tab:examples_restaurant}, we organize annotations of atypical aspects on two layers:
\begin{itemize}
    \item A {\it primary} layer that contains atypical aspects that are clearly not connected to any core feature of that domain.
    \item A {\it secondary} layer that contains atypical aspects that are related to a typical aspect, such as ambiance or location, but that stand out and are interesting on their own, separate from the core features of the domain.
\end{itemize}
For example, \textit{'I was even encouraged to visit their \textbf{petting zoo} in the back'} would be considered a primary atypical aspect in any of the 3 categories, whereas \textit{'There is an \textbf{interesting giant stuffed spider} that goes up and down when the door leading to the bathrooms opens and closes'} would be annotated as a secondary atypical aspect.

Table~\ref{tab:dataset} shows summary statistics for the 3 datasets, one for each domain (category), split between data used for training and testing, and data used for development. Under the Primary column, we show the number of atypical reviews and atypical aspect annotations contained in them. The next column shows the same statistics for when both primary and secondary atypical aspects are considered. The total number of reviews in each dataset, shown in the third column, is about double the number of primary atypical reviews, reflecting a balanced dataset where the number of typical reviews was selected to be about the same as the number of  atypical reviews. We computed inter-annotator agreement (ITA) on both the extractive and abstractive annotations in the development sets of the Restaurant and Hair Salon domains. The ITA metrics are shown in Table~\ref{tab:atypical_extraction_evaluation} and are calculated by assuming one annotator provides the ground truth while the other annotator is considered as the system.

\begin{table*}[t]
\centering
 \caption{Statistics for the 3 ExtrAtA datasets, split between Train+Test and Development (Dev). The number of atypical reviews and atypical aspects are presented separately for primary (Primary) vs. both primary and secondary atypical aspects ($+$ Secondary).}
\begin{tabular}{@{}ccccccc@{}}
\toprule
\multicolumn{1}{l}{\textbf{Domain}} & \multicolumn{1}{l}{\textbf{Dataset split}} & \multicolumn{2}{c}{\begin{tabular}{c} \textbf{Primary}\end{tabular}} & \multicolumn{2}{c}{\begin{tabular}{c} $+$ \textbf{Secondary}\end{tabular}} & \multicolumn{1}{c}{\begin{tabular}{c} \textbf{Total}\end{tabular}}  \\
 \multicolumn{1}{l}{} & \multicolumn{1}{l}{} & \multicolumn{1}{c}{\# reviews} & \multicolumn{1}{c}{\# aspects} & \multicolumn{1}{c}{\# reviews} & \multicolumn{1}{c}{\# aspects} & \multicolumn{1}{c}{\textbf{reviews}}\\
\midrule
\multirow{2}{*}{Restaurant} & Train+Test & 100 & 253 & 107 & 340 & 200 \\
 & Dev & 14 & 32 & 16 & 46 & 28 \\
 \cmidrule(lr){2-7}
\multirow{2}{*}{Hotel} & Train+Test & 69 & 274 & 85 & 401 & 150 \\
 & Dev & 10 & 33 & 11 & 49 & 20\\
\cmidrule(lr){2-7}
\multirow{2}{*}{Hair Salon} & Train+Test & 45 & 147 & 48 & 181 &  90 \\
 & Dev & 5 & 24 & 5 & 29 & 10 \\
\bottomrule
\end{tabular}
\label{tab:dataset}
\end{table*}

\subsection{LLM-Based Extraction of Atypical Aspects}
\label{sec:extrata-llm}

In \cite{nannaware2023extraction}, we approached the task of atypical aspect extraction using 2 language models: Flan-T5 \cite{chung2022scaling} and GPT-3.5~\cite{chatgpt}. FLAN-T5 is a 3 billion parameter encoder-decoder Transformer based on the T5 model~\cite{raffel2020exploring} that was further instruction-tuned on the FLAN dataset~\cite{wei2022finetuned}. For atypical aspect extraction, Flan-T5 was fine-tuned and tested using a 10-fold evaluation scenario, where the Train+Test reviews dataset was partitioned into 10 folds, 9 folds were used for training and 1 fold was used for testing. This process was repeated 10 times until each fold in the dataset was used as a test fold. GPT-3.5 is a 175 billion parameters encoder Transformer from the GPT family that was used for atypical aspect extraction through a prompt that was optionally augmented with a fixed set of 5 in-context learning examples.

In this work, we developed an improved approach to atypical aspect extraction that leverages the much larger GPT-4 model, which has about 10 times the number of parameters of GPT-3.5. This new approach does extraction in 2 main steps, as follows:
\begin{itemize}
    \item \textbf{Step 1}: Process the review into a list of {\it aspect sentences} that capture all the aspects of the restaurant, typical or atypical. Other types of information, such as opinions or incidental information that does not pertain to the restaurant, are to be ignored (prompt template in Table~\ref{tab:prompt_reviewreformuation}).
    \item \textbf{Step 2}: Extract atypical aspects from the list of aspect sentences, as shown below (prompt template in Table~\ref{tab:prompt_ataextraction}):
    \begin{itemize}
        \item \textbf{Step 2.1}: Classify each aspect sentence as positive if it mentions an atypical aspect, otherwise classify it as negative.
        \item \textbf{Step 2.2}: Extract all atypical aspects from each of the sentences classified as positive.
    \end{itemize}  
\end{itemize}

\begin{table}[t]
    \small
    \centering
    \caption{Step 1 prompt for processing a review into a list of aspect sentences that only highlight features or properties of the restaurant, using in-context learning.} 
    \label{tab:prompt_reviewreformuation}
    \begin{tabular}{p{\textwidth}}
        \toprule
        \noindent\textbf{\textit{Instructions:}} \textit{You will be given a restaurant review. Read the provided review thoroughly and generate a list of sentences that only talk about features or properties of the restaurant. You can remove redundant statements and opinions as long as the generated list of sentences is comprehensive in that it should include all the restaurant features that were mentioned in the original review. The generated sentences should stay as close as possible to the sentences in the original review. For example:}\\
    
        \hspace{1em} \textbf{\textit{Example 1:}} 
        Really good sushi - in a strip mall in Bristol. Go figure. They have hibachi as well, but I feel like hibachi food is hibachi food no matter where you go; it's more about the performance. Do it once and then keep coming back for the sushi! This is a mom-and-pop type restaurant with a friendly atmosphere. If you're there for a special occasion, they'll take a polaroid and add it to their wall. I have fond memories of many family meals here. Look for my picture on the wall; I'm the adorable one.\\
        
        \hspace{1em} \textbf{\textit{Output 1}}: \textit{The restaurant has really good sushi. The restaurant is in a strip mall in Bristol. The restaurant has hibachi. This is a mom-and-pop type restaurant with a friendly atmosphere. If you're there for a special occasion, they'll take a polaroid and add it to their wall.}\\
        
        \hspace{1em} \textbf{\textit{Example 2:}} $\langle$ {\tt review} $\rangle$ \textbf{\textit{Output 2:}} $\langle$ {\tt output} $\rangle$ \hspace{1em} \textbf{\textit{Example 3:}} $\langle$ {\tt review} $\rangle$ \textbf{\textit{Output 3:}} $\langle$ {\tt output} $\rangle$  \\
        \ \\
        \textit{Now, read and process the following  review:} $\langle$ {\tt review} $\rangle$\\
        \bottomrule
    \end{tabular}
\end{table}

\begin{table}[h]
    \caption{Step 2 prompt for identifying which review sentences contains atypical aspects (Step 2.1) and for extracting atypical aspects (Step 2.2), optionally augmented with examples for in-context learning.} 
    \label{tab:prompt_ataextraction}
    \small
    \centering
    \begin{tabular}{p{\textwidth}}
        \toprule
        \noindent\textbf{\textit{Instructions:}} \textit{You will be given as input a sentence from a restaurant review. Read the sentence and determine if it contains aspects that are atypical for a restaurant. An aspect is considered atypical if it is not related to restaurant business, otherwise the aspect is considered typical. For example, typical aspects include service, food, drinks, location, price, menu, discounts, policies, staff, customer satisfaction, parking, typical restaurant ambience or any other feature commonly associated with a restaurant. No matter how extraordinary or novel an aspect is, if it is an instance or a type of typical aspect, it should not be considered atypical. For example, an exquisite cocktail named "sugary hurricane" is not an atypical aspect, because it is still a type of drink, and drinks are typical aspects of a restaurant. On the contrary, an "game playing table" in the waiting area would be an atypical aspect, because games are not part of the restaurant business.}\\
        \hspace{1em}\textit{Atypical aspects need to be features of the restaurant. Therefore, sentiment or opinions expressed by customers in the review do not qualify as atypical aspects, nor do similes, analogies, metaphors, or random events involving patrons. Examine not only the aspect itself but also its context within the review to determine whether it is genuinely an atypical feature of the restaurant. Atypical aspects can be events that the restaurant is known to offer from infrequently or on special occasions, such as standup routines or concerts.}\\
        \hspace{1em}\textit{Output <pos> for sentences that contain one or more atypical aspects, then list the atypical aspects. Output <neg> for sentences that do not contain any atypical aspect, then write <None>. For example:}\\
        \textit{\bf Examples 1 to 8:} $\langle$ {\tt review sentence} $\rangle$ \\
        \hspace{1em} \textit{Classification:} $\langle$ {\tt pos or neg} $\rangle$ \textit{Atypical Aspects:} $\langle$ {\tt extracted aspects} $\rangle$ \\
        \ \\
        \textit{Now, read and process the following review sentence:} $\langle$ {\tt sentence} $\rangle$ \\
        \bottomrule
    \end{tabular}
\end{table}
Table~\ref{tab:prompt_reviewreformuation} shows the prompt and one ICL example used for the review processing at Step 1, for the {\sc Restaurants} domain, whereas Appendix~\ref{app:reviewreformulation-examples} shows the complete text of the 3 in-context examples used at Step 1, for each of the 3 domains. Table~\ref{tab:prompt_ataextraction} shows the overall prompt instructions and ICL template for the classification and atypical aspect extraction sub-steps performed at Step 2, whereas Appendix~\ref{app:ataextraction-examples} shows the complete text of the 8 in-context examples, for each domain. Each set of 8 ICL examples is comprised of 4 positively classified review sentences with their corresponding extracted atypical aspects and 4 negatively classified review sentences. Step 2 was evaluated with 0-shot prompting, as well as two few-shot prompting approaches: \textit{Fixed} few-shot and {\it Dynamic} few-shot. In the Fixed few-shot setting, the 4 positive and 4 negative in-context examples were selected manually and kept fixed, meaning each review was processed using the same set of 8 ICL examples. In the Dynamic few-shot setting, we use a leave-one-group-out cross validation approach, where all review sentences associated with one item are put in the test fold, while the remaining sentences form the training fold. Each sentence is mapped to its JINA embedding \cite{gunther-etal-2023-jina}, and the similarity between sentences is calculated as the cosine similarity of their corresponding embeddings. For each test review sentence,  we identify the 4 most similar review sentences from the training fold that were classified as having atypical aspects (i.e. positive), and the 4 most similar review sentences that were classified as not having atypical aspects (i.e. negative). The 8 sentences are then used as the 8 in-context learning examples for the test review sentence.

\subsection{Experimental Evaluation}
\label{sec:extrata_evaluation}

\begin{table}[t]
    \centering
    \small
    \caption{Results (\%) for aspect sentence classification (Step 2.1) across the three domains, in terms of Precision (P), Recall (R), and F1, reported separately for Primary  vs. Primary $+$ Secondary atypical aspects.}
    \begin{tabular}{l|rrr|rrr}
        \toprule
        \multicolumn{1}{l|}{\textbf{Domain}} & \multicolumn{3}{c|}{\textbf{Primary}} & \multicolumn{3}{c}{$+$ \textbf{Secondary}} \\
        \multicolumn{1}{l|}{\textbf{\ \ Scenario}} & \multicolumn{1}{c}{P} & \multicolumn{1}{c}{R} & \multicolumn{1}{c|}{F1} & \multicolumn{1}{c}{P} & \multicolumn{1}{c}{R} & \multicolumn{1}{c}{F1} \\
        \midrule
        \multicolumn{7}{l}{\sc Restaurants} \\
        \ \ GPT-4 (0-shot)         & 57.6 & 81.1 & 67.4 & 71.5 & 73.7 & 72.6 \\
        \ \ GPT-4 (Fixed 8-shot)   & 58.4 & \textbf{89.2} & 70.6 & {\bf 73.6} & \textbf{84.8} & \textbf{78.8} \\
        \ \ GPT-4 (Dynamic 8-shot)     & \textbf{61.4} & 83.8 & \textbf{70.8} & 72.9 & 78.8 & 75.7 \\
        \multicolumn{7}{l}{\sc Hotels} \\
        \ \ GPT-4 (0-shot)         & 51.3 & 55.7 & 53.4 & 67.1 & 46.7 & 55.1 \\
        \ \ GPT-4 (Fixed 8-shot)   & 55.8 & \textbf{75.7} & 64.2 & 73.9 & 64.8 & 69.0 \\
        \ \ GPT-4 (Dynamic 8-shot)     & \textbf{65.8} & 74.3 & \textbf{69.8} & \textbf{74.3} & \textbf{77.1} & \textbf{75.7} \\
        \multicolumn{7}{l}{\sc Hair Salons} \\
        \ \ GPT-4 (0-shot)         & 64.5 & 94.2 & 76.6 & 75.0 & 91.9 & 82.6 \\
        \ \ GPT-4 (Fixed 8-shot)   & 71.0 & 94.2 & 81.0 & 78.6 & 88.7 & 83.3 \\
        \ \ GPT-4 (Dynamic 8-shot)     & \textbf{76.1} & \textbf{98.1} & \textbf{85.7} & \textbf{83.3} & \textbf{96.8} & \textbf{89.6} \\
        \bottomrule
    \end{tabular}
    \label{tab:step2.1}
\end{table}

\begin{algorithm}[t]
\small
   \SetKwInOut{KwIn}{Input}
   \SetKwInOut{KwOut}{Output}

   \KwIn{\ Gold Phrases $GP$, \ Extracted Phrases $EP$}
   \KwOut{\ Precision (P), \ Recall (R), \ and $F_1$ measure}
    \tcp{\small $TPe$ = \# True Positives w.r.t $EP$, \ $TPg$ = \# True Positives w.r.t $GP$}
    \tcp{\small $FP$ = \# False Positives, \ $FN$ = \# False Negatives}
   \For{gp in GP}{
       \eIf{EP is empty}{
           $FN \gets FN + |gp| / |gp|$
        }{ 
        Find $ep \in EP$ that has maximum Jaccard similarity with $gp$\\
            $TPe \gets TPe + |ep \cap gp| / |ep|$, \ \ $TPg \gets TPg + |ep \cap gp| / |gp|$ \\
            $FP \gets FP + |ep - gp| / |ep|$, \ \ $FN \gets FN + |gp - ep| / |gp|$ \\
        }
        Remove $ep$ from the set $EP$
    }
    \For{ep in EP}{
        $FP \gets FP + |ep| / |ep|$
    }
    Precision $P \gets TPe / (TPe + FP)$, \ \ Recall $R \gets TPg / (TPg + FN)$, \ \ $F_1 \gets 2 P R / (P + R)$\\
    \KwRet{$P, \; R, \; F_1$}
    \caption{PartialMatchMetrics($GP$, $EP$)}\label{alg:cap}
 \end{algorithm}

\begin{table}[t]
    \centering
    \small
    \caption{Precision (P), Recall (R), and F1 scores (\%) for extractive extraction, across the 3 domains.}
    \begin{tabular}{l|rrr|rrr}
        \toprule
        \multicolumn{1}{l|}{\textbf{Domain}} & \multicolumn{3}{c|}{\textbf{Exact Match}} & \multicolumn{3}{c}{\textbf{Partial Match}} \\
        \multicolumn{1}{l|}{\textbf{  Method}} & \multicolumn{1}{c}{P} & \multicolumn{1}{c}{R} & \multicolumn{1}{c|}{F1} & \multicolumn{1}{c}{P} & \multicolumn{1}{c}{R} & \multicolumn{1}{c}{F1} \\
        \midrule
        \multicolumn{7}{l}{\sc Restaurants} \\
        \ \ Flan T5 (fine-tuned)       & 67.5 & 60.2 & 63.4 & 72.9 & 65.3 & 68.6 \\
        \ \ \textit{+ secondary}       & \textit{57.7} & \textit{58.6} & \textit{56.4} & \textit{64.2} & \textit{64.4} & \textit{62.3} \\
        \ \ GPT-4 (0-shot)             & 71.2 & 72.1 & 71.6 & 75.8 & 80.8 & 78.3 \\
        \ \ \textit{+ secondary}       & \textit{70.0} & \textit{69.1} & \textit{69.6} & \textit{76.1} & \textit{79.9} & \textit{78.0} \\
        \ \ GPT-4 (Fixed 8-shot)       & 70.3 & 78.0 & 73.9 & 74.1 & 84.2 & 78.8 \\
        \ \ \textit{+ secondary}       & \textit{71.5} & \textit{77.6} & \textit{74.4} & \textit{76.4} & \textit{85.2} & \textit{80.6} \\
        \ \ GPT-4 (Dynamic 8-shot)         & \textbf{76.1} & \textbf{78.9} & \textbf{77.5} & \textbf{79.1} & \textbf{84.4} & \textbf{81.7} \\
        \ \ \textit{+ secondary}       & \textit{72.9} & \textit{74.9} & \textit{73.9} & \textit{77.6} & \textit{82.7} & \textit{80.1} \\        
        \ \ Human ITA & 80.0 & 78.3 & 79.1 & 85.6 & 82.6 & 84.1 \\
        \ \ \textit{+ secondary}   &  {\it 75.4} & {\it 79.3} & {\it 77.3} & {\it 82.4} & {\it 86.8} & {\it 84.5} \\
        \multicolumn{7}{l}{\sc Hotels} \\
        \ \ Flan T5 (fine-tuned)       & 60.2 & 54.9 & 55.9 & 63.8 & 57.6 & 59.0 \\
        \ \ GPT-4 (0-shot)             & 72.8 & 71.1 & 71.9 & 75.4 & 76.9 & 76.2 \\
        \ \ \textit{+ secondary}       & \textit{69.0} & \textit{61.7} & \textit{65.1} & \textit{72.8} & \textit{68.8} & \textit{70.7} \\
        \ \ GPT-4 (Fixed 8-shot)       & 70.4 & 75.6 & 72.9 & 72.8 & 81.5 & 76.9 \\
        \ \ \textit{+ secondary}       & \textit{69.0} & \textit{68.1} & \textit{68.5} & \textit{73.0} & \textit{77.2} & \textit{75.0} \\
        \ \ GPT-4 (Dynamic 8-shot)         & \textbf{79.2} & \textbf{79.2} & \textbf{79.2} & \textbf{81.6} & \textbf{83.5} & \textbf{82.6} \\
        \ \ \textit{+ secondary}       & \textit{75.0} & \textit{71.4} & \textit{73.1} & \textit{78.8} & \textit{77.9} & \textit{78.4} \\
        \multicolumn{7}{l}{\sc Hair Salons} \\
        \ \ Flan T5 (fine-tuned)       & 66.6 & 62.9 & 63.9 & 75.1 & 69.1 & 71.1 \\
        \ \ GPT-4 (0-shot)             & 52.7 & 59.9 & 56.1 & 58.7 & 76.2 & 66.3 \\
        \ \ \textit{+ secondary}       & \textit{56.4} & \textit{61.8} & \textit{59.0} & \textit{63.2} & \textit{78.2} & \textit{69.9} \\
        \ \ GPT-4 (Fixed 8-shot)       & 65.7 & 76.9 & 70.9 & 68.0 & 82.9 & 74.7 \\
        \ \ \textit{+ secondary}       & \textit{65.6} & \textit{74.6} & \textit{69.8} & \textit{68.1} & \textit{80.8} & \textit{73.9} \\
        \ \ GPT-4 (Dynamic 8-shot)         & \textbf{72.8} & \textbf{80.7} & \textbf{76.5} & \textbf{75.7} & \textbf{85.4} & \textbf{80.2} \\
        \ \ \textit{+ secondary}       & \textit{72.9} & \textit{79.6} & \textit{76.1} & \textit{75.8} & \textit{84.8} & \textit{80.1} \\
        \bottomrule
    \end{tabular}
    \label{tab:atypical_extraction_evaluation}
\end{table}

The accuracy of the classification of aspect sentences done in Step 2.1 can have a significant impact on the downstream extraction done in Step 2.2. Therefore, we first evaluated the performance of this step in terms of Precision (P), Recall (R), and F1-measure when using GPT-4 with either 0-shot prompting or in-context learning with fixed vs. dynamically selected 8-shot examples, as described in Section~\ref{sec:extrata-llm}. The results are shown in Table~\ref{tab:step2.1} and demonstrate that in-context learning leads to substantial improvements in performance. Furthermore, dynamic RAG-based selection of the 8-shot examples is the winner, with the exception of the {\sc Restaurants} domain, where fixed 8-shot performs on par with (primary) or better than (primary + secondary) the dynamic approach.

To evaluate the end-to-end performance of the entire atypical extraction pipeline, we compute the precision, recall, and $F_1$ scores for the {\bf exact} and {\bf partial matches} of the extracted base noun phrase (BNP) with the ground truth phrase. In the exact match method, an extracted phrase is considered correct if and only if it matches exactly a ground truth (gold) phrase. Precision (P), Recall (R), and $F_1$ measure are then computed as follows:
\begin{equation}
    P = \frac{\# \mbox{\it{correct extracted BNPs}}}{\# \mbox{\it{extracted BNPs}}}, \; \; \; R = \frac{\# \mbox{\it{correct extracted BNPs}}}{\# \mbox{\it{gold BNPs}}}, \; \; \;  F_1 = \frac{2 P R}{P + R} \nonumber
\end{equation}
In the partial match evaluation, partial credit is given to extracted phrases that overlap with ground truth phrases. For this, we use the greedy method shown in Algorithm~\ref{alg:cap} to compute a bipartite matching between gold phrases $gp$ and extracted phrases $ep$ that is aimed at maximizing their word overlap $|ep \cap gp|$ in total. The overlaps between extracted and gold phrases are then used to compute precision, recall, and $F_1$.

In Table~\ref{tab:atypical_extraction_evaluation}, we show the end-to-end evaluation of the entire atypical extraction pipeline (Steps 1 and 2) and contrast its performance with the fine-tuned Flan T5, which was the best performing model in the experiments reported in \cite{nannaware2023extraction}. The results show that even in the 0-shot setting, GPT-4 outperforms the fine-tuned Flan T5 model substantially. This is interesting because in \cite{nannaware2023extraction} Flan T5 outperformed GPT-3.5 by a large margin, be it in 0-shot or 5-shot ICL settings. Given that atypical item extraction generally does not require deep understanding of the text or sophisticated reasoning, we believe the much better performance of the GPT-4 model is partly explained by the decomposition into separate tasks in the two-step pipeline approach, compared to the original monolithic approach followed for GPT-3.5. The results presented in Appendix~\ref{app:extrata_comparison} show that, indeed, GPT-3.5 does much better when using the two-step approach, however it is still outperformed by GPT-4.

Lastly, similar to the aspect sentence classification results, the dynamic RAG-based approach for selecting the 8-shot in-context learning examples improves performance substantially, outperforming all methods across all evaluation scenarios. Therefore, we will use this best performing approach for extracting atypical aspects when computing the serendipity-based system rankings in Section~\ref{sec:reranking}.

Error analysis of the best performing method, GPT-4 (Dynamic 8-shot), reveals that sometimes the model incorrectly considers unusual-sounding typical aspects as atypical. For example, in the sentence, {\it "The restaurant offers a \textcolor{violet}{blue ribbon tasting experience}"}, the model incorrectly extracts the highlighted phrase as atypical, despite it being about the food service. The model also misclassifies one-time events as atypical features, such as the highlighted phrase in the sentence, {\it "During the grand opening, WDAS had set up \textcolor{violet}{speakers in front blasting music} and inviting folks to come in"}.



\section{MAPPING ATYPICAL ASPECTS TO USER-DEPENDENT UTILITY}
\label{sec:utility}

\begin{figure}[t]
    \centering
    \includegraphics[width=\linewidth]{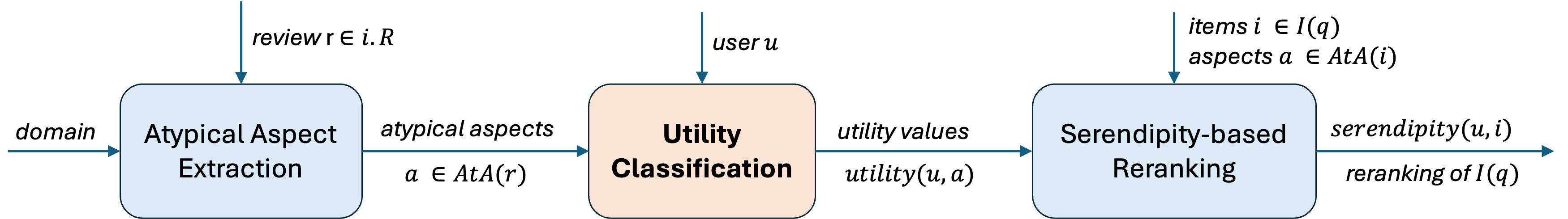}
    \caption{ATARS system architecture: Utility Classification.}
    \label{fig:pipeline2}
\end{figure}

The second step in the system architecture pipeline, as highlighted in Figure~\ref{fig:pipeline2}, estimates a user-dependent utility for each of the atypical aspects extracted at the first step. More formally, given a user profile $u$ and a review $r$ containing a set of atypical aspects $a \in AtA(r)$, the task is to assign each atypical aspect a utility value $utility(u, a)$ that reflects how useful or relevant that atypical aspect is for the user. With an eye towards both ease of manual annotation and discriminative power, we defined 4 levels of utility as $None = 0.0$, $Low = 0.5$, $Medium = 0.75$, and $High = 1.0$. With these 4 values, the utility estimation can be cast as a utility classification task.

The development and evaluation of the utility module requires 3 types of data: a set of user profile $u \in \mathcal{U}$, a set of reviews $r$ annotated with atypical aspects $a \in AtA(r)$, and a set of $\langle u, a \rangle$ pairs that are annotated with their $utility(u, a)$ values. Publicly available user profiles, however, are very rare, especially profiles that indicate a user's hobbies, passions, or lifestyles, as illustrated in Figure \ref{fig:atars-example}. Correspondingly, in Section~\ref{sec:profiles-dataset} we introduce an LLM-based approach for generating a diverse set of user profiles that can be used in combination with the dataset of reviews described in Section~\ref{sec:extrata}. Based on this dataset of user profile and the dataset of customer reviews, in Section~\ref{sec:utility-dataset} we introduce a crowd-sourced dataset of $utility(u, a)$ mappings. These mappings are then used to provide in-context learning examples for the LLM-based utility classification approach that will be introduced in Section~\ref{sec:utility-llm}, as well as an evaluation test bed for the experimental evaluations reported in Section~\ref{sec:utility-evaluation}.

\subsection{Generation of Synthetic User Profiles}
\label{sec:profiles-dataset}

To ensure that the synthetic user profiles sound as natural as possible, we first crafted 9 profiles in a narrative, biographical style, incorporating a range of personal attributes that include personal or family history, social connections, educational and professional background, hobbies, passions, lifestyle needs, accessibility requirements, and significant recent life events. We did not distinguish between primary and secondary atypical aspects, as both can have equal utility in triggering serendipity, as such we used both types when designing the personal attributes. Starting from the extractive versions of atypical aspects in the ExtrAtA dataset (Section~\ref{sec:extrata-dataset}), we sampled between one and five atypical aspects and integrated them into the personal attributes of a user profile either directly, as shown in the example at (\ref{item:explicitu}), or indirectly through related concepts, as shown in the example at (\ref{item:implicitu}) below:
\begin{enumerate}
    \item \label{item:explicitu} For the atypical aspect '\textcolor{teal}{yoga and meditation}', we crafted the profile '\textit{Jordan Lee is a cultural studies lecturer in the college, where he graduated with a major in anthropology. He grew up in Portland with his 2 other siblings, with whom he maintains a very close relationship. Jordan discovered \textcolor{teal}{yoga and meditation} while doing a study on South Asian cultures. It quickly evolved from a hobby to more of a lifestyle}. He enjoys spending time with his friends and family.'
    \begin{itemize}
        \item Jordan's profile \textit{explicitly} references '\textcolor{teal}{yoga and meditation}' as both a hobby and a notable aspect of his lifestyle. 
    \end{itemize}
    \item \label{item:implicitu} For the atypical aspect '\textcolor{teal}{locks of love}', I crafted the profile '\textit{Emma is a high school teacher from Austin, Texas, who has always prioritized community involvement and support. She teaches her students the importance of empathy and giving back by leading by example. Drawn to causes that make a tangible difference, Emma especially appreciates grassroots philanthropic activities.}' 
    \begin{itemize}
        \item The charity '\textcolor{teal}{locks of love}' helps children with rare diseases such as alopecia areata, loose anagen syndrome by providing custom-made hair prosthetics. Emma's affinity for the charity is implicit in her appreciation or passion for \textit{grassroots philanthropic activities}.
    \end{itemize}
\end{enumerate}
Appendix~\ref{app:profile-examples} contains the complete text of the 9 profiles, showing for each of them the list of input topics, the chain-of-thought rationale, and the output user profile that aligns with the topics. 

Handcrafting the 9 profiles proved to be a significantly time-consuming and cognitively demanding effort. To build a much larger dataset more efficiently, we determined to use the manually generated profiles as in-context learning examples for an LLM-based generation approach that generates user profiles starting from a given list of atypical aspects. Given the overall set if items $\mathcal{I}$ contained in the ExtrAtA dataset for each domain, their atypical aspects were first aggregated into an overall list $AtA(\mathcal{I}) = \bigcup_{i \in \mathcal{I}} AtA(i)$. Before generating a user profile $u$, a number of $n$ atypical aspects were selected at random from the list  $AtA(\mathcal{I})$, where $n \in \{1, 2, 3, 4, 5\}$ was sampled according to the categorical distribution $p(1) = 0.1$, $p(2) = 0.3$, $p(3) = 0.3$, $p(4) = 0.2$, and $p(5) = 0.1$. The $n$ sampled atypical aspects were then provided to GPT-4 to generate a user profile with personal attributes aligned to these aspects. Each of the 9 handcrafted user profiles was used as an in-context learning (ICL) example, augmented with chain-of-thought (CoT) reasoning illustrating the thought process behind a profile creation. The sampling temperature was set to 1.0 to enhance the diversity of the generated profiles while maintaining their quality. Table~\ref{tab:prompt_userprofilegeneration} shows the overall prompt template, which was used to generate user profiles separately for each domain. 
\begin{table}[t]
    \addtolength{\tabcolsep}{2pt}
    \centering
    \caption{User profiles in each of the 3 user profile datasets, across Train + Test and Development.}
    \begin{tabular}{@{}cccc@{}}
        \toprule
        \textbf{Dataset split} & {\sc Restaurant} & {\sc Hotel} & {\sc Hair Salon} \\
        \midrule
        Train $+$ Test & 100 & 50 & 50 \\
        Development & 10 & 10 & 10 \\
        \bottomrule
    \end{tabular}
\label{tab:updataset}
\end{table}

\begin{table}[t]
    \caption{Prompt for generating user profiles using in-context learning examples and CoT reasoning.} 
    \label{tab:prompt_userprofilegeneration}
    \small
    \centering
    \begin{tabular}{p{\textwidth}}
        \toprule
        \noindent\textbf{\textit{Instructions:}} \textit{Write a brief and focused biography for an individual that contains a combination of personal attributes such as personal or family history, social connections, educational and professional background, hobbies, passions, lifestyle needs, accessibility needs, and significant recent life events. Your responses must be all in one paragraph and written in plain, unembellished language. You will be given a topic or a comma-separated list of topics that is within angle brackets. The biography must coherently integrate the list of topics, explicitly as given, or implicitly through related concepts, or both, into its narrative, aligning them with the individual's hobbies, passions, lifestyle needs, or accessibility needs. In other words, by reading the biography, one should be able to infer that the individuals would find the topic interesting, appealing, or useful to them. It's important to note that some topics may be more effectively presented as hobbies or passions, while others might be better portrayed as objects or services that facilitate a particular lifestyle or meet accessibility needs.}\\
        
        \hspace{1em} \textit{A given topic may be expressed in the biography narrative as given, through a more general category, e.g., if the given topic is 'live indie music', it can be broadened to 'live music', or through a more specific category, e.g., if the given topic is 'pets', it can be narrowed to 'dogs'. If there are multiple topics with similar themes, we can express those through a more general category, e.g., if given topics are 'life size beer pong, pool table, shuffleboard', it can be broadened to 'indoor games' or express it through a specific category, e.g., if given topics are 'watercolors, local artist, liberal arts camp', it can be narrowed to 'portrait painting'. For example:}\\
        \textit{{\bf Examples 1 to 9}: The given topics are} $\langle$ {\tt input topic list} $\rangle$ \\
        \textit{Let's think step by step} $\langle$ {\tt CoT rationale} $\rangle$ \\
        \textit{So, a good biography would be}: $\langle$ {\tt output user profile} $\rangle$\\
        \ \\
        {\it Now, using the instructions and examples above, generate a user profile for the topics} $\langle$ {\tt input topic list} $\rangle$. \\
        \bottomrule
    \end{tabular}
\end{table}

Table~\ref{tab:updataset} shows the summary statistics of the three user profile datasets, one for each domain category, split between data used for training and testing, and data used for development.

\subsection{Dataset of Utility Value Annotations}
\label{sec:utility-dataset}

Atypical aspects, which are inherently surprising, have varying degrees of usefulness or utility for a user. For example, the fictional user Sheldon from Figures~\ref{fig:amt_interface-example} and~\ref{fig:atars-example} is likely be surprised to see a cafe or a restaurant featuring an origami making station in its waiting area; however, given the interests described in his profile, it is unlikely he will be particularly interested in this atypical aspect. In contrast, if he visited Julia's Cafe and Books, it is likely that he would be delighted to discover the \textit{ Batman Adventures issue from 96}, due to his being a fan of vintage comic books, particularly of Batman comics, as is \textit{directly} mentioned in his profile. Similarly, if he visited Undercurrent Coffee, he would likely appreciate the presence of \textit{autographed memorabilia from movies such The Dark Knight}, an interest that is \textit{indirectly} related to his interest in comic book movies which he reviews on his YouTube channel.

We refer to objects or activities stated in the profile that a user finds interesting, appealing, or useful, as {\it attractors}; these typically relate to the user's hobbies, passions, or lifestyle. As described in the annotation guidelines from Section~\ref{sec:utility-guidelines} below, determining the user-dependent utility of an atypical aspect is done by considering its relationship to semantically related attractors in the user profile.

\subsubsection{Utility Annotation Guidelines}
\label{sec:utility-guidelines}
\ \\
In order to balance the effort and reliability of manual annotation of utility values vs. their discriminative power, we chose to define 4 levels of utility $None = 0.0$, $Low = 0.5$, $Medium = 0.75$, and $High = 1.0$, as described below.\\

\noindent  The {\sc None} utility value is assigned to an atypical aspect in any of the cases below:
    \begin{enumerate}
        \item \label{item:noneone}The \textit{atypical aspect} lacks any relevance to the \textit{attractors} in the user profile.
        \item \label{item:nonetwo}The atypical aspect, despite appearing relevant, is too specific for the user's very broad interests or hobbies in the profile. For example, an atypical aspect such as \textit{live music}, is too specific for the very broad attractor creating a \textit{playful and family-friendly} environment.
        \item \label{item:nonethree}The atypical aspect involves materials or items that, despite appearing relevant, belong to a completely different context or are used for a completely different purpose than those in the user profile. For example: 
        \begin{itemize}
            \item[-] In the review the atypical aspect of \textcolor{teal}{high-end look} pertains to architecture or interior design, whereas the attractor \textit{high-end look} is about fashion choices.
            \item[-] If according to the profile a user is interested in \textit{modern plastic furniture} for home design, this does not necessarily translate to an interest in all plastic items. Therefore, the utility of an atypical aspect like \textcolor{teal}{plastic ponchos} that are sold by a restaurant should be annotated as None.
        \end{itemize}
    \end{enumerate}

\noindent The {\sc Low} utility value is assigned to an atypical aspect in any of the cases below:
    \begin{enumerate}
        \item \label{item:lowone}The attractor and the atypical aspect are distinct and specific concepts with only a small degree of conceptual similarity between them. For example, if a user is passionate about \textit{surfing}, the utility of an atypical aspect such as a \textcolor{teal}{sand sculpture} or a \textcolor{teal}{swimming pool} should be annotated as Low, as they have only a small degree of conceptual similarity to surfing.
        \item \label{item:lowtwo}The attractor and the atypical aspect are very broad concepts with a very limited conceptual overlap. For example, if a user has a general interest in \textit{visual aesthetics}, which can include \textit{photography}, \textit{painting}, and \textit{architectural design}, but the atypical aspect is \textcolor{teal}{art}, which can include \textcolor{teal}{handicrafts}, \textcolor{teal}{paintings}, and \textcolor{teal}{sculptures}, then even though there is some overlap between these two very broad concepts, the overlap is small, as such the utility of the atypical aspect should be annotated as Low.
    \end{enumerate}

\noindent The {\sc Medium} utility value is assigned to an atypical aspect in any of the cases below:
    \begin{enumerate}
        \item \label{item:mediumone}The attractor is a broad concept, while the atypical aspect is a more specific concept subsumed by that broad concept. For example, if a user is passionate about a broad topic such as \textit{interior design}, an atypical aspect like a \textcolor{teal}{lime green counter} would have Medium utility, as it may be very appealing to the user.
        \item \label{item:mediumtwo}Conversely, the attractor is a specific concept that is subsumed by the broader atypical aspect. For example, if a user enjoys listening to \textit{live pop music performances}, they might also be interested in listening to music in general, hence the atypical aspect \textcolor{teal}{music} would have Medium utility for the user.
    \end{enumerate}

\noindent The {\sc High} utility value is assigned to an atypical aspect if:
    \begin{enumerate}
        \item \label{item:high}The atypical aspect closely matches an attractor in the user profile. Both the attractor and the atypical aspect must involve the same specific context. For example, if the highlight of a user's year is seeing \textit{the Fourth of July fireworks display over the ocean}, then a restaurant that offers \textcolor{teal}{a vantage point over the ocean for watching fireworks} would have High utility. In this example, it was important that the fireworks happen in the same context, i.e. over the ocean.
    \end{enumerate}
In Appendix~\ref{app:utility-examples}, we provide detailed illustrative examples of utility annotations for each of the 4 utility values. Each example shows the user profile and the review containing the atypical aspect labeled with a utility value.

\subsubsection{Crowd-sourcing Utility Annotations with Amazon Mechanical Turk}
\label{sec:crowdsourcing}
\ \\
Using the annotation guidelines above, we designed a set of instructions and examples that were used for crowd-sourcing the ground truth utility annotations through the Amazon Mechanical Turk (MTurk) platform \cite{mturk}.  MTurk is a web service that enables a requester to create data annotation tasks, referred to as Human Intelligence Tasks (HITs), which workers then complete for a predetermined reward. For the utility annotation task, each HIT provided workers with a user profile and a review in which an atypical aspect was highlighted, and they were required to select the utility of the atypical aspect according to the provided guidelines. If a review contained more than one atypical aspect, workers could move between them in order to specify a utility value for each. The overall design of the MTurk user interface is detailed in Appendix~\ref{app:mturk}.

For each HIT, namely for each user profile, review and atypical aspect triplet $\langle u, r, a\rangle$, annotations were collected from 3 independent MTurk workers. For the experimental evaluations reported in Section~\ref{sec:utility-evaluation}, we use only HIT triplets where at least 2 annotators agreed on the utility value, e.g. accepted HITS. To estimate the inter-annotator agreement (ITA) on utility annotations, we report the mean, median and maximum standard deviation across each domain-specific dataset. Thus, for each HIT triplet $t_n = \langle u, r, a\rangle$, let \(v^n_k\) be the numerically encoded utility value provided by the $k^{th}$ annotator, with \(K = 3\) annotations per triplet. The mean $\hat{\mu}_n$ and standard deviation $\hat{\sigma}_n$ for the HIT $t_n$ are then computed as:
\begin{eqnarray}
   \hat{\mu}_n = \frac{1}{K} \sum_{k=1}^K v^n_k \quad \quad  \hat{\sigma}_n = \sqrt{\frac{1}{K} \sum_{k=1}^K (v^n_k - \hat{\mu}_n)^2} \label{eq:6}
\end{eqnarray}
An overall mean standard deviation \(\hat{\sigma}\) is then calculated over all $N$ HITS in the dataset, together with median $\sigma_{1/2}$ and maximum $\sigma_{\text{max}}$ values:
\begin{eqnarray}
    \hat{\sigma} = \frac{1}{N} \sum_{n=1}^N \hat{\sigma}_n \quad \quad \sigma_{1/2} = \med_{1 \leq n \leq N}  \hat{\sigma}_n \quad \quad \sigma_{\text{max}} = \max_{1 \leq n \leq N}  \hat{\sigma}_n \label{eq:7}
\end{eqnarray}
The inter-annotator agreement (ITA) for all HITS vs. accepted HITS is shown in Table~\ref{tab:mturk-statistics}. Overall, the highest agreement is obtained for the {\sc Restaurant} domain, where the median standard deviation in utility annotation drops below half a point (note that consecutive utility values are separated by one point).

\begin{table}[ht]
    \centering
    \caption{Statistics for All vs. Accepted HITs $\langle u,r,a\rangle$ by domain: total HIT counts and inter-annotator agreement (ITA) on utility annotations.}
    \begin{tabular}{l|cccc|cccc}
        \toprule
        \textbf{Domain} & \textbf{All} & $\hat{\sigma}$ & $\sigma_{1/2}$ & $\sigma_{\text{max}}$ & \textbf{Accepted} & $\hat{\sigma}$ & $\sigma_{1/2}$ & $\sigma_{\text{max}}$ \\
        \midrule
        Restaurants  & 2,770 & 0.68 & 0.47 & 1.41 & 2,105 & 0.58 & 0.47 & 1.41 \\
        Hotels       & 2,512 & 0.80 & 0.81 & 1.41 & 1,854 & 0.71 & 0.94 & 1.41 \\
        Hair Salons  & 1,092 & 0.78 & 0.81 & 1.41 & \ \ \ 811 & 0.68 & 0.47 & 1.41 \\
        \bottomrule
    \end{tabular}
    \label{tab:mturk-statistics}
\end{table}

\subsection{LLM-Based Estimation of User-Dependent Aspect Utility}
\label{sec:utility-llm}

For the task of utility classification, we experimented with the GPT-4 model using zero-shot prompting as well as in-context learning with either fixed or dynamically selected few-shot examples. The LLM was provided as input with both a user profile and a positive aspect sentence, where exactly one atypical aspect is tagged. The LLM was instructed to estimate the utility value of that atypical aspect relative to the user profile. To ensure a fair evaluation of the LLM-based approach, the same annotation guidelines that were provided to the MTurk workers were also incorporated in the prompt instructions, as detailed in the prompt template shown in Tables~\ref{tab:prompt_utilityclassification}.

\begin{table}[t]
    \caption{Prompt template for classifying the utility of an atypical aspect as None, Low, Medium, or High with respect to a user profile, using in-context learning.} 
    \label{tab:prompt_utilityclassification}
    \centering
    \footnotesize
    \begin{tabular}{p{\textwidth}}
        \toprule
        \noindent\textbf{\textit{Instructions:}} \textit{You will be given a user profile U and a restaurant review sentence R as input. First, you will read the profile U to identify objects or activities that the user finds interesting, appealing, or useful, henceforth referred to as 'attractors.' These typically relate to the user's hobbies, passions, or lifestyle.} \\
        \hspace{1em}\textit{Next, you will read the restaurant review sentence R, which mentions an aspect A that is atypical for a restaurant. An atypical aspect is marked between the tags <ata> and </ata> in the text of the review. Determine if the atypical aspect would be interesting, appealing, or useful to the user based on the profile U, i.e., whether the atypical aspect is relevant to the attractors identified in the user profile, by examining not only the atypical aspect itself but also its context within R and the context of the attractors in U. Ensure that you read the entire sentence that mentions the atypical aspect. For an atypical aspect, estimate its relevance as 'None,' 'Low,' 'Medium,' or 'High' as follows:}\\
        $\langle$ {\tt Utility annotation guidelines from Section~\ref{sec:utility-guidelines}} $\rangle$\\
        \hspace{1em}\textit{Process: To ensure accuracy in your relevance estimation, draw correlations conservatively and avoid drawing connections between attractors and atypical aspects that are not explicitly related. Cross-reference each relevance decision with the explicit guidelines provided, verifying that the chosen label adheres to the rules and is contextually appropriate concerning the atypical aspect in R and the attractors in U. Include the guideline applied verbatim in the explanation.}\\   
        \hspace{1em}\textit{The output should list the atypical aspect that you label with a 'None,' 'Low,' 'Medium,' or 'High' relevance value, together with an explanation, using the format below:}\\   
        \textit{A' = [("<atypical aspect>", "<label>")]}   
        \textit{Explanation: Provide a detailed explanation for the 'None,' 'Low,' 'Medium,' or 'High' relevance labels you assigned to the atypical aspects listed above and how you used the 'Process' to validate the label.}\\   
        \hspace{1em}\textit{Please be precise in your response and strictly follow the format given. Here are some examples below:}\\
        \textit{{\bf Examples 1 to 4}}: \\    
        \textit{U:} $\langle$ {\tt user profile} $\rangle$ \\   
        \textit{R:} $\langle$ {\tt review sentence with one atypical aspect tagged} $\rangle$ \\
        \textit{Output:} $\langle$ {\tt A' = ..} $\rangle$ \\
        \textit{Explanation:} $\langle$ {\tt reasoning for selecting utility for atypical aspect} $\rangle$\\
        \hspace{1em}\textit{Now, read the user profile U and the restaurant review sentence R with the tagged the atypical aspect below, and assign its relevance value, as explained earlier.}\\
        \textit{\textbf{U:}} $\langle$ {\tt user profile} $\rangle$ \\
        \textit{\textbf{R:}} $\langle$ {\tt review sentence with one atypical aspect tagged} $\rangle$\\
        \bottomrule
    \end{tabular}
\end{table}

For the in-context learning scenarios, we experimented with two different approaches for few-shot prompting:
\begin{enumerate}
    \item {\it Fixed} few-shot ICL with chain-of-thought (CoT) rationale.
    \item {\it Dynamic} few-shot ICL prompting with RAG.
\end{enumerate}

In the {\it Fixed} approach, each of the 4 few-shot examples is augmented with a step-by-step reasoning explanation of why a particular utility value was assigned to the input atypical aspect. Appendix~\ref{app:utilityclassificationcot-examples} provides the complete text of the 4 in-context examples, together with the CoT rationales.

\begin{figure}[t]
    \centering
    \includegraphics[width=\linewidth]{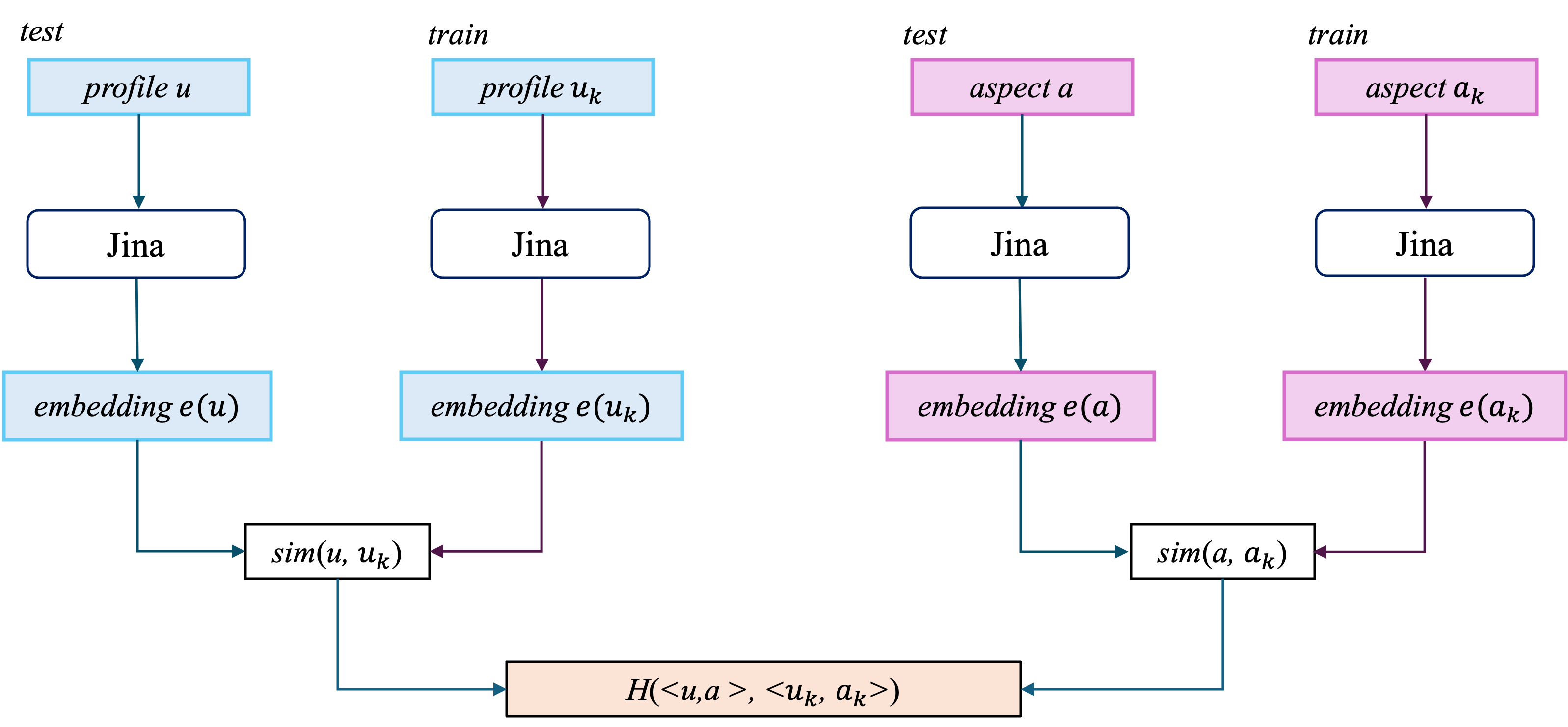}
    \caption{Computation of retrieval score for Dynamic selection of few-shot examples.}
    \label{fig:rag-architechture}
\end{figure}

The {\it Dynamic} approach uses a leave-one-out cross validation procedure, where given a test triplet \(\langle u, r, a \rangle\), all triplets that have the user or the item in common with the test example are removed from the training set. The few-shot ICL examples are then selected using the procedure illustrated in Figure~\ref{fig:rag-architechture}, based on running the following computation steps for each remaining example $\langle u_k, r_k, a_k \rangle$ in the training set:
\begin{enumerate}
    \item Compute the cosine similarity between the JINA embeddings $e(\cdot)$ of the user profiles:
    \[
    sim(u, u_k) = cos(e(u), e(u_k))
    \]
    \item Compute the cosine similarity between the JINA embeddings $e(\cdot)$ of the abstractive formulations of the atypical aspects:
    \[
    sim(a, a_k) = cos(e(a), e(a_k))
    \]
    \item Compute the harmonic mean of the user-level similarity and atypical aspect-level similarity:
    \[
    H(\langle u, a \rangle, \langle u_k, a_k \rangle) = \frac{2 \cdot sim(u, u_k) \cdot sim(a, a_k)}{sim(u, u_k) + sim(a, a_k)}
    \]
\end{enumerate}
Once the $H$ scores are computed for all training examples \(\langle u, r, a \rangle\), the 4 examples with the highest $H$ score are selected as few-shot ICL examples in the Dynamic setting.

\subsection{Experimental Evaluations}
\label{sec:utility-evaluation}

\begin{table}[t]
\addtolength{\tabcolsep}{-2pt}
    \centering
    \caption{Precision (P), Recall (R), and F1 performance (\%) of the LLM-based utility classification.}
    \begin{tabular}{l|cccc|c}
        \toprule
        \textbf{Domain} & \multicolumn{4}{c|}{\textbf{2-way (binary)}} & \textbf{4-way} \\
        \textbf{  Methods} & \textbf{Precision} & 
        \textbf{Recall} & 
        \textbf{F1} & 
        \textbf{Accuracy} &  \textbf{Accuracy} \\
       \midrule
        \multicolumn{1}{l}{\textsc{Restaurants}} & & & & \multicolumn{1}{l}{} & \\
        \ \ GPT-4 (0-shot)        & 88.0 & 74.8 & 80.8 & 79.8 & 66.5 \\
        \ \ GPT-4 (CoT 4-shot)    & 89.4 & 75.9 & 82.1 & 80.9 & 67.8 \\
        \ \ GPT-4 (Dynamic 4-shot)    & 89.1 & 76.8 & 82.5 & 80.8 & 67.4 \\[0.1cm]
        \multicolumn{1}{l}{\textsc{Hotels}} & & & & \multicolumn{1}{l}{} & \\
        \ \ GPT-4 (0-shot)        & 69.5 & 69.8 & 69.7 & 77.3 & 66.5 \\
        \ \ GPT-4 (CoT 4-shot)    & 68.7 & 76.5 & 72.4 & 78.0 & 66.5 \\
        \ \ GPT-4 (Dynamic 4-shot)    & 72.9 & 72.7 & 72.8 & 80.1 & 69.5 \\[0.15cm]
        \multicolumn{1}{l}{\textsc{Hair Salons}} & & & & \multicolumn{1}{l}{} &  \\
        \ \ GPT-4 (0-shot)        & 73.5 & 88.1 & 80.1 & 81.6 & 71.6 \\
        \ \ GPT-4 (CoT 4-shot)    & 78.5 & 88.0 & 83.0 & 85.3 & 75.2 \\
        \ \ GPT-4 (Dynamic 4-shot)    & 81.5 & 83.3 & 82.4 & 85.8 & 76.4 \\
        \bottomrule
    \end{tabular}
    \label{tab:utility_classification_evaluation}
\end{table}
We evaluate the LLM-based approaches to utility classification and report 4-way (all utility classes) and 2-way (two utility classes) performance measures. In the 4-way evaluation setting, all 4 utility classes are used as labels. Perfect matches such as <H,H>, <M,M>, <L,L>, and <N,N> incur a cost of 0, mismatches between adjacent labels such as <H,M>, <M, L>, or <L,N> incur a cost of 0.5, and all other mismatches incur a cost of 1.

For the 2-way (binary) evaluation setting, the High (H) and Medium (M) utility values are treated as {\it positive} labels, whereas the Low (L) and None (N) utility values are considered {\it negative} labels. Symmetric misclassifications cost are then assigned according to the distance between utility values: perfect matches such as <H,H>, <M,M>, <L,L>, and <N,N>, as well as adjacent matches within the same label group such as <H,M>, <L,N>, incur a 0 cost. Adjacent mismatches across the two label groups, such as <M,L>, are assigned a cost of 0.25, whereas distant mismatches, including <H,L>, <H,N>, and <M,N> are assigned a maximum cost of 1. 

Table~\ref{tab:utility_classification_evaluation} reports Precision (P), Recall (R), F1, and Accuracy for the 2-way and 4-way evaluation settings. Overall, the results show that in-context learning examples improve utility classification across all datasets, with Dynamic selection obtaining further improvements for the {\sc Hotels} and {\sc Hair Salons} domains.

Error analysis of the best-performing method, GPT-4 (Dynamic 8-shot), revealed that in some cases the LLM draws superficial correlations between the {\it attractors} mentioned in a user profile and the atypical aspect. For example, the model labeled the atypical aspect \textcolor{teal}{\it hand-carved wooden replicas of the native birds of Louisiana} with a High utility value for a user by spuriously connecting it to the attractor {\it wildlife photography} in the user profile.

\section{SERENDIPITY-BASED RE-RANKING OF ITEMS}
\label{sec:reranking}

\begin{figure}[t]
    \centering
    \includegraphics[width=\linewidth]{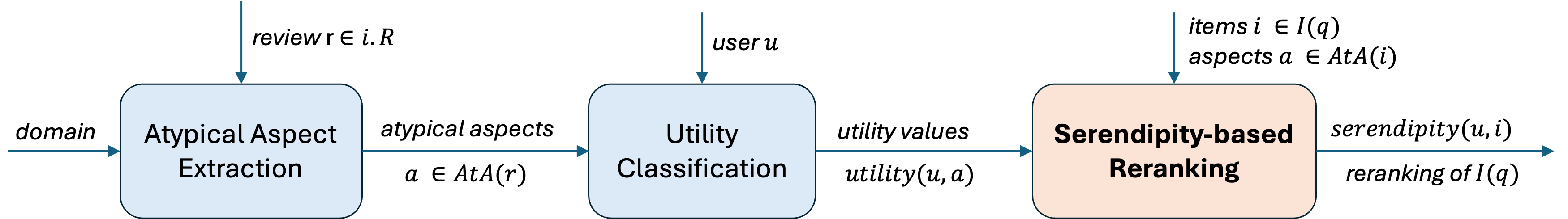}
    \caption{ATARS system architecture: Serendipity-based Re-ranking.}
    \label{fig:pipeline3}
\end{figure}

So far, we have described automatic procedures for the first two components of the system pipeline, as follows:
\begin{enumerate}
    \item Extraction of atypical aspects $a \in AtA(i)$ from the reviews $i.R$ of an item $i \in \mathcal{I}(q)$ that is relevant to a user query $q$ (Section~\ref{sec:extrata}).
    \item Classification of the $utility(a, u)$ of an atypical aspect $a$ for a user profile $u$ (Section~\ref{sec:utility}),
\end{enumerate}
Given the output of the two modules above, the final module of the system pipeline, as highlighted in Figure~\ref{fig:pipeline3}, performs a serendipity-based re-ranking of the items in $\mathcal{I}(q)$.

We define the user-dependent $serendipity(u, i)$ of an item $i$ as the sum of the utility values of all the atypical aspects found across all the reviews of that item, where each utility value is divided by the total similarity of that aspect with all atypical aspects, including itself:
\begin{equation}
    serendipity(u, i) = \sum_{a \in AtA(i)} \frac{utility(u, a)}{\displaystyle\sum_{\hat{a} \in AtA(i)} sim(a, \hat{a})}
    \label{eq:serendipity}
\end{equation}
Assuming a normalized aspect to aspect similarity measure $sim(a, \hat{a}) \in [0,1]$, with a value of 1 if and only if $\hat{a} = a$, this means that:
\begin{itemize}
    \item When all aspects are identical, each utility value in the sum above is divided by the total number of atypical aspects for that item, effectively producing a mean utility.
    \item When each aspect has zero similarity with other aspects, each utility value in the sum above is divided by 1, effectively producing a sum of utilities.
\end{itemize}
Note that by summing up user-dependent utilities for atypical aspects, which by definition have strong potential for surprise, the measure computed in Equation~\ref{eq:serendipity} captures the two core features of serendipity, namely surprise (through atypical aspects) and relevance (through utility).

We also define a user-independent measure of $surprise(i)$ for item $i$ by setting the utility value to $utility(u, a) = 1$ for all atypical aspects of that item in Equation~\ref{eq:serendipity}, which results in:
\begin{equation} 
    surprise(i) = \sum_{a\in AtA(i)} \frac{1}{\displaystyle\sum_{\hat{a}\in AtA(i)} sim(a, \hat{a})}
\end{equation} 
By setting the utility to a constant, the resulting surprise measure reflects only the aggregate surprise caused by all of the item's atypical aspects, and ignores their potential relevance to the user.

\subsection{Experimental Evaluations of End-to-End System Pipeline}
\label{sec:serendipity-evaluation}

In this section, we report end-to-end evaluations of the entire system pipeline. In Table~\ref{tab:sample_queries} we shows 5 sample user queries that were used for the re-ranking experiments for each of the 3 domains, where each query targets a specific category of item in that domain. Each query was associated with 10 user profiles, in order to simulate a scenario in which 10 users submit that query to the system, as illustrated in Figure~\ref{fig:atars-interaction} in the introduction. For each user, a ranking was computed over a list of items that matched the query category, of which at list 3 items $i$ had a positive user-dependent serendipity score $serendipity(u,i) > 0$ (for serendipity-based re-ranking) or a positive user-independent surprise score $surprise(i) > 0$ (for surprise-based re-ranking). Thus, for each ranking experiment, there are in total 3 {\it domains} $\times$ 5 {\it queries} $\times$ {\it 10 users} = 150 item rankings.

\begin{table}[t]
    \addtolength{\tabcolsep}{2pt}
    \centering
    \small
    \caption{Sample user queries mapped to Yelp categories.}
    \begin{tabular}{@{}l|l|l@{}}
        \toprule
        \textbf{Domain} & \textbf{Query} & \textbf{Category} \\
        \midrule
        \multirow{5}{*}{\scshape Restaurants}
           & $(q_1)$ Can you recommend a bar nearby? & Bars \\
           & $(q_2)$ Where are the best burgers in town? & Burgers \\
           & $(q_3)$ Suggest a cozy cafe to meet friends. & Cafes \\
           & $(q_4)$ Suggest a place with vibrant nightlife. & Nightlife \\
           & $(q_5)$ Can you recommend a classic American place. & American \\ 
        \cmidrule(lr){1-3}
        \multirow{5}{*}{\scshape Hotels}
           & $(q_1)$ Can you recommend a good hotel in the city? & Hotels \\
           & $(q_2)$ Suggest some good travel accommodations. & Hotels \& Travel \\
           & $(q_3)$ Where can I find a resort nearby? & Resorts \\
           & $(q_4)$ Suggest a spa for a relaxing weekend. & Beauty \& Spas\\
           & $(q_5)$ Planning an event, suggest a venue. & Event Planning \\ 
        \cmidrule(lr){1-3}
        \multirow{5}{*}{\scshape Hair Salons}
           & $(q_1)$ Can you recommend a hair salon for a haircut? & Hair Salons \\
           & $(q_2)$ Suggest a good barber shop for my son. & Barbers \\
           & $(q_3)$ Suggest a good hair stylist. & Hair Stylists \\
           & $(q_4)$ Suggest a makeup artist for a special occasion. & Makeup Artists \\
           & $(q_5)$ Where can I get hair extensions? & Hair Extensions \\
        \bottomrule
    \end{tabular}
    \label{tab:sample_queries}
\end{table}

The experimental evaluations compare the system rankings that leverage serendipity or surprise with the reference, ground truth rankings. The ground truth rankings use serendipity scores computed based on ground truth (manual) atypical aspects and ground truth (manual) utility scores. Depending on whether the final ranking uses serendipity on its own vs. combined with star ratings, we define two types of Ground Truth (GT) rankings:
\begin{itemize}
    \item {\bf GT$\circ$Seren}: Items are ranked solely based on their ground truth $serendipity(u, i)$ score, ignoring their star ratings.
    \item {\bf GT$\star$Seren}: Items are ranked based on both their ground truth $serendipity(u, i)$ score and their star ratings. The star rating is first partitioned into 4 ranges: \([4.0,5.0]\), \([3.0,4.0)\), \([2.0,3.0)\), and \([1.0,2.0)\). The serendipity scores are then used to re-rank items {\it only} inside each range, thus ensuring that items from a higher range, e.g., \([4.0,5.0]\), are always ranked above items from a lower range, e.g., \([3.0,4.0)\).
\end{itemize}

\begin{table}[t]
\addtolength{\tabcolsep}{-2pt}
    \centering
    \caption{Kendall $\tau$ correlations between System (Sys) rankings and the Ground Truth (GT) ranking GT$\circ$Seren, when ignoring star ratings.}
    \small
    \begin{tabular}{l|ccccc|c}
        \toprule
        {\sc Restaurants} & $q_1$ (bars) & $q_2$ (burgers) & $q_3$ (cafe) & $q_4$ (nightlife) & $q_5$ (american) & Mean \\
        \midrule
        Sys$\circ$Seren     & $\mathbf{+0.57}$ & $\mathbf{+0.77}$ & $\mathbf{+0.53}$ & $\mathbf{+0.66}$ & $\mathbf{+0.67}$ & $\mathbf{+0.64}$ \\
        Sys$\circ$Sur      & $+0.53$ & $+0.70$ & $+0.47$ & $+0.26$ & $+0.40$ & $+0.47$ \\
        GT$\star$         & $+0.36$ & $+0.16$ & $-0.06$ & $-0.26$ & $-0.42$ & $-0.04$ \\
        \midrule
        {\sc Hotels} & 
        $q_1$ (hotel) & 
        $q_2$ (travel) & 
        $q_3$ (resorts) & 
        $q_4$ (spas) & 
        $q_5$ (events) & 
        Mean \\
        \midrule
        Sys$\circ$Seren     & $\mathbf{+0.87}$ & $\mathbf{+0.87}$ & $\mathbf{+0.40}$ & $\mathbf{+0.60}$ & $\mathbf{+0.53}$ & $\mathbf{+0.65}$ \\
        Sys$\circ$Sur        & $+0.27$ & $+0.73$ & $+0.27$ & $+0.46$ & $+0.40$ & $+0.43$ \\
        GT$\star$          & $-0.40$ & $-0.10$ & $+0.20$ & $-0.44$ & $+0.29$ & $-0.09$ \\
        \midrule
        {\sc Hair Salons} & 
        $q_1$ (salon) & 
        $q_2$ (barber) & 
        $q_3$ (stylist) & 
        $q_4$ (makeup) & 
        $q_5$ (extensions) & 
        Mean \\
        \midrule
        Sys$\circ$Seren    & $\mathbf{+0.67}$ & $\mathbf{+0.67}$ & $\mathbf{+0.60}$ & $\mathbf{+0.46}$ & $\mathbf{+0.73}$ & $\mathbf{+0.63}$ \\
        Sys$\circ$Sur      & $+0.27$ & $+0.13$ & $+0.33$ & $+0.13$ & $+0.40$ & $+0.25$ \\
        GT$\star$       & $-0.08$ & $+0.03$ & $+0.09$ & $-0.15$ & $-0.52$ & $-0.13$ \\
        \bottomrule
    \end{tabular}
    \label{tab:serendipity_evaluation}
\end{table}

\begin{table}[t]
\addtolength{\tabcolsep}{-2pt}
    \centering
    \caption{Kendall $\tau$ correlations between System (Sys) rankings and the Ground Truth (GT) ranking GT$\star$Seren, when combining star ratings with surprise or serendipity.}
    \small
    \begin{tabular}{l|ccccc|c}
        \toprule
        {\sc Restaurants} & 
        $q_1$ (bars) & 
        $q_2$ (burgers) & 
        $q_3$ (cafe) & 
        $q_4$ (nightlife) & 
        $q_5$ (american) & 
        Mean \\
        \midrule   
        Sys$\star$Seren      & $\mathbf{+0.77}$ & $\mathbf{+0.93}$ & $\mathbf{+0.53}$ & $\mathbf{+0.87}$ & $\mathbf{+0.80}$ & $\mathbf{+0.78}$ \\
        Sys$\star$Sur       & $+0.70$ & $+0.87$ & $+0.40$ & $+0.80$ & $+0.60$ & $+0.67$ \\
        GT$\star$        & $+0.69$ & $+0.59$ & $+0.05$ & $+0.71$ & $+0.56$ & $+0.52$ \\
        \midrule
        {\sc Hotels} & 
        $q_1$ (hotel) & 
        $q_2$ (travel) & 
        $q_3$ (resorts) & 
        $q_4$ (spas) & 
        $q_5$ (events) & 
        Mean \\
        \midrule
        Sys$\star$Seren     & $\mathbf{+0.80}$ & $\mathbf{+0.93}$ & $\mathbf{+0.67}$ & $\mathbf{+0.87}$ & $\mathbf{+0.93}$ & $\mathbf{+0.84}$ \\
        Sys$\star$Sur       & $+0.53$ & $+0.87$ & $+0.53$ & $+0.60$ & $+0.93$ & $+0.69$ \\
        GT$\star$         & $+0.42$ & $+0.77$ & $+0.49$ & $+0.55$ & $+0.79$ & $+0.60$ \\
        \midrule
        {\sc Hair Salons} &
        $q_1$ (salon) & 
        $q_2$ (barber) & 
        $q_3$ (stylist) & 
        $q_4$ (makeup) & 
        $q_5$ (extensions) & 
        Mean \\
        \midrule
        Sys$\star$Seren      & $\mathbf{+0.80}$ & $\mathbf{+0.80}$ & $\mathbf{+0.60}$ & $\mathbf{+0.46}$ & $\mathbf{+0.73}$ & $\mathbf{+0.68}$ \\
        Sys$\star$Sur        & $+0.53$ & $+0.47$ & $+0.40$ & $+0.13$ & $+0.40$ & $+0.38$ \\
        GT$\star$          & $+0.16$ & $+0.26$ & $+0.32$ & $-0.15$ & $-0.21$ & $+0.08$ \\
        \bottomrule
    \end{tabular}
    \label{tab:star_serendipity_evaluation}
\end{table}

For all the system rankings, we use the Dynamic 8-shot approach for extracting atypical aspects described in Section~\ref{sec:extrata-llm}, whereas the user-dependent utility values are estimated using the Dynamic 4-shot approach described in Section~\ref{sec:utility-llm}. We compute system rankings using either $serendipity(u, i)$ or $surprise(i)$ for re-ranking, and similar to the ground truth scenario we report results for two types of System (Sys) rankings:
\begin{itemize}
    \item {\bf Sys$\circ$Seren} or {\bf Sys$\circ$Sur}: Items are ranked solely based on their system $serendipity(u, i)$ or $surprise(i)$ score, respectively, ignoring their star ratings.
    \item {\bf Sys$\star$Seren} or {\bf Sys$\star$Sur}: Items are ranked based on both their star ratings and their system $serendipity(u, i)$ or $surprise(i)$ score, respectively.
\end{itemize}

Table~\ref{tab:serendipity_evaluation} shows the Kendall $\tau$ correlations between the system generated rankings and the ground truth serendipity-based rankings, for each domain. Additionally, we also show the correlation between the original star-based ranking {\bf GT$\star$} and the ground truth serendipity-based ranking {\bf GT$\circ$Seren}. Analogous results are presented in Table~\ref{tab:star_serendipity_evaluation} for when star-based and serendipity-based or surprise-based scores are combined to produce the final ranking.

\subsection{Discussion of Results}

The results in Tables~\ref{tab:serendipity_evaluation} and~\ref{tab:star_serendipity_evaluation} show that rankings based solely on star ratings (GT$\star$) obtain a very low correlation with the ground truth serendipity-based rankings. Customers may provide high ratings to restaurants that lead to memorable experiences caused by surprising aspects, which means that theoretically star ratings may indirectly reflect the potential serendipity of an item. However, the low correlations between the star-based rankings and ground truth serendipity-based rankings show that the serendipity signal contained in star ratings is weak, and thus star ratings are insufficient for promoting items with high potential for serendipity. This demonstrates the importance of designing a ranking function that captures explicit, direct measures of surprise or serendipity.

As expected, the results also show that surprise-based system rankings (Sys$\circ$Sur and Sys$\star$Sur) are much less correlated with the ground truth serendipity-based rankings (GT$\circ$Seren and GT$\star$Seren) than the serendipity-based system rankings (Sys$\circ$Seren and Sys$\star$Seren), demonstrating the importance of estimating user-dependent utility and integrating it in the serendipity measure. The assumption behind the ground truth serendipity-based rankings is that users are not habituated with ATARS, i.e. they do not expect a recommender system that promotes items with atypical aspects that match their interests. However, as noted in Section~\ref{sec:introduction}, the serendipitous effect may diminish if the approach is employed repeatedly across multiple recommendation sessions. In such cases, one solution is to insert more randomness in the process, e.g. by randomly choosing between serendipity-based and surprise-based re-rankings at each recommendation session.

Finally, re-rankings based on a combination of star ratings with serendipity scores lead to higher correlations with the ground truth rankings, as observed by looking at the Mean results in Table~\ref{tab:star_serendipity_evaluation} vs. Table~\ref{tab:serendipity_evaluation}. This is expected, as including star rating information in both system and ground truth re-rankings provides a base level of correlation between the two.


\section{Conclusion and Future Work}
\label{sec:conclusion}

In this paper, we described a new method for engineering serendipity in recommender systems by leveraging items with atypical aspects that are relevant to the user. To this aim, we introduced an Atypical Aspect-based Recommender Systems (ATARS) that is implemented as a three-stage pipeline, where the first component processes item reviews in order to identify aspects that are atypical of the item's category, and thus have high potential for surprise. The second component estimates the user-dependent utility of an atypical aspect based on information in a user profile, whereas the third component integrates the utility values across all atypical aspects of an item into a serendipity score that is used to re-rank the list of items. We introduced datasets of customer reviews that were manually annotated with atypical aspects, and crowd-sourced datasets of utility annotations. These datasets were used to provide in-context learning (ICL) examples and evaluation benchmarks for the corresponding components. In particular, we introduced a new LLM-based approach for extracting domain-atypical aspects from customer reviews, which was shown to substantially improve over a previous fine-tuned approach. Similarly, we developed LLM-based approaches for mapping atypical aspects to user-dependent utility scores, ensuring that aspects are not only surprising, but also relevant to an individual's hobbies and passions. A dynamic approach for selection of ICL examples was introduced and shown to improve performance. Extensive experimental evaluations of the system-generated rankings based on surprise or serendipity scores show high correlations with ground-truth rankings where serendipity is computed from manual annotations of atypical aspects and utility values. 
Results also demonstrate that traditional star ratings are insufficient to promote serendipitous recommendations, thus justifying the need for the explicit modeling of surprise and utility.

In future work, we plan to explore methods for maintaining the serendipity effect across recommendation sessions that employ randomness in the serendipity generation mechanisms, such as randomly selecting between user-dependent serendipity measures and user-independent surprise measures for re-ranking.

\section*{Acknowledgments}

This research project has benefited from the Microsoft Accelerating Foundation Models Research (AFMR) grant program.

\bibliographystyle{ACM-Reference-Format}
\bibliography{tors25}


\begin{thebibliography}{54}


\ifx \showCODEN    \undefined \def \showCODEN     #1{\unskip}     \fi
\ifx \showISBNx    \undefined \def \showISBNx     #1{\unskip}     \fi
\ifx \showISBNxiii \undefined \def \showISBNxiii  #1{\unskip}     \fi
\ifx \showISSN     \undefined \def \showISSN      #1{\unskip}     \fi
\ifx \showLCCN     \undefined \def \showLCCN      #1{\unskip}     \fi
\ifx \shownote     \undefined \def \shownote      #1{#1}          \fi
\ifx \showarticletitle \undefined \def \showarticletitle #1{#1}   \fi
\ifx \showURL      \undefined \def \showURL       {\relax}        \fi
\providecommand\bibfield[2]{#2}
\providecommand\bibinfo[2]{#2}
\providecommand\natexlab[1]{#1}
\providecommand\showeprint[2][]{arXiv:#2}

\bibitem[Berlyne(1973)]%
        {berlyne1973aesthetics}
\bibfield{author}{\bibinfo{person}{Daniel~E Berlyne}.} \bibinfo{year}{1973}\natexlab{}.
\newblock \showarticletitle{Aesthetics and psychobiology}.
\newblock \bibinfo{journal}{\emph{Journal of Aesthetics and Art Criticism}} \bibinfo{volume}{31}, \bibinfo{number}{4} (\bibinfo{year}{1973}).
\newblock


\bibitem[Bollen et~al\mbox{.}(2010)]%
        {bollen_overchoice_2010}
\bibfield{author}{\bibinfo{person}{Dirk Bollen}, \bibinfo{person}{Bart~P. Knijnenburg}, \bibinfo{person}{Martijn~C. Willemsen}, {and} \bibinfo{person}{Mark Graus}.} \bibinfo{year}{2010}\natexlab{}.
\newblock \showarticletitle{Understanding Choice Overload in Recommender Systems}. In \bibinfo{booktitle}{\emph{Proceedings of the Fourth ACM Conference on Recommender Systems}} (Barcelona, Spain) \emph{(\bibinfo{series}{RecSys '10})}. \bibinfo{publisher}{Association for Computing Machinery}, \bibinfo{address}{New York, NY, USA}, \bibinfo{pages}{63--70}.
\newblock
\showISBNx{9781605589060}
\href{https://doi.org/10.1145/1864708.1864724}{doi:\nolinkurl{10.1145/1864708.1864724}}


\bibitem[Carlson and Pelletier(1995)]%
        {carlson_generic_1995}
\bibfield{author}{\bibinfo{person}{G.N. Carlson} {and} \bibinfo{person}{F.J. Pelletier}.} \bibinfo{year}{1995}\natexlab{}.
\newblock \bibinfo{booktitle}{\emph{The {Generic} {Book}}}.
\newblock \bibinfo{publisher}{University of Chicago Press}.
\newblock
\showISBNx{978-0-226-09292-8}
\showLCCN{94032400}


\bibitem[Chernev et~al\mbox{.}(2015)]%
        {CHERNEV2015333}
\bibfield{author}{\bibinfo{person}{Alexander Chernev}, \bibinfo{person}{Ulf Bockenholt}, {and} \bibinfo{person}{Joseph Goodman}.} \bibinfo{year}{2015}\natexlab{}.
\newblock \showarticletitle{Choice overload: A conceptual review and meta-analysis}.
\newblock \bibinfo{journal}{\emph{Journal of Consumer Psychology}} \bibinfo{volume}{25}, \bibinfo{number}{2} (\bibinfo{year}{2015}), \bibinfo{pages}{333--358}.
\newblock
\showISSN{1057-7408}
\href{https://doi.org/10.1016/j.jcps.2014.08.002}{doi:\nolinkurl{10.1016/j.jcps.2014.08.002}}


\bibitem[Chung et~al\mbox{.}(2022)]%
        {chung2022scaling}
\bibfield{author}{\bibinfo{person}{Hyung~Won Chung}, \bibinfo{person}{Le Hou}, \bibinfo{person}{Shayne Longpre}, \bibinfo{person}{Barret Zoph}, \bibinfo{person}{Yi Tay}, \bibinfo{person}{William Fedus}, \bibinfo{person}{Yunxuan Li}, \bibinfo{person}{Xuezhi Wang}, \bibinfo{person}{Mostafa Dehghani}, \bibinfo{person}{Siddhartha Brahma}, \bibinfo{person}{Albert Webson}, \bibinfo{person}{Shixiang~Shane Gu}, \bibinfo{person}{Zhuyun Dai}, \bibinfo{person}{Mirac Suzgun}, \bibinfo{person}{Xinyun Chen}, \bibinfo{person}{Aakanksha Chowdhery}, \bibinfo{person}{Alex Castro-Ros}, \bibinfo{person}{Marie Pellat}, \bibinfo{person}{Kevin Robinson}, \bibinfo{person}{Dasha Valter}, \bibinfo{person}{Sharan Narang}, \bibinfo{person}{Gaurav Mishra}, \bibinfo{person}{Adams Yu}, \bibinfo{person}{Vincent Zhao}, \bibinfo{person}{Yanping Huang}, \bibinfo{person}{Andrew Dai}, \bibinfo{person}{Hongkun Yu}, \bibinfo{person}{Slav Petrov}, \bibinfo{person}{Ed~H. Chi}, \bibinfo{person}{Jeff Dean}, \bibinfo{person}{Jacob Devlin},
  \bibinfo{person}{Adam Roberts}, \bibinfo{person}{Denny Zhou}, \bibinfo{person}{Quoc~V. Le}, {and} \bibinfo{person}{Jason Wei}.} \bibinfo{year}{2022}\natexlab{}.
\newblock \bibinfo{title}{Scaling Instruction-Finetuned Language Models}.
\newblock
\showeprint[arxiv]{2210.11416}~[cs.LG]


\bibitem[Crowston(2012)]%
        {mturk}
\bibfield{author}{\bibinfo{person}{Kevin Crowston}.} \bibinfo{year}{2012}\natexlab{}.
\newblock \showarticletitle{Amazon Mechanical Turk: A Research Tool for Organizations and Information Systems Scholars}. In \bibinfo{booktitle}{\emph{Shaping the Future of ICT Research. Methods and Approaches}}, \bibfield{editor}{\bibinfo{person}{Anol Bhattacherjee} {and} \bibinfo{person}{Brian Fitzgerald}} (Eds.). \bibinfo{publisher}{Springer Berlin Heidelberg}, \bibinfo{address}{Berlin, Heidelberg}, \bibinfo{pages}{210--221}.
\newblock
\showISBNx{978-3-642-35142-6}


\bibitem[Fan and Niu(2018)]%
        {niu_serendipity_2018}
\bibfield{author}{\bibinfo{person}{Xiangyu Fan} {and} \bibinfo{person}{Xi Niu}.} \bibinfo{year}{2018}\natexlab{}.
\newblock \showarticletitle{Implementing and Evaluating Serendipity in Delivering Personalized Health Information}.
\newblock \bibinfo{journal}{\emph{ACM Trans. Manage. Inf. Syst.}} \bibinfo{volume}{9}, \bibinfo{number}{2}, Article \bibinfo{articleno}{7} (\bibinfo{date}{aug} \bibinfo{year}{2018}), \bibinfo{numpages}{19}~pages.
\newblock
\showISSN{2158-656X}
\href{https://doi.org/10.1145/3205849}{doi:\nolinkurl{10.1145/3205849}}


\bibitem[Fu and Niu(2024)]%
        {niu_serenprompt_2024}
\bibfield{author}{\bibinfo{person}{Zhe Fu} {and} \bibinfo{person}{Xi Niu}.} \bibinfo{year}{2024}\natexlab{}.
\newblock \showarticletitle{The Art of Asking: Prompting Large Language Models for Serendipity Recommendations}. In \bibinfo{booktitle}{\emph{Proceedings of the 2024 ACM SIGIR International Conference on Theory of Information Retrieval}} (Washington DC, USA) \emph{(\bibinfo{series}{ICTIR '24})}. \bibinfo{publisher}{Association for Computing Machinery}, \bibinfo{address}{New York, NY, USA}, \bibinfo{pages}{157--166}.
\newblock
\showISBNx{9798400706813}
\href{https://doi.org/10.1145/3664190.3672521}{doi:\nolinkurl{10.1145/3664190.3672521}}


\bibitem[Fu et~al\mbox{.}(2023)]%
        {xi_2023_serendipity}
\bibfield{author}{\bibinfo{person}{Zhe Fu}, \bibinfo{person}{Xi Niu}, {and} \bibinfo{person}{Li Yu}.} \bibinfo{year}{2023}\natexlab{}.
\newblock \showarticletitle{Wisdom of Crowds and Fine-Grained Learning for Serendipity Recommendations}. In \bibinfo{booktitle}{\emph{Proceedings of the 46th International ACM SIGIR Conference on Research and Development in Information Retrieval}} (Taipei, Taiwan) \emph{(\bibinfo{series}{SIGIR '23})}. \bibinfo{publisher}{Association for Computing Machinery}, \bibinfo{address}{New York, NY, USA}, \bibinfo{pages}{739--748}.
\newblock
\showISBNx{9781450394086}
\href{https://doi.org/10.1145/3539618.3591787}{doi:\nolinkurl{10.1145/3539618.3591787}}


\bibitem[G{\"u}nther et~al\mbox{.}(2023)]%
        {gunther-etal-2023-jina}
\bibfield{author}{\bibinfo{person}{Michael G{\"u}nther}, \bibinfo{person}{Louis Milliken}, \bibinfo{person}{Jonathan Geuter}, \bibinfo{person}{Georgios Mastrapas}, \bibinfo{person}{Bo Wang}, {and} \bibinfo{person}{Han Xiao}.} \bibinfo{year}{2023}\natexlab{}.
\newblock \showarticletitle{{J}ina Embeddings: A Novel Set of High-Performance Sentence Embedding Models}. In \bibinfo{booktitle}{\emph{Proceedings of the 3rd Workshop for Natural Language Processing Open Source Software (NLP-OSS 2023)}}, \bibfield{editor}{\bibinfo{person}{Liling Tan}, \bibinfo{person}{Dmitrijs Milajevs}, \bibinfo{person}{Geeticka Chauhan}, \bibinfo{person}{Jeremy Gwinnup}, {and} \bibinfo{person}{Elijah Rippeth}} (Eds.). \bibinfo{publisher}{Association for Computational Linguistics}, \bibinfo{address}{Singapore}, \bibinfo{pages}{8--18}.
\newblock
\href{https://doi.org/10.18653/v1/2023.nlposs-1.2}{doi:\nolinkurl{10.18653/v1/2023.nlposs-1.2}}


\bibitem[Hasan and Bunescu(2023)]%
        {tonmoy_bunescu_serendipity_2023}
\bibfield{author}{\bibinfo{person}{Tonmoy Hasan} {and} \bibinfo{person}{Razvan Bunescu}.} \bibinfo{year}{2023}\natexlab{}.
\newblock \showarticletitle{Topic-Level Bayesian Surprise and Serendipity for Recommender Systems}. In \bibinfo{booktitle}{\emph{Proceedings of the 17th ACM Conference on Recommender Systems}} (Singapore, Singapore) \emph{(\bibinfo{series}{RecSys '23})}. \bibinfo{publisher}{Association for Computing Machinery}, \bibinfo{address}{New York, NY, USA}, \bibinfo{pages}{933--939}.
\newblock
\showISBNx{9798400702419}
\href{https://doi.org/10.1145/3604915.3608851}{doi:\nolinkurl{10.1145/3604915.3608851}}


\bibitem[Horstmann(2015)]%
        {horstmann_surpriseattention_2015}
\bibfield{author}{\bibinfo{person}{Gernot Horstmann}.} \bibinfo{year}{2015}\natexlab{}.
\newblock \showarticletitle{The surprise–attention link: a review}.
\newblock \bibinfo{journal}{\emph{Annals of the New York Academy of Sciences}} \bibinfo{volume}{1339}, \bibinfo{number}{1} (\bibinfo{year}{2015}), \bibinfo{pages}{106--115}.
\newblock
\showISSN{1749-6632}
\href{https://doi.org/10.1111/nyas.12679}{doi:\nolinkurl{10.1111/nyas.12679}}
\newblock
\shownote{\_eprint: https://onlinelibrary.wiley.com/doi/pdf/10.1111/nyas.12679}.


\bibitem[Huber and Klein(1991)]%
        {huber_adapting_cutoffs_1991}
\bibfield{author}{\bibinfo{person}{Joel Huber} {and} \bibinfo{person}{Noreen~M Klein}.} \bibinfo{year}{1991}\natexlab{}.
\newblock \showarticletitle{{Adapting Cutoffs to the Choice Environment: The Effects of Attribute Correlation and Reliability}}.
\newblock \bibinfo{journal}{\emph{Journal of Consumer Research}} \bibinfo{volume}{18}, \bibinfo{number}{3} (\bibinfo{year}{1991}), \bibinfo{pages}{346--357}.
\newblock
\href{https://doi.org/10.1086/209264}{doi:\nolinkurl{10.1086/209264}}


\bibitem[Inbar et~al\mbox{.}(2011)]%
        {inbar_decision_2011}
\bibfield{author}{\bibinfo{person}{Yoel Inbar}, \bibinfo{person}{Simona Botti}, {and} \bibinfo{person}{Karlene Hanko}.} \bibinfo{year}{2011}\natexlab{}.
\newblock \showarticletitle{Decision speed and choice regret: {When} haste feels like waste}.
\newblock \bibinfo{journal}{\emph{Journal of Experimental Social Psychology}} \bibinfo{volume}{47}, \bibinfo{number}{3} (\bibinfo{year}{2011}), \bibinfo{pages}{533--540}.
\newblock
\showISSN{0022-1031}
\href{https://doi.org/10.1016/j.jesp.2011.01.011}{doi:\nolinkurl{10.1016/j.jesp.2011.01.011}}


\bibitem[Iyengar and Lepper(2001)]%
        {article2}
\bibfield{author}{\bibinfo{person}{Sheena Iyengar} {and} \bibinfo{person}{Mark Lepper}.} \bibinfo{year}{2001}\natexlab{}.
\newblock \showarticletitle{When Choice is Demotivating: Can One Desire Too Much of a Good Thing?}
\newblock \bibinfo{journal}{\emph{Journal of personality and social psychology}}  \bibinfo{volume}{79} (\bibinfo{date}{01} \bibinfo{year}{2001}), \bibinfo{pages}{995--1006}.
\newblock
\href{https://doi.org/10.1037/0022-3514.79.6.995}{doi:\nolinkurl{10.1037/0022-3514.79.6.995}}


\bibitem[Jakob et~al\mbox{.}(2009)]%
        {jakob_reviewrec_2009}
\bibfield{author}{\bibinfo{person}{Niklas Jakob}, \bibinfo{person}{Stefan~Hagen Weber}, \bibinfo{person}{Mark~Christoph M\"{u}ller}, {and} \bibinfo{person}{Iryna Gurevych}.} \bibinfo{year}{2009}\natexlab{}.
\newblock \showarticletitle{Beyond the Stars: Exploiting Free-Text User Reviews to Improve the Accuracy of Movie Recommendations}. In \bibinfo{booktitle}{\emph{Proceedings of the 1st International CIKM Workshop on Topic-Sentiment Analysis for Mass Opinion}} (Hong Kong, China) \emph{(\bibinfo{series}{TSA '09})}. \bibinfo{publisher}{Association for Computing Machinery}, \bibinfo{address}{New York, NY, USA}, \bibinfo{pages}{57--64}.
\newblock
\showISBNx{9781605588056}
\href{https://doi.org/10.1145/1651461.1651473}{doi:\nolinkurl{10.1145/1651461.1651473}}


\bibitem[Jing(2002)]%
        {jing-2002-using}
\bibfield{author}{\bibinfo{person}{Hongyan Jing}.} \bibinfo{year}{2002}\natexlab{}.
\newblock \showarticletitle{Using Hidden {M}arkov Modeling to Decompose Human-Written Summaries}.
\newblock \bibinfo{journal}{\emph{Computational Linguistics}} \bibinfo{volume}{28}, \bibinfo{number}{4} (\bibinfo{year}{2002}), \bibinfo{pages}{527--543}.
\newblock
\href{https://doi.org/10.1162/089120102762671972}{doi:\nolinkurl{10.1162/089120102762671972}}


\bibitem[Jing and McKeown(1999)]%
        {hongyan:sigir99}
\bibfield{author}{\bibinfo{person}{Hongyan Jing} {and} \bibinfo{person}{Kathleen~R. McKeown}.} \bibinfo{year}{1999}\natexlab{}.
\newblock \showarticletitle{The Decomposition of Human-Written Summary Sentences}. In \bibinfo{booktitle}{\emph{Proceedings of the 22nd Annual International ACM SIGIR Conference on Research and Development in Information Retrieval}} (Berkeley, California, USA) \emph{(\bibinfo{series}{SIGIR '99})}. \bibinfo{publisher}{Association for Computing Machinery}, \bibinfo{address}{New York, NY, USA}, \bibinfo{pages}{129–136}.
\newblock
\showISBNx{1581130961}
\href{https://doi.org/10.1145/312624.312666}{doi:\nolinkurl{10.1145/312624.312666}}


\bibitem[Kaimann et~al\mbox{.}(2018)]%
        {kaiman:mde18}
\bibfield{author}{\bibinfo{person}{Daniel Kaimann}, \bibinfo{person}{Nadja Stroh-Maraun}, {and} \bibinfo{person}{Joe Cox}.} \bibinfo{year}{2018}\natexlab{}.
\newblock \showarticletitle{Variety in the video game industry: An empirical study of the Wundt curve}.
\newblock \bibinfo{journal}{\emph{Managerial and Decision Economics}} \bibinfo{volume}{39}, \bibinfo{number}{3} (\bibinfo{year}{2018}), \bibinfo{pages}{354--362}.
\newblock
\href{https://doi.org/10.1002/mde.2909}{doi:\nolinkurl{10.1002/mde.2909}}


\bibitem[Kim et~al\mbox{.}(2021)]%
        {kim_serendipity_2021}
\bibfield{author}{\bibinfo{person}{Aekyoung Kim}, \bibinfo{person}{Felipe~M. Affonso}, \bibinfo{person}{Juliano Laran}, {and} \bibinfo{person}{Kristina~M. Durante}.} \bibinfo{year}{2021}\natexlab{}.
\newblock \showarticletitle{Serendipity: {Chance} {Encounters} in the {Marketplace} {Enhance} {Consumer} {Satisfaction}}.
\newblock \bibinfo{journal}{\emph{Journal of Marketing}} \bibinfo{volume}{85}, \bibinfo{number}{4} (\bibinfo{date}{July} \bibinfo{year}{2021}), \bibinfo{pages}{141--157}.
\newblock
\showISSN{0022-2429}
\href{https://doi.org/10.1177/00222429211000344}{doi:\nolinkurl{10.1177/00222429211000344}}
\newblock
\shownote{Publisher: SAGE Publications Inc}.


\bibitem[Knight and Marcu(2002)]%
        {knight_summarization_2002}
\bibfield{author}{\bibinfo{person}{Kevin Knight} {and} \bibinfo{person}{Daniel Marcu}.} \bibinfo{year}{2002}\natexlab{}.
\newblock \showarticletitle{Summarization beyond sentence extraction: {A} probabilistic approach to sentence compression}.
\newblock \bibinfo{journal}{\emph{Artificial Intelligence}} \bibinfo{volume}{139}, \bibinfo{number}{1} (\bibinfo{year}{2002}), \bibinfo{pages}{91--107}.
\newblock
\showISSN{0004-3702}
\href{https://doi.org/10.1016/S0004-3702(02)00222-9}{doi:\nolinkurl{10.1016/S0004-3702(02)00222-9}}


\bibitem[Kornish and Lynch(2003)]%
        {diehl_smart_agents_2003}
\bibfield{author}{\bibinfo{person}{Laura Kornish} {and} \bibinfo{person}{John Lynch}.} \bibinfo{year}{2003}\natexlab{}.
\newblock \showarticletitle{Smart Agents: When Lower Search Costs for Quality Information Increase Price Sensitivity}.
\newblock \bibinfo{journal}{\emph{Journal of Consumer Research}}  \bibinfo{volume}{30} (\bibinfo{date}{06} \bibinfo{year}{2003}), \bibinfo{pages}{56--71}.
\newblock
\href{https://doi.org/10.2139/ssrn.340040}{doi:\nolinkurl{10.2139/ssrn.340040}}


\bibitem[Kotkov et~al\mbox{.}(2018)]%
        {kotkov_serDataset_2018}
\bibfield{author}{\bibinfo{person}{Denis Kotkov}, \bibinfo{person}{Joseph~A. Konstan}, \bibinfo{person}{Qian Zhao}, {and} \bibinfo{person}{Jari Veijalainen}.} \bibinfo{year}{2018}\natexlab{}.
\newblock \showarticletitle{Investigating Serendipity in Recommender Systems Based on Real User Feedback}. In \bibinfo{booktitle}{\emph{Proceedings of the 33rd Annual ACM Symposium on Applied Computing}} (Pau, France) \emph{(\bibinfo{series}{SAC '18})}. \bibinfo{publisher}{Association for Computing Machinery}, \bibinfo{address}{New York, NY, USA}, \bibinfo{pages}{1341--1350}.
\newblock
\showISBNx{9781450351911}
\href{https://doi.org/10.1145/3167132.3167276}{doi:\nolinkurl{10.1145/3167132.3167276}}


\bibitem[Kotkov et~al\mbox{.}(2016)]%
        {kotkov_serendipity_survey_2016}
\bibfield{author}{\bibinfo{person}{Denis Kotkov}, \bibinfo{person}{Shuaiqiang Wang}, {and} \bibinfo{person}{Jari Veijalainen}.} \bibinfo{year}{2016}\natexlab{}.
\newblock \showarticletitle{A survey of serendipity in recommender systems}.
\newblock \bibinfo{journal}{\emph{Know.-Based Syst.}} \bibinfo{volume}{111}, \bibinfo{number}{C} (\bibinfo{date}{Nov.} \bibinfo{year}{2016}), \bibinfo{pages}{180--192}.
\newblock
\showISSN{0950-7051}
\href{https://doi.org/10.1016/j.knosys.2016.08.014}{doi:\nolinkurl{10.1016/j.knosys.2016.08.014}}


\bibitem[Li et~al\mbox{.}(2020b)]%
        {li_unexrec_2020}
\bibfield{author}{\bibinfo{person}{Pan Li}, \bibinfo{person}{Maofei Que}, \bibinfo{person}{Zhichao Jiang}, \bibinfo{person}{Yao Hu}, {and} \bibinfo{person}{Alexander Tuzhilin}.} \bibinfo{year}{2020}\natexlab{b}.
\newblock \showarticletitle{PURS: Personalized Unexpected Recommender System for Improving User Satisfaction}. In \bibinfo{booktitle}{\emph{Proceedings of the 14th ACM Conference on Recommender Systems}} (Virtual Event, Brazil) \emph{(\bibinfo{series}{RecSys '20})}. \bibinfo{publisher}{Association for Computing Machinery}, \bibinfo{address}{New York, NY, USA}, \bibinfo{pages}{279--288}.
\newblock
\showISBNx{9781450375832}
\href{https://doi.org/10.1145/3383313.3412238}{doi:\nolinkurl{10.1145/3383313.3412238}}


\bibitem[Li and Tuzhilin(2022)]%
        {li_latentfeat_2022}
\bibfield{author}{\bibinfo{person}{Pan Li} {and} \bibinfo{person}{Alexander Tuzhilin}.} \bibinfo{year}{2022}\natexlab{}.
\newblock \showarticletitle{Learning Latent Multi-Criteria Ratings From User Reviews for Recommendations}.
\newblock \bibinfo{journal}{\emph{IEEE Transactions on Knowledge and Data Engineering}} \bibinfo{volume}{34}, \bibinfo{number}{8} (\bibinfo{year}{2022}), \bibinfo{pages}{3854--3866}.
\newblock
\href{https://doi.org/10.1109/TKDE.2020.3030623}{doi:\nolinkurl{10.1109/TKDE.2020.3030623}}


\bibitem[Li et~al\mbox{.}(2019)]%
        {li_elastic_serendipity_2019}
\bibfield{author}{\bibinfo{person}{Xueqi Li}, \bibinfo{person}{Wenjun Jiang}, \bibinfo{person}{Weiguang Chen}, \bibinfo{person}{Jie Wu}, {and} \bibinfo{person}{Guojun Wang}.} \bibinfo{year}{2019}\natexlab{}.
\newblock \showarticletitle{HAES: A New Hybrid Approach for Movie Recommendation with Elastic Serendipity}. \bibinfo{pages}{1503--1512}.
\newblock
\showISBNx{978-1-4503-6976-3}
\href{https://doi.org/10.1145/3357384.3357868}{doi:\nolinkurl{10.1145/3357384.3357868}}


\bibitem[Li et~al\mbox{.}(2020a)]%
        {li_desr_2020}
\bibfield{author}{\bibinfo{person}{Xueqi Li}, \bibinfo{person}{Wenjun Jiang}, \bibinfo{person}{Weiguang Chen}, \bibinfo{person}{Jie Wu}, \bibinfo{person}{Guojun Wang}, {and} \bibinfo{person}{Kenli Li}.} \bibinfo{year}{2020}\natexlab{a}.
\newblock \showarticletitle{Directional and Explainable Serendipity Recommendation}. In \bibinfo{booktitle}{\emph{Proceedings of The Web Conference 2020}} (Taipei, Taiwan) \emph{(\bibinfo{series}{WWW '20})}. \bibinfo{publisher}{Association for Computing Machinery}, \bibinfo{address}{New York, NY, USA}, \bibinfo{pages}{122--132}.
\newblock
\showISBNx{9781450370233}
\href{https://doi.org/10.1145/3366423.3380100}{doi:\nolinkurl{10.1145/3366423.3380100}}


\bibitem[Liu et~al\mbox{.}(2023)]%
        {liu_neural_reranking_rs_2023}
\bibfield{author}{\bibinfo{person}{Weiwen Liu}, \bibinfo{person}{Yunjia Xi}, \bibinfo{person}{Jiarui Qin}, \bibinfo{person}{Xinyi Dai}, \bibinfo{person}{Ruiming Tang}, \bibinfo{person}{Shuai Li}, \bibinfo{person}{Weinan Zhang}, {and} \bibinfo{person}{Rui Zhang}.} \bibinfo{year}{2023}\natexlab{}.
\newblock \showarticletitle{Personalized Diversification for Neural Re-ranking in Recommendation}. \bibinfo{pages}{802--815}.
\newblock
\href{https://doi.org/10.1109/ICDE55515.2023.00067}{doi:\nolinkurl{10.1109/ICDE55515.2023.00067}}


\bibitem[Makri et~al\mbox{.}(2014)]%
        {makri_making_2014}
\bibfield{author}{\bibinfo{person}{Stephann Makri}, \bibinfo{person}{Ann Blandford}, \bibinfo{person}{Mel Woods}, \bibinfo{person}{Sarah Sharples}, {and} \bibinfo{person}{Deborah Maxwell}.} \bibinfo{year}{2014}\natexlab{}.
\newblock \showarticletitle{“{Making} my own luck”: {Serendipity} strategies and how to support them in digital information environments}.
\newblock \bibinfo{journal}{\emph{Journal of the Association for Information Science and Technology}} \bibinfo{volume}{65}, \bibinfo{number}{11} (\bibinfo{year}{2014}), \bibinfo{pages}{2179--2194}.
\newblock
\showISSN{2330-1643}
\href{https://doi.org/10.1002/asi.23200}{doi:\nolinkurl{10.1002/asi.23200}}
\newblock
\shownote{\_eprint: https://onlinelibrary.wiley.com/doi/pdf/10.1002/asi.23200}.


\bibitem[McGrew(1970)]%
        {doi:10.1177/019263657005434912}
\bibfield{author}{\bibinfo{person}{Jean~B. McGrew}.} \bibinfo{year}{1970}\natexlab{}.
\newblock \showarticletitle{Future Shock. Alvin Toffler. New York: Random House, 1970. 505 pp. \$7.95}.
\newblock \bibinfo{journal}{\emph{The bulletin of the National Association of Secondary School Principals}} \bibinfo{volume}{54}, \bibinfo{number}{349} (\bibinfo{year}{1970}), \bibinfo{pages}{123--129}.
\newblock
\href{https://doi.org/10.1177/019263657005434912}{doi:\nolinkurl{10.1177/019263657005434912}}
\showeprint{https://doi.org/10.1177/019263657005434912}


\bibitem[Musto et~al\mbox{.}(2017)]%
        {musto_absarec_2017}
\bibfield{author}{\bibinfo{person}{Cataldo Musto}, \bibinfo{person}{Marco de Gemmis}, \bibinfo{person}{Giovanni Semeraro}, {and} \bibinfo{person}{Pasquale Lops}.} \bibinfo{year}{2017}\natexlab{}.
\newblock \showarticletitle{A Multi-Criteria Recommender System Exploiting Aspect-Based Sentiment Analysis of Users' Reviews}. In \bibinfo{booktitle}{\emph{Proceedings of the Eleventh ACM Conference on Recommender Systems}} (Como, Italy) \emph{(\bibinfo{series}{RecSys '17})}. \bibinfo{publisher}{Association for Computing Machinery}, \bibinfo{address}{New York, NY, USA}, \bibinfo{pages}{321--325}.
\newblock
\showISBNx{9781450346528}
\href{https://doi.org/10.1145/3109859.3109905}{doi:\nolinkurl{10.1145/3109859.3109905}}


\bibitem[Nannaware et~al\mbox{.}(2023)]%
        {nannaware2023extraction}
\bibfield{author}{\bibinfo{person}{Smita Nannaware}, \bibinfo{person}{Erfan Al-Hossami}, {and} \bibinfo{person}{Razvan Bunescu}.} \bibinfo{year}{2023}\natexlab{}.
\newblock \bibinfo{title}{Extraction of Atypical Aspects from Customer Reviews: Datasets and Experiments with Language Models}.
\newblock
\showeprint[arxiv]{2311.02702}~[cs.CL]


\bibitem[Niu et~al\mbox{.}(2018)]%
        {niu:acm18}
\bibfield{author}{\bibinfo{person}{Xi Niu}, \bibinfo{person}{Fakhri Abbas}, \bibinfo{person}{Mary~Lou Maher}, {and} \bibinfo{person}{Kazjon Grace}.} \bibinfo{year}{2018}\natexlab{}.
\newblock \showarticletitle{Surprise Me If You Can: Serendipity in Health Information}. In \bibinfo{booktitle}{\emph{Proceedings of the 2018 CHI Conference on Human Factors in Computing Systems}} (Montreal QC, Canada) \emph{(\bibinfo{series}{CHI '18})}. \bibinfo{publisher}{Association for Computing Machinery}, \bibinfo{address}{New York, NY, USA}, \bibinfo{pages}{1--12}.
\newblock
\showISBNx{9781450356206}
\href{https://doi.org/10.1145/3173574.3173597}{doi:\nolinkurl{10.1145/3173574.3173597}}


\bibitem[OpenAI(2022)]%
        {chatgpt}
\bibfield{author}{\bibinfo{person}{OpenAI}.} \bibinfo{year}{2022}\natexlab{}.
\newblock \bibinfo{title}{Introducing ChatGPT}.
\newblock
\urldef\tempurl%
\url{https://openai.com/blog/chatgpt}
\showURL{%
\tempurl}


\bibitem[Raffel et~al\mbox{.}(2020)]%
        {raffel2020exploring}
\bibfield{author}{\bibinfo{person}{Colin Raffel}, \bibinfo{person}{Noam Shazeer}, \bibinfo{person}{Adam Roberts}, \bibinfo{person}{Katherine Lee}, \bibinfo{person}{Sharan Narang}, \bibinfo{person}{Michael Matena}, \bibinfo{person}{Yanqi Zhou}, \bibinfo{person}{Wei Li}, {and} \bibinfo{person}{Peter~J Liu}.} \bibinfo{year}{2020}\natexlab{}.
\newblock \showarticletitle{Exploring the limits of transfer learning with a unified text-to-text transformer}.
\newblock \bibinfo{journal}{\emph{The Journal of Machine Learning Research}} \bibinfo{volume}{21}, \bibinfo{number}{1} (\bibinfo{year}{2020}), \bibinfo{pages}{5485--5551}.
\newblock


\bibitem[Reisenzein et~al\mbox{.}(2019)]%
        {reisenzein_cognitive-evolutionary_2019}
\bibfield{author}{\bibinfo{person}{Rainer Reisenzein}, \bibinfo{person}{Gernot Horstmann}, {and} \bibinfo{person}{Achim Schützwohl}.} \bibinfo{year}{2019}\natexlab{}.
\newblock \showarticletitle{The {Cognitive}-{Evolutionary} {Model} of {Surprise}: {A} {Review} of the {Evidence}}.
\newblock \bibinfo{journal}{\emph{Topics in Cognitive Science}} \bibinfo{volume}{11}, \bibinfo{number}{1} (\bibinfo{year}{2019}), \bibinfo{pages}{50--74}.
\newblock
\showISSN{1756-8765}
\href{https://doi.org/10.1111/tops.12292}{doi:\nolinkurl{10.1111/tops.12292}}
\newblock
\shownote{\_eprint: https://onlinelibrary.wiley.com/doi/pdf/10.1111/tops.12292}.


\bibitem[Rush et~al\mbox{.}(2015)]%
        {rush-etal-2015-neural}
\bibfield{author}{\bibinfo{person}{Alexander~M. Rush}, \bibinfo{person}{Sumit Chopra}, {and} \bibinfo{person}{Jason Weston}.} \bibinfo{year}{2015}\natexlab{}.
\newblock \showarticletitle{A Neural Attention Model for Abstractive Sentence Summarization}. In \bibinfo{booktitle}{\emph{Proceedings of the 2015 Conference on Empirical Methods in Natural Language Processing}}. \bibinfo{publisher}{Association for Computational Linguistics}, \bibinfo{address}{Lisbon, Portugal}, \bibinfo{pages}{379--389}.
\newblock
\href{https://doi.org/10.18653/v1/D15-1044}{doi:\nolinkurl{10.18653/v1/D15-1044}}


\bibitem[Scheibehenne et~al\mbox{.}(2010)]%
        {scheibehenne_can_2010}
\bibfield{author}{\bibinfo{person}{Benjamin Scheibehenne}, \bibinfo{person}{Rainer Greifeneder}, {and} \bibinfo{person}{Peter~M. Todd}.} \bibinfo{year}{2010}\natexlab{}.
\newblock \showarticletitle{Can {There} {Ever} {Be} {Too} {Many} {Options}? {A} {Meta}-{Analytic} {Review} of {Choice} {Overload}}.
\newblock \bibinfo{journal}{\emph{Journal of Consumer Research}} \bibinfo{volume}{37}, \bibinfo{number}{3} (\bibinfo{date}{Feb.} \bibinfo{year}{2010}), \bibinfo{pages}{409--425}.
\newblock
\showISSN{0093-5301}
\href{https://doi.org/10.1086/651235}{doi:\nolinkurl{10.1086/651235}}
\newblock
\shownote{\_eprint: https://academic.oup.com/jcr/article-pdf/37/3/409/5173186/37-3-409.pdf}.


\bibitem[Scherer(2005)]%
        {scherer2005emotions}
\bibfield{author}{\bibinfo{person}{Klaus~R Scherer}.} \bibinfo{year}{2005}\natexlab{}.
\newblock \showarticletitle{What are emotions? And how can they be measured?}
\newblock \bibinfo{journal}{\emph{Social science information}} \bibinfo{volume}{44}, \bibinfo{number}{4} (\bibinfo{year}{2005}), \bibinfo{pages}{695--729}.
\newblock


\bibitem[Schwartz(2004)]%
        {schwartz2004paradox}
\bibfield{author}{\bibinfo{person}{Barry Schwartz}.} \bibinfo{year}{2004}\natexlab{}.
\newblock \bibinfo{booktitle}{\emph{The paradox of choice: Why more is less}}.
\newblock \bibinfo{publisher}{Harper Perennial}, \bibinfo{address}{New York}.
\newblock


\bibitem[Selmene and Kodia(2020)]%
        {selmene_sentanalysis_2020}
\bibfield{author}{\bibinfo{person}{Safa Selmene} {and} \bibinfo{person}{Zahra Kodia}.} \bibinfo{year}{2020}\natexlab{}.
\newblock \showarticletitle{Recommender System Based on User's Tweets Sentiment Analysis}. In \bibinfo{booktitle}{\emph{Proceedings of the 4th International Conference on E-Commerce, E-Business and E-Government}} (Arenthon, France) \emph{(\bibinfo{series}{ICEEG '20})}. \bibinfo{publisher}{Association for Computing Machinery}, \bibinfo{address}{New York, NY, USA}, \bibinfo{pages}{96--102}.
\newblock
\showISBNx{9781450388030}
\href{https://doi.org/10.1145/3409929.3414744}{doi:\nolinkurl{10.1145/3409929.3414744}}


\bibitem[SemEval(2014)]%
        {Semeval-2014}
\bibfield{author}{\bibinfo{person}{SemEval}.} \bibinfo{year}{2014}\natexlab{}.
\newblock \bibinfo{booktitle}{\emph{Task Description: Aspect Based Sentiment Analysis (ABSA) $<$ SemEval-2014 Task 4}}.
\newblock
\urldef\tempurl%
\url{https://alt.qcri.org/semeval2014/task4/}
\showURL{%
\tempurl}


\bibitem[Shah and Wolford(2007)]%
        {Shah2007BuyingBA}
\bibfield{author}{\bibinfo{person}{Avni~M Shah} {and} \bibinfo{person}{George~L. Wolford}.} \bibinfo{year}{2007}\natexlab{}.
\newblock \showarticletitle{Buying Behavior as a Function of Parametric Variation of Number of Choices}.
\newblock \bibinfo{journal}{\emph{Psychological Science}}  \bibinfo{volume}{18} (\bibinfo{year}{2007}), \bibinfo{pages}{369 -- 370}.
\newblock


\bibitem[spaCy(2023)]%
        {spacy}
\bibfield{author}{\bibinfo{person}{spaCy}.} \bibinfo{year}{2023}\natexlab{}.
\newblock \bibinfo{title}{Industrial-Strength Natural Language Processing}.
\newblock
\urldef\tempurl%
\url{https://spacy.io/}
\showURL{%
\tempurl}


\bibitem[Tokutake and Okamoto(2024)]%
        {tokutake_zsserendipity_2024}
\bibfield{author}{\bibinfo{person}{Yu Tokutake} {and} \bibinfo{person}{Kazushi Okamoto}.} \bibinfo{year}{2024}\natexlab{}.
\newblock \showarticletitle{Can Large Language Models Assess Serendipity in Recommender Systems?}
\newblock \bibinfo{journal}{\emph{Journal of Advanced Computational Intelligence and Intelligent Informatics}}  \bibinfo{volume}{28} (\bibinfo{date}{11} \bibinfo{year}{2024}), \bibinfo{pages}{1263--1272}.
\newblock
\href{https://doi.org/10.20965/jaciii.2024.p1263}{doi:\nolinkurl{10.20965/jaciii.2024.p1263}}


\bibitem[Vohs et~al\mbox{.}(2008)]%
        {article}
\bibfield{author}{\bibinfo{person}{Kathleen Vohs}, \bibinfo{person}{Roy Baumeister}, \bibinfo{person}{Brandon Schmeichel}, \bibinfo{person}{Jean Twenge}, \bibinfo{person}{Noelle Nelson}, {and} \bibinfo{person}{Dianne Tice}.} \bibinfo{year}{2008}\natexlab{}.
\newblock \showarticletitle{Making Choices Impairs Subsequent Self-Control: A Limited-Resource Account of Decision Making, Self-Regulation, and Active Initiative}.
\newblock \bibinfo{journal}{\emph{Journal of personality and social psychology}}  \bibinfo{volume}{94} (\bibinfo{date}{05} \bibinfo{year}{2008}), \bibinfo{pages}{883--98}.
\newblock
\href{https://doi.org/10.1037/0022-3514.94.5.883}{doi:\nolinkurl{10.1037/0022-3514.94.5.883}}


\bibitem[Wang et~al\mbox{.}(2023)]%
        {wang_serendipity_2023}
\bibfield{author}{\bibinfo{person}{Zongyi Wang}, \bibinfo{person}{Yanyan Zou}, \bibinfo{person}{Anyu Dai}, \bibinfo{person}{Linfang Hou}, \bibinfo{person}{Nan Qiao}, \bibinfo{person}{Luobao Zou}, \bibinfo{person}{Mian Ma}, \bibinfo{person}{Zhuoye Ding}, {and} \bibinfo{person}{Sulong Xu}.} \bibinfo{year}{2023}\natexlab{}.
\newblock \showarticletitle{An Industrial Framework for Personalized Serendipitous Recommendation in E-commerce}. In \bibinfo{booktitle}{\emph{Proceedings of the 17th ACM Conference on Recommender Systems}} (Singapore, Singapore) \emph{(\bibinfo{series}{RecSys '23})}. \bibinfo{publisher}{Association for Computing Machinery}, \bibinfo{address}{New York, NY, USA}, \bibinfo{pages}{1015--1018}.
\newblock
\showISBNx{9798400702419}
\href{https://doi.org/10.1145/3604915.3610234}{doi:\nolinkurl{10.1145/3604915.3610234}}


\bibitem[Wei et~al\mbox{.}(2022)]%
        {wei2022finetuned}
\bibfield{author}{\bibinfo{person}{Jason Wei}, \bibinfo{person}{Maarten Bosma}, \bibinfo{person}{Vincent Zhao}, \bibinfo{person}{Kelvin Guu}, \bibinfo{person}{Adams~Wei Yu}, \bibinfo{person}{Brian Lester}, \bibinfo{person}{Nan Du}, \bibinfo{person}{Andrew~M. Dai}, {and} \bibinfo{person}{Quoc~V Le}.} \bibinfo{year}{2022}\natexlab{}.
\newblock \showarticletitle{Finetuned Language Models are Zero-Shot Learners}. In \bibinfo{booktitle}{\emph{International Conference on Learning Representations}}.
\newblock
\urldef\tempurl%
\url{https://openreview.net/forum?id=gEZrGCozdqR}
\showURL{%
\tempurl}


\bibitem[Xi et~al\mbox{.}(2025)]%
        {xi_allm_2025}
\bibfield{author}{\bibinfo{person}{Yunjia Xi}, \bibinfo{person}{Muyan Weng}, \bibinfo{person}{Wen Chen}, \bibinfo{person}{Chao Yi}, \bibinfo{person}{Dian Chen}, \bibinfo{person}{Gaoyang Guo}, \bibinfo{person}{Mao Zhang}, \bibinfo{person}{Jian Wu}, \bibinfo{person}{Yuning Jiang}, \bibinfo{person}{Qingwen Liu}, \bibinfo{person}{Yong Yu}, {and} \bibinfo{person}{Weinan Zhang}.} \bibinfo{year}{2025}\natexlab{}.
\newblock \bibinfo{title}{Bursting Filter Bubble: Enhancing Serendipity Recommendations with Aligned Large Language Models}.
\newblock
\href{https://doi.org/10.48550/arXiv.2502.13539}{doi:\nolinkurl{10.48550/arXiv.2502.13539}}


\bibitem[Yelp(2023)]%
        {yelp}
\bibfield{author}{\bibinfo{person}{Yelp}.} \bibinfo{year}{2023}\natexlab{}.
\newblock \bibinfo{title}{The Yelp Dataset}.
\newblock
\urldef\tempurl%
\url{https://www.yelp.com/dataset}
\showURL{%
\tempurl}


\bibitem[Zhang et~al\mbox{.}(2021)]%
        {zhang_snpr_2021}
\bibfield{author}{\bibinfo{person}{Mingwei Zhang}, \bibinfo{person}{Yang Yang}, \bibinfo{person}{Rizwan Abbas}, \bibinfo{person}{Ke Deng}, \bibinfo{person}{Jianxin Li}, {and} \bibinfo{person}{Bin Zhang}.} \bibinfo{year}{2021}\natexlab{}.
\newblock \showarticletitle{SNPR: A Serendipity-Oriented Next POI Recommendation Model}. In \bibinfo{booktitle}{\emph{Proceedings of the 30th ACM International Conference on Information \& Knowledge Management}} (Virtual Event, Queensland, Australia) \emph{(\bibinfo{series}{CIKM '21})}. \bibinfo{publisher}{Association for Computing Machinery}, \bibinfo{address}{New York, NY, USA}, \bibinfo{pages}{2568--2577}.
\newblock
\showISBNx{9781450384469}
\href{https://doi.org/10.1145/3459637.3482394}{doi:\nolinkurl{10.1145/3459637.3482394}}


\bibitem[Zhang et~al\mbox{.}(2014)]%
        {zhang_reviewrec_2014}
\bibfield{author}{\bibinfo{person}{Yongfeng Zhang}, \bibinfo{person}{Guokun Lai}, \bibinfo{person}{Min Zhang}, \bibinfo{person}{Yi Zhang}, \bibinfo{person}{Yiqun Liu}, {and} \bibinfo{person}{Shaoping Ma}.} \bibinfo{year}{2014}\natexlab{}.
\newblock \showarticletitle{Explicit Factor Models for Explainable Recommendation Based on Phrase-Level Sentiment Analysis}. In \bibinfo{booktitle}{\emph{Proceedings of the 37th International ACM SIGIR Conference on Research \& Development in Information Retrieval}} (Gold Coast, Queensland, Australia) \emph{(\bibinfo{series}{SIGIR '14})}. \bibinfo{publisher}{Association for Computing Machinery}, \bibinfo{address}{New York, NY, USA}, \bibinfo{pages}{83--92}.
\newblock
\showISBNx{9781450322577}
\href{https://doi.org/10.1145/2600428.2609579}{doi:\nolinkurl{10.1145/2600428.2609579}}


\bibitem[Ziarani and Ravanmehr(2021)]%
        {ziarani_serendipity_rs_survey_2021}
\bibfield{author}{\bibinfo{person}{Reza Ziarani} {and} \bibinfo{person}{Reza Ravanmehr}.} \bibinfo{year}{2021}\natexlab{}.
\newblock \showarticletitle{Serendipity in Recommender Systems: A Systematic Literature Review}.
\newblock \bibinfo{journal}{\emph{Journal of Computer Science and Technology}}  \bibinfo{volume}{36} (\bibinfo{date}{04} \bibinfo{year}{2021}), \bibinfo{pages}{375--396}.
\newblock
\href{https://doi.org/10.1007/s11390-020-0135-9}{doi:\nolinkurl{10.1007/s11390-020-0135-9}}


\end{thebibliography}

\appendix

\section{Examples of Extractive and Abstractive Atypical Aspect Annotations}
\label{app:annotatedata-examples}

In addition to the examples of atypical aspect annotations of restaurants shown in Table~\ref{tab:examples_restaurant}, Table~\ref{tab:examples_hotels-lt} and Table~\ref{tab:examples_hairsalons-lt} present annotations for hotels and hair salons, respectively.

\subsection{Hotels}

\LTcapwidth=14.7cm
\begin{longtable}{p{14.5cm}}
    \caption{Examples of {\bf extractive} and {\it \textbf{abstractive}} annotations of customer reviews in the hotel domain. Secondary or optional annotations are shown in \textcolor{teal}{\textbf{blue}}.}
    \label{tab:examples_hotels-lt}\\
    \toprule
    \endfirsthead
    \multicolumn{1}{p{14.5cm}}%
    {\tablename\ \thetable\ -- \textit{Continued from previous page}} \\
    \toprule
    \endhead
    \bottomrule \multicolumn{1}{r}{\textit{Continued on next page}} \\
    \endfoot
    \bottomrule
    \endlastfoot
    \noindent $\blacktriangleright$ A group of us stayed here for our annual 'moms trip' in late May and stayed in the two story suite. It was plenty of room for everyone with great location and friendly staff. I would highly recommend staying here especially if you are looking for the historical feel. This is not a modern elevator sky-rise so keep that in mind. They helped organize us with a \textbf{gator tou}r and \textbf{plantation tour} and we had a great trip. You can walk anywhere in the French Quarter. \\
    \hdashrule{14.7cm}{1pt}{2.5pt}
    {\it \textbf{The hotel helped organize customers with a gator tour and plantation tour.}} \\
    \midrule
    \noindent $\blacktriangleright$ Awesome place to get away for the weekend. The government rate for state or federal employees is a great deal if you qualify (that includes UA and military employees - so much of Tucson is covered!). However, keep in mind that the cheap government rate - often between \$80 and \$90, does NOT include the "resort fee", which is an additional \$48 a night! Granted, you do get a lot of "resort" amenities to take advantage of. Included in the "fee" are access to the pools (a given) and a \textbf{pretty fun waterslide}, free use of the \textbf{driving range} and \textbf{putting greens} with a bag of clubs to use, \textbf{mountain bikes} and \textbf{helmets} to use, and the fitness center. There's also lots of \textbf{nightly activities}, like a "mixology" presentation at the bar (which is basically just the bartender describing how to make a drink and giving you about a shot worth of it for free). There's \textbf{nightly stargazing} on one of the lawns with a telescope and professional guide. On Saturday night they also had a "\textbf{dive in movie}" in the spa pool (it was "\textbf{Frozen}"). \\
    \hdashrule{14.7cm}{1pt}{2.5pt}
    {\it \textbf{The hotel gives access to a waterslide, free use of the driving range and putting greens with a bag of clubs, mountain bikes and helmets. The hotel hosts lots of nightly activities. There's nightly stargazing on one of the lawns of the hotel with a telescope and professional guide. One Saturday night they also had a "dive in movie" in the spa pool. They showed Frozen movie.}} \\
    \midrule
    \noindent \textcolor{teal}{\textbf{Beautiful beach}} away from big crowds. Smaller hotel. Simple and clean rooms. Nice pool and courtyard with beautiful plants and trees. Sink, microwave, and fridge in room. Loved the feel of this hotel.\\
    \hdashrule{14.7cm}{1pt}{2.5pt}
    \textcolor{teal}{\textbf{The hotel is near the beach away from the big crowds.}} \\
\end{longtable}

\subsection{Hair Salons}

\LTcapwidth=14.7cm
\begin{longtable}{p{14.5cm}}
    \caption{Examples of {\bf extractive} and {\it \textbf{abstractive}} annotations of customer reviews in the hair salon domain. Secondary or optional annotations are shown in \textcolor{teal}{\textbf{blue}}.}
    \label{tab:examples_hairsalons-lt}\\
    \toprule
    \endfirsthead
    \multicolumn{1}{p{14.5cm}}%
    {\tablename\ \thetable\ -- \textit{Continued from previous page}} \\
    \toprule
    \endhead
    \bottomrule \multicolumn{1}{r}{\textit{Continued on next page}} \\
    \endfoot
    \bottomrule
    \endlastfoot
    \noindent $\blacktriangleright$ Absolutely love this \textbf{shop}, \textbf{studio}, and salon! The flow of creative energy in here is palpable and inspiring. When they say Hair, Fashion, Art, they mean it. I've been looking for a salon I'd like to come back to for a few years and I finally found it! I'm so sad it took me so long. This is the perfect place for a whole new 'do or just a bang trim. I have felt very comfortable sitting in the chair and letting Kristine, Salina, or Gloriana update my look.
    As for the art on the walls, I love how they are always changing. This is one of the \textbf{best little art galleries} and I've discovered so many cool \textbf{local artists} by also looking at the walls while I was shopping. I met Beth, the gallery director, and she really knows her stuff. 
    And as for the \textbf{fashion part} of the store, it is also art! Leigh Anne is such a talented designer. She has \textbf{beautiful dresses} with sleek lines that fit an array of bodies. Her original designs are what cocktail party dreams are made of.
    AND my favorite part about The Disco Dolls Studio is that it's sustainable! These women have made a commitment to helping our planet and each other and it shows in their ethics and in the way they conduct business. And it's soooo nice to shop at a place where I know the products are ethically sourced and I can feel good about buying them. 
    Lastly, this place is taking great precautions during these strange COVID times and I am very thankful to see business owners actually care about my safety as well as their own. \\
    \hdashrule{14.7cm}{1pt}{2.5pt}
    {\it \textbf{The salon is a shop as well as a studio. The salon has a little art gallery with art on the walls done by local artists. The designer of the fashion part of the store has dresses with sleek lines.}} \\
    \midrule
    \noindent $\blacktriangleright$ MY wife went here to donate her hair to \textbf{locks of love}, if you donate 10 inches of hair your haircut is free. can't get better than that. Everyone there was nice and professional. my wife was in and out in under and hour despite the fact that there were a bunch of people there ahead of us. If you want to check in to their store on your phone you can get into the haircut line before you get there and it lasts for up to an hour. it will save you wait time. \\
    \hdashrule{14.7cm}{1pt}{2.5pt}
    {\it \textbf{A customer went to the salon to donate her hair to Locks of Love, if you donate 10 inches of hair your haircut is free. }} \\
    \midrule
    \noindent Last minute appointment for blowout and curl. They were very accommodating and extremely friendly. Was my first time here and I accidentally showed up at the wrong location (Central) and the girl called the downtown location for me and explained I would be a little late. Ari shampooed me and styled my hair. She is sweet and friendly as well as knows what she's doing. All smiles from everyone when I left. I would recommend this place to anyone. The ambiance is awesome as it's established in an \textcolor{teal}{\textbf{old art gallery}}. It's the building with the \textcolor{teal}{\textbf{Tucson mural}} on it. \\
    \hdashrule{14.7cm}{1pt}{2.5pt}
    \textcolor{teal}{\textbf{The salon is established in an old art gallery. The salon building is one with a Tucson mural on it.}} 
    \\
\end{longtable}

\section{In-context Learning Examples for Processing Reviews into a list of Aspect Sentences}
\label{app:reviewreformulation-examples}

We used 3 in-context learning examples in the prompt for processing reviews into a list of aspect sentences that capture all aspects of an item, typical and atypical, while ignoring other types of information, such as opinions or incidental information that does not pertain to the item, as shown in Table~\ref{tab:prompt_reviewreformuation}. Here, we present the full text of the 3 in-context examples used for restaurants, hotels and hair salons in Table~\ref{tab:restreviewreformulation-examples-lt}, Table~\ref{tab:hotelreviewreformulation-examples-lt}, and Table~\ref{tab:hsreviewreformulation-examples-lt}, respectively.

\subsection{Restaurants}

\begin{longtable}{p{14.5cm}}
  \caption{In-context learning examples for processing reviews into list of aspect sentences that capture all the aspects (typical or atypical) of the restaurant.} 
  \label{tab:restreviewreformulation-examples-lt} \\
  \toprule
  \endfirsthead

  \multicolumn{1}{p{14.5cm}}{\tablename\ \thetable\ -- \textit{Continued from previous page}}\\
  \toprule
  \endhead

  \bottomrule
  \multicolumn{1}{r}{\textit{Continued on next page}}\\
  \endfoot

  \bottomrule
  \endlastfoot

  \noindent $\triangleright$ {\bf Example 1:}
  
      Really good sushi - in a strip mall in Bristol. Go figure. They have hibachi as well, but I feel like hibachi food is hibachi food no matter where you go; it's more about the performance. Do it once and then keep coming back for the sushi! This is a mom-and-pop type restaurant with a friendly atmosphere. If you're there for a special occasion, they'll take a polaroid and add it to their wall. I have fond memories of many family meals here. Look for my picture on the wall; I'm the adorable one.\\

      \hspace{2em} \textbf{Output}: \textit{The restaurant has really good sushi. The restaurant is in a strip mall in Bristol. The restaurant has hibachi. This is a mom-and-pop type restaurant with a friendly atmosphere. If you're there for a special occasion, they'll take a polaroid and add it to their wall.}\\

  \midrule

  \noindent $\triangleright$ {\bf Example 2:}
  
      A bit of self care and indulgence for when you want to relax and de-stress and if you don't want to go to the bar but have another option. Relax and enjoy. There's a lounge area behind. So you can take your drink in the lounge and kick back. Caffeinated choices are available. If you like your tastebuds teased in a different way, this is a treat. Perfect for yogis and if you like incense. (there's no incense, but it may remind you ...). Can be a spiritual experience as there are symbolism and custom associations provided if you want to try them. Just ask the very knowledgeable Server. It is what you make it. Whatever style makes you comfortable. You have to go inside the West Market to find it! Do not miss.\\
    
      \hspace{2em} \textbf{Output}: \textit{The restaurant provides a bit of self care and indulgence for when you want to relax and de-stress. There's a lounge area behind the restaurant, where you can take your drink in the lounge and kick back. Caffeinated choices are available at the restaurant. The restaurant is perfect for yogis and if you like incense. The restaurant provides a spiritual experience as there are symbolism and custom associations provided, for which you can ask the very knowledgeable Server. The restaurant is located inside the West Market.}\\

  \midrule

  \noindent $\triangleright$ {\bf Example 3:}
  
      A bit high priced, and the food is overall good, but not amazing. Still, I really enjoyed myself here probably because I just love sitting at a place on the water (even if the road is right nearby). The place seems large and mostly outdoors. It's a casual fun atmosphere and boaters can come right up to dock. You can also bring your pets! In fact, I saw one guy who brought his parakeet on his shoulder! lol we had some tasty frozen drinks, that either hid the alcohol taste very well with all the sweetness or it was lacking enough alcohol. I'm not sure which, but kinda leaning towards the latter. They have a few beers on tap, but may not be familiar with them. For example, our waitress didn't know that there are several types of Sam Adams beers when I asked which one was on tap. I had the stuffed flounder for my main course and ate every bite. However, I could have been better with less sauce on it. My stomach hurt from all that afterwards I think. We tried the fried cheesecake for dessert...it was okay. Nothing exciting. If you're looking for a leisurely dinner outside while enjoying the Florida atmosphere, this is your place. I wouldn't say no to going back!\\

      \hspace{2em} \textbf{Output}: \textit{The restaurant is a bit high priced, and the food is overall good, but not amazing. The restaurant is on the water and the road is right nearby. The place seems large and mostly outdoors. The restaurant has a casual fun atmosphere and boaters can come right up to dock. You can also bring your pets to the restaurant. The restaurant offers tasty frozen drinks, that either hide the alcohol taste very well with all the sweetness or that lack enough alcohol. They have a few beers on tap, but the servers may not be familiar with them. The restaurant offers stuffed flounder as a main course. The restaurant offers fried cheesecake for dessert. The restaurant is suitable for a leisurely dinner outside while enjoying the Florida atmosphere. }\\
\end{longtable}

\subsection{Hotels}

\begin{longtable}{p{14.5cm}}
  \caption{In-context learning examples for processing reviews into list of aspect sentences that capture all the aspects (typical or atypical) of the hotel.} 
  \label{tab:hotelreviewreformulation-examples-lt} \\
  \toprule
  \endfirsthead

  \multicolumn{1}{p{14.5cm}}{\tablename\ \thetable\ -- \textit{Continued from previous page}}\\
  \toprule
  \endhead

  \bottomrule
  \multicolumn{1}{r}{\textit{Continued on next page}}\\
  \endfoot

  \bottomrule
  \endlastfoot

  \noindent $\triangleright$ {\bf Example 1:}
  
      This boutique is very cute and quite a steal for the price. They give you water when you check in which is an added bonus. The lobby is very chic and the decor is 5 stars. The room was clean and very nice well worth the price. The best is that they have a free shuttle to take you to all of the local sites. They also have a beautiful resort style pool. The only cons were the free parking is great but the lot gets crowded which can be difficult to find a spot. Secondly they have a coffee pot in the room but dont offer tea or decaf. Everything thing else makes this a 4 star hotel at a 3 star price.\\
    
      \hspace{2em} \textbf{Output}: \textit{This boutique is very cute and quite a steal for the price. They give you water when you check in. The lobby is very chic and the decor is 5 stars. The room was clean and very nice. The hotel has a free shuttle to take you to all of the local sites. The hotel also has a beautiful resort style pool. The hotel offers free parking but the lot gets crowded which can make it difficult to find a spot. They have a coffee pot in the room but dont offer tea or decaf. Everything thing else makes this a 4 star hotel at a 3 star price.}\\

  \midrule

  \noindent $\triangleright$ {\bf Example 2:}
  
      The views are beautiful. Room has no microwave. Bathroom has no fan. Grill by pool closes at 6 so good luck feeding a picky kid. 
      
      The activities for the kids are great. Cookie baking, bingo, hula hoop contest, and simon says. Pool is nice but the pool with the slide needs to be much deeper. Staff is. Just personal preference. Food is way overpriced and equivalent to street fair quality.\\

      \hspace{2em} \textbf{Output}: \textit{The views from the hotel are beautiful. Room has no microwave. Bathroom has no fan. Grill by pool closes at 6. The activities for the kids in the hotel are great. The activities for the kids in the hotel include Cookie baking, bingo, hula hoop contest, and simon says. The pool is nice. Staff is . Just personal preference. Food is way overpriced.}\\

  \midrule

  \noindent $\triangleright$ {\bf Example 3:}
  
      Hidden gem! What a glorious property! Very reasonable for what you get , made to order organic breakfast, nice pool and a veranda to check out the butterfly garden. The staff was friendly and accommodating and it's close to everything! Would definitely stay here again!\\

      \hspace{2em} \textbf{Output}: \textit{The hotel is very reasonable priced for what you get. They have made to order organic breakfast, nice pool and a veranda to check out the butterfly garden. The staff was friendly and accommodating and it's close to everything!}\\
\end{longtable}

\subsection{Hair Salons}

\begin{longtable}{p{14.5cm}}
  \caption{In-context learning examples for processing reviews into list of aspect sentences that capture all the aspects (typical or atypical) of the hair salon.} 
  \label{tab:hsreviewreformulation-examples-lt} \\
  \toprule
  \endfirsthead

  \multicolumn{1}{p{14.5cm}}{\tablename\ \thetable\ -- \textit{Continued from previous page}}\\
  \toprule
  \endhead

  \bottomrule
  \multicolumn{1}{r}{\textit{Continued on next page}}\\
  \endfoot

  \bottomrule
  \endlastfoot

  \noindent $\triangleright$ {\bf Example 1:}
  
      Cookie Cutters is perfect for kids and especially those with sensory issues. There are play activities in the waiting area. The chairs are all themed like Thomas, Disney princess, fire truck, etc. Each station has a smart TV so kids can pick out their favorite show on Netflix, Youtube, etc. The best part about the chairs is that the kids sit deep in there and get buckled in so that they can't kick anybody or anything. My only real issues have been that they tend to run late on appointments which can get the kids agitated and the open ceiling and flat walls make it an echo chamber. I don't think the kids mind but it is even louder than you would expect in a building of screaming kids. The hair cutting has been good each time and the price isn't terrible for what you are getting.\\
    
      \hspace{2em} \textbf{Output}: \textit{Cookie Cutters is perfect for kids and especially those with sensory issues. The hair salon has play activities in the waiting area. The chairs in the hair salon are all themed like Thomas, Disney princess, fire truck, etc. The hair salon has a smart TV at each station so kids can pick out their favorite show on Netflix, Youtube, etc. The kids can sit deep in the chairs and get buckled in so that they can't kick anybody or anything. The hair salon can run late on appointments which can get the kids agitated and the open ceiling and flat walls make it an echo chamber. }\\

  \midrule

  \noindent $\triangleright$ {\bf Example 2:}
  
      One of the nicest salons I have ever been to. I loved the interior design and the wide selection of refreshments. The pricing was very fair got what I received. This is the best haircut I have had in a very long time. If also helps that is walking distance from my apartment, and they were also hosting a blanket drive, and I love when businesses are involved in the local community. There are two reasons this isn't a five star review: first, the online system on which you book appointments is clunky and often doesn't load correctly. I wasn't sure if my appointment was confirmed until the next day. Secondly, my appointment started almost half an hour late, making my haircut take much longer than normal.\\

      \hspace{2em} \textbf{Output}: \textit{The hair salon is one of the nicest salons. They have a lovely interior design and a wide selection of refreshments. The pricing at the hair salon is very fair given what one receives. They also host blanket drives, demonstrating their involvement in the local community. There are two reasons this isn't a five star review: first, the online system on which you book appointments is clunky and often doesn't load correctly. The customer might not be sure if the appointment was confirmed until the next day. Secondly, appointments can start almost half an hour late, making the haircut take much longer than normal.}\\

  \midrule

  \noindent $\triangleright$ {\bf Example 3:}
  
      The salon is a recent addition to the Duross and Langel shop and they were smart enough to bring in my favorite stylist of all time, Rene. In May I decided to change my look with white-blonde hair instead of the usual highlights. Rene did an awesome job with the cut (as usual) and color. I got it touched up recently and it came out even better. The owners are lovely and host industry/customer appreciation nights where there is free yoga and they cook us dinner. The space is beautiful, the prices are very reasonable for center city and the products are great. Highly recommended.\\

      \hspace{1em} \textbf{Output}: \textit{The salon is a recent addition to the Duross and Langel shop and one of the stylist is Rene. Rene does an awesome job with the cut and color. The owners are lovely and host industry/customer appreciation nights where there is free yoga and they cook dinner. The space is beautiful, the prices are very reasonable for center city and the products are great. Highly recommended.}\\
\end{longtable}

\section{In-context Learning Examples for Extracting Atypical Aspects from Aspect Sentences}
\label{app:ataextraction-examples}

We used 8 in-context learning examples in the prompt for extracting atypical aspects from aspect sentences. As shown in Table~\ref{tab:prompt_ataextraction}, the prompt instructs the model to classify an aspect sentence as positive if it mentions an atypical aspect and negative otherwise; then for sentences classified as positive, it instructs to extract all mentioned atypical aspects. As described in Section~\ref{sec:extrata-llm}, we adopted two distinct ICL strategies: {\it Fixed} few-shot and {\it Dynamic} few-shot. Here, we present the full text of 8 ICL examples used in the {\it Fixed} few-shot setting for restaurants, hotels and hair salons in Table~\ref{tab:restataextraction-examples-lt}, Table~\ref{tab:hotelataextraction-examples-lt}, and Table~\ref{tab:hsataextraction-examples-lt}, respectively.

\subsection{Restaurants}

\begin{longtable}{p{14.5cm}}
  \caption{In-context learning examples for classifying restaurant aspect sentences as <pos> or <neg> and extracting atypical aspect from sentences classified as <pos>.} 
  \label{tab:restataextraction-examples-lt} \\
  \toprule
  \endfirsthead

  \multicolumn{1}{p{14.5cm}}{\tablename\ \thetable\ -- \textit{Continued from previous page}}\\
  \toprule
  \endhead

  \bottomrule
  \multicolumn{1}{r}{\textit{Continued on next page}}\\
  \endfoot

  \bottomrule
  \endlastfoot

  \noindent $\triangleright$ {\bf Example 1:}
  
      The restaurant features a long hallway leading to a garden, adorned with numerous pieces of art\\
    
      \hspace{2em} \textit{\textbf{Classification}}:  \textbf{<pos>}\\
      \hspace{2em} \textit{\textbf{Atypical Aspects}}:  garden, art\\

  \midrule

  \noindent $\triangleright$ {\bf Example 2:}
  
      The restaurant also has a gift shop with reasonably priced items, including cute wallets, shawls and necklaces\\
    
      \hspace{2em} \textit{\textbf{Classification}}:  \textbf{<pos>}\\
      \hspace{2em} \textit{\textbf{Atypical Aspects}}:  gift shop, cute wallets, shawls, necklaces\\

  \midrule

  \noindent $\triangleright$ {\bf Example 3:}
  
      The restaurant, Fat Straw, has a futuristic style with a lime green counter, shiny plastic furniture, and stained concrete floors\\
    
      \hspace{2em} \textit{\textbf{Classification}}:  \textbf{<pos>}\\
      \hspace{2em} \textit{\textbf{Atypical Aspects}}:  lime green counter, shiny plastic furniture, stained concrete floors\\

  \midrule

  \noindent $\triangleright$ {\bf Example 4:}
  
      It features a wall of polaroids of customers, which adds a personal touch to the decor of geometric shapes and synthetic materials\\
    
      \hspace{2em} \textit{\textbf{Classification}}:  \textbf{<pos>}\\
      \hspace{2em} \textit{\textbf{Atypical Aspects}}:  polaroids, geometric shapes, synthetic materials\\

  \midrule

  \noindent $\triangleright$ {\bf Example 5:}
  
      The restaurant is located at Edgehill Village\\
    
      \hspace{2em} \textit{\textbf{Classification}}:  \textbf{<neg>}\\
      \hspace{2em} \textit{\textbf{Atypical Aspects}}:  <None>\\

  \midrule

  \noindent $\triangleright$ {\bf Example 6:}
  
      The prices are standard, with a 16oz bubble tea priced at \$3.50\\
    
      \hspace{2em} \textit{\textbf{Classification}}:  \textbf{<neg>}\\
      \hspace{2em} \textit{\textbf{Atypical Aspects}}:  <None>\\

  \midrule

  \noindent $\triangleright$ {\bf Example 7:}
  
      The restaurant offers a variety of bubble tea flavors, including red bean, honeydew, lychee, pomegranate, ginger, elvis (banana + peanut butter), and sesame\\
    
      \hspace{2em} \textit{\textbf{Classification}}:  \textbf{<neg>}\\
      \hspace{2em} \textit{\textbf{Atypical Aspects}}:  <None>\\

  \midrule

  \noindent $\triangleright$ {\bf Example 8:}
  
      The restaurant has a colorful ambiance, with cups for frozen drinks that change colors\\
    
      \hspace{1em} \textit{\textbf{Classification}}:  \textbf{<neg>}\\
      \hspace{1em} \textit{\textbf{Atypical Aspects}}:  <None>\\
\end{longtable}

\subsection{Hotels}

\begin{longtable}{p{14.5cm}}
  \caption{In-context learning examples for classifying hotel aspect sentences as <pos> or <neg> and extracting atypical aspect from sentences classified as <pos>.} 
  \label{tab:hotelataextraction-examples-lt} \\
  \toprule
  \endfirsthead

  \multicolumn{1}{p{14.5cm}}{\tablename\ \thetable\ -- \textit{Continued from previous page}}\\
  \toprule
  \endhead

  \bottomrule
  \multicolumn{1}{r}{\textit{Continued on next page}}\\
  \endfoot

  \bottomrule
  \endlastfoot

  \noindent $\triangleright$ {\bf Example 1:}
  
      The B\&B is located near the departure points for Trolley tours and Land Shark tours\\
    
      \hspace{2em} \textit{\textbf{Classification}}:  \textbf{<pos>}\\
      \hspace{2em} \textit{\textbf{Atypical Aspects}}:  Trolley tours, Land Shark tour\\

  \midrule

  \noindent $\triangleright$ {\bf Example 2:}
  
      The hotel provides a free shuttle service to local sites\\
    
      \hspace{2em} \textit{\textbf{Classification}}:  \textbf{<pos>}\\
      \hspace{2em} \textit{\textbf{Atypical Aspects}}:  free shuttle\\

  \midrule

  \noindent $\triangleright$ {\bf Example 3:}
  
      The hotel offers activities for kids, including cookie baking, bingo, hula hoop contests, and Simon says\\
    
      \hspace{2em} \textit{\textbf{Classification}}:  \textbf{<pos>}\\
      \hspace{2em} \textit{\textbf{Atypical Aspects}}:  activities, cookie baking, bingo, hula hoop contests, Simon says\\

  \midrule

  \noindent $\triangleright$ {\bf Example 4:}
  
      The hotel grounds feature a koi pond with a water feature, a tortoise rescue, a butterfly garden, and a trail leading to a natural waterfall\\
    
      \hspace{2em} \textit{\textbf{Classification}}:  \textbf{<pos>}\\
      \hspace{2em} \textit{\textbf{Atypical Aspects}}:  koi pond with water feature, tortoise rescue, butterfly garden, trail, natural waterfall\\

  \midrule

  \noindent $\triangleright$ {\bf Example 5:}
  
      The front desk staff is friendly, helpful, and polite\\
    
      \hspace{2em} \textit{\textbf{Classification}}:  \textbf{<neg>}\\
      \hspace{2em} \textit{\textbf{Atypical Aspects}}:  <None>\\

  \midrule

  \noindent $\triangleright$ {\bf Example 6:}
  
      The B\&B has ample parking, especially during off-peak seasons\\
    
      \hspace{2em} \textit{\textbf{Classification}}:  \textbf{<neg>}\\
      \hspace{2em} \textit{\textbf{Atypical Aspects}}:  <None>\\

  \midrule

  \noindent $\triangleright$ {\bf Example 7:}
  
      The hotel appears to have undergone recent renovations, with comfortable beds, freshly painted walls, and new carpeting\\
    
     \hspace{2em} \textit{\textbf{Classification}}:  \textbf{<neg>}\\
     \hspace{2em} \textit{\textbf{Atypical Aspects}}:  <None>\\

  \midrule

  \noindent $\triangleright$ {\bf Example 8:}
  
      The hotel beds are full-sized, not queen-sized\\
    
      \hspace{2em} \textit{\textbf{Classification}}:  \textbf{<neg>}\\
      \hspace{2em} \textit{\textbf{Atypical Aspects}}:  <None>\\
\end{longtable}

\subsection{Hair Salons}

\begin{longtable}{p{14.5cm}}
  \caption{In-context learning examples for classifying hair salon aspect sentences as <pos> or <neg> and extracting atypical aspect from sentences classified as <pos>.} 
  \label{tab:hsataextraction-examples-lt} \\
  \toprule
  \endfirsthead

  \multicolumn{1}{p{14.5cm}}{\tablename\ \thetable\ -- \textit{Continued from previous page}}\\
  \toprule
  \endhead

  \bottomrule
  \multicolumn{1}{r}{\textit{Continued on next page}}\\
  \endfoot

  \bottomrule
  \endlastfoot

  \noindent $\triangleright$ {\bf Example 1:}
  
      The salon also sells costume jewelry, handbags, and scarves\\
    
      \hspace{2em} \textit{\textbf{Classification}}:  \textbf{<pos>}\\
      \hspace{2em} \textit{\textbf{Atypical Aspects}}:  costume jewelry, handbags, scarves\\

  \midrule

  \noindent $\triangleright$ {\bf Example 2:}
  
      The salon chairs are themed with popular children's characters and vehicles like Thomas, Disney princess, and fire truck\\
    
      \hspace{2em} \textit{\textbf{Classification}}:  \textbf{<pos>}\\
      \hspace{2em} \textit{\textbf{Atypical Aspects}}:  Thomas, Disney princess, fire truck\\

  \midrule

  \noindent $\triangleright$ {\bf Example 3:}
  
      Each station is equipped with a smart TV where children can watch their favorite shows on Netflix, Youtube, etc\\
    
      \hspace{2em} \textit{\textbf{Classification}}:  \textbf{<pos>}\\
      \hspace{2em} \textit{\textbf{Atypical Aspects}}:  smart TV, their favorite show, Netflix, Youtube\\

  \midrule

  \noindent $\triangleright$ {\bf Example 4:}
  
      The salon provides makeovers, including haircuts, manicures, pedicures, and makeup, for patients in hospitals\\
    
      \hspace{2em} \textit{\textbf{Classification}}:  \textbf{<pos>}\\
      \hspace{2em} \textit{\textbf{Atypical Aspects}}:  manicures, pedicures, makeup\\

  \midrule

  \noindent $\triangleright$ {\bf Example 5:}
  
      The salon is housed in a stucco building and has been recognized as one of the top 200 fastest growing salons in the USA by Salon Today magazine\\
    
      \hspace{2em} \textit{\textbf{Classification}}:  \textbf{<neg>}\\
      \hspace{2em} \textit{\textbf{Atypical Aspects}}:  <None>\\

  \midrule

  \noindent $\triangleright$ {\bf Example 6:}
  
      The salon staff, including Lisa and Kathy, are professional and attentive to customer needs\\
    
      \hspace{2em} \textit{\textbf{Classification}}:  \textbf{<neg>}\\
      \hspace{2em} \textit{\textbf{Atypical Aspects}}:  <None>\\

  \midrule

  \noindent $\triangleright$ {\bf Example 7:}
  
      The salon offers discounts through Groupon Now\\
    
      \hspace{2em} \textit{\textbf{Classification}}:  \textbf{<neg>}\\
      \hspace{2em} \textit{\textbf{Atypical Aspects}}:  <None>\\

  \midrule

  \noindent $\triangleright$ {\bf Example 8:}
  
      The salon staff provide personalized consultations to understand the customer's style and preferences\\
    
      \hspace{2em} \textit{\textbf{Classification}}:  \textbf{<neg>}\\
      \hspace{2em} \textit{\textbf{Atypical Aspects}}:  <None>\\
\end{longtable}

\section{Comparative Analysis of Atypical Aspect Extraction Across Models and Approaches}
\label{app:extrata_comparison}

In Table~\ref{tab:extrata_comparison_f1}, we present a comparative analysis of F1 scores in Exact‐Match and Partial-Match scenarios for Atypical Aspect Extraction across different models and extraction approaches. GPT-3.5 shows improved performance when using the two-step pipeline approach compared to the monolithic approach introduced in \cite{nannaware2023extraction}. GPT-4, using the two-step pipeline approach, consistently achieves the highest F1 scores in all experimental settings.

\begin{table}[h]
    \centering
    \scriptsize
    \caption{F1 scores (\%) for extractive extraction, comparing monolithic (GPT-3.5) and pipeline approaches (GPT-3.5 \& GPT-4) under Exact‐Match and Partial-Match scenarios.}
    \begin{tabular}{l|rr|l|rr|l|rr}
        \toprule
        \multicolumn{1}{l|}{\textbf{Method}} & \multicolumn{2}{c|}{\textbf{F1}} 
        & \multicolumn{1}{l|}{\textbf{Method}} & \multicolumn{2}{c|}{\textbf{F1}} 
        & \multicolumn{1}{l|}{\textbf{Method}} & \multicolumn{2}{c}{\textbf{F1}} \\
        \multicolumn{1}{l|}{} & \textbf{Exact} & \textbf{Partial}
        & \multicolumn{1}{l|}{} & \textbf{Exact} & \textbf{Partial}
        & \multicolumn{1}{l|}{} & \textbf{Exact} & \textbf{Partial} \\
        \midrule
        \multicolumn{3}{l|}{\sc Monolithic} 
        & \multicolumn{3}{l|}{\sc Pipeline} 
        & \multicolumn{3}{l}{\sc Pipeline} \\
        \ \ GPT-3.5 (0-shot)          & 27.7 & 37.4 
        & GPT-3.5 (0-shot)            & 29.3 & 41.7 
        & GPT-4  (0-shot)             & 71.6 & 78.3 \\
        \ \ \textit{+ secondary}      & 29.3 & 41.2 
        & \textit{+ secondary}        & 30.5 & 45.7 
        & \textit{+ secondary}        & 69.6 & 78.0 \\
        \ \ GPT-3.5 (5-shot)          & 37.6 & 48.0
        & GPT-3.5 (Fixed 8-shot)      & 62.7 & 67.3 
        & GPT-4  (Fixed 8-shot)       & 73.9 & 78.8 \\
        \ \ \textit{+ secondary}      & 31.7 & 43.2 
        & \textit{+ secondary}        & 64.6 & 70.8
        & \textit{+ secondary}        & 74.4 & 80.6 \\
        &      &      
        & GPT-3.5 (Dynamic 8-shot)    & 55.5 & 57.0 
        & GPT-4  (Dynamic 8-shot)     & $\mathbf{77.5}$ & $\mathbf{81.7}$ \\
        &      &      
        & \textit{+ secondary}        & 64.6 & 70.1 
        & \textit{+ secondary}        & 73.9 & 80.1 \\
        \bottomrule
    \end{tabular}
    \label{tab:extrata_comparison_f1}
\end{table}

\section{In-context Learning Examples of Chain-of-Thought User Profile Generation}
\label{app:profile-examples}

In Table~\ref{tab:profilegeneration-examples-lt}, we present the full text of the 9 in-context learning examples of manually crafted user profiles, along with their respective chain-of-thought rationales, which were used in the prompt for generating synthetic user profiles shown in Table~\ref{tab:prompt_userprofilegeneration}.

\begin{longtable}{p{14.5cm}}
  \caption{In-context learning examples for generating synthetic user profiles.} 
  \label{tab:profilegeneration-examples-lt} \\
  \toprule
  \endfirsthead

  \multicolumn{1}{p{14.5cm}}{\tablename\ \thetable\ -- \textit{Continued from previous page}}\\
  \toprule
  \endhead

  \bottomrule
  \multicolumn{1}{r}{\textit{Continued on next page}}\\
  \endfoot

  \bottomrule
  \endlastfoot

  \noindent $\triangleright$ {\bf Example 1} for user profile generation with chain of thought reasoning:

    \textit{\textbf{The given topics}} are $\langle$\textbf{\it yoga}$\rangle$.\\
    
    \textit{\textbf{Let's think step by step}} about how to write a good biography for an individual using the given topics. Here, the given topic can be more effectively integrated as a hobby. We express the topic 'yoga' as given. We can begin by discussing the individual's educational and professional background. Let's place it within the domain of social studies, adding that this led him to study South Asian cultures and be introduced to yoga. We can then highlight that this interest evolved into a hobby, thus explicitly integrating the given topic into the narrative and aligning it with the individual's hobbies and passions. This approach not only enriches the biography but also underscores the interest or appeal that the individual would feel toward the given topic. We can also mention where he grew up and that he enjoys spending time with his friends and family, which is something unrelated to the given topic, but it contributes to a natural-sounding biography.\\
    
    \textit{\textbf{So, a good biography would be:}} "Jordan Lee is a cultural studies lecturer in the college, where he graduated with a major in anthropology. He grew up in Portland with his 2 other siblings, with whom he maintains a very close relationship. Jordan discovered yoga \& meditation while doing a study on South Asian cultures. It quickly evolved from a hobby to more of a lifestyle. He enjoys spending time with his friends and family."\\

  \midrule

  \noindent $\triangleright$ {\bf Example 2} for user profile generation with chain of thought reasoning:

    \textit{\textbf{The given topics}} are $\langle$\textbf{\it professional babysitting service}$\rangle$.\\
    
    \textit{\textbf{Let's think step by step}} about how to write a good biography for an individual using the given topic. Here, the given topic can be more effectively integrated as a lifestyle need. We can begin by discussing the individual's family background. As the topic relates to 'babysitting', it's pertinent to mention that the individual has a young child. Next, we can highlight her profession. To make the biography sound well-rounded and natural, we can choose an educational and professional background unrelated to the given topic. Let's place it within the domain of business management and add her current profession to be an operations manager. Now, we can outline situations that logically necessitate the need for a babysitter or babysitting service, such as her desire to go out but being unable to find a babysitter. Thus, we explicitly integrate the given topic into the narrative, aligning it with the individual's lifestyle needs, and also underscoring the usefulness of the given topic for the individual.\\
    
    \textit{\textbf{So, a good biography would be:}} "Maria Gonzalez resides in San Diego, California, with her husband and their 7-year-old son. She works as an operations manager. Maria met her husband in college, where she graduated with a major in business management. Maria loves being social and enjoys spending time with her husband and friends, often sharing gossip over a glass of wine. But finding time to get together is hard because of her job. Even weekends, which could be a break, are difficult. Her son is too young to be left by himself, and finding reliable babysitters in her area is tough."\\

  \midrule

  \noindent $\triangleright$ {\bf Example 3} for user profile generation with chain of thought reasoning:

    \textit{\textbf{The given topics}} are $\langle$\textbf{\it life size beer pong, pool table, arcade games, nfl blitz 99, madden 15}$\rangle$.\\

    \textit{\textbf{Let's think step by step}} about how to write a good biography for an individual using the given topics. Here, since there are multiple topics with a similar theme, we express these through a more general category i.e, 'video games and indoor games'. This can be more effectively integrated as hobbies. We can begin by exploring the individual's upbringing and family history, highlighting that as the eldest of four children, he took over his family's plumbing business. Although these details are unrelated to the given topic, they provide essential background and contribute to a natural-sounding biography. Next, we mention his hobbies, which are playing video games and indoor games, also emphasizing that these activities are not only personal interests but also part of traditional game nights in his household. By doing so, we explicitly integrate the given topic into the narrative, aligning it with the individual's hobbies and also underscoring the interest or appeal that the individual would feel toward the given topics.\\

    \textit{\textbf{So, a good biography would be:}} "John comes from Omaha and grew up in a big family as the oldest of four kids. He recently took over his family's plumbing business and loves his job. One of his favorite hobbies is playing video games and indoor games like ping pong, shuffleboard and foosball. Game nights on Saturdays are a tradition in the household."\\

  \midrule

  \noindent $\triangleright$ {\bf Example 4} for user profile generation with chain of thought reasoning:

    \textit{\textbf{The given topics}} are $\langle$\textbf{\it water, boaters, your pets, peacocks, sunset, park woods, long public dock}$\rangle$.\\

    \textit{\textbf{Let's think step by step}} about how to write a good biography for an individual using the given topics. Here, since there are multiple topics with similar themes, we express these through a more general category i.e., the topics 'water, boaters, peacocks, sunset, park woods, long public dock' with 'nature and the outdoors'. This can be more effectively integrated as hobbies. We can start by exploring the individual's upbringing, choosing a location close to nature to implicitly integrate the given topics into the narrative and aligning it with the potential hobbies and passions of the individual, such as hiking and trekking. To add depth, we'll incorporate a significant recent life event that is relevant, further underscoring the interest or appeal that the individual would feel toward the given topics. Additionally, for the given topic 'your pets', we can mention that she owns a cat, implicitly integrating her affection for pets into the biography narrative as well. We can also note her enjoyment of spending time with her friends and family and watching movies, details that are unrelated to the given topics, but contribute to a natural-sounding biography.\\

    \textit{\textbf{So, a good biography would be:}} "Sam was raised in a small mountain town near the northern limits of Asheville, North Carolina. She grew up loving nature and the outdoors, hiking and trekking from a young age. Her most recent adventure was a solo trek through the Appalachian trails where she met and bonded with other hikers over shared stories around a campfire. She enjoys hanging out with her friends and family, watching films, with her cat, Whiskers, curled up beside her."\\

  \midrule

  \noindent $\triangleright$ {\bf Example 5} for user profile generation with chain of thought reasoning:

    \textit{\textbf{The given topics}} are $\langle$\textbf{\it watercolors, local artist, big community impact, local artists, liberal arts camps}$\rangle$.\\

    \textit{\textbf{Let's think step by step}} about how to write a good biography for an individual using the given topics. Here again, we can see there are multiple topics with similar themes i.e., art and artist. This can be more effectively integrated as a hobby. This time, we will specify the category as 'painting portraits'. We begin by exploring the individual's upbringing, mentioning that he resides in San Francisco where he and his mother run a coffee shop, a local favorite within the community. This detail ties his current profession into the narrative. We then highlight his passion for painting portraits, thus explicitly integrating the given topic into the narrative, aligning it with the individual's hobbies and passions, and also underscoring the interest or appeal that the individual would feel toward the given topics. Additionally, we can delve further into his family history, mentioning that he lost his father at a young age and is very close to his childhood friends. We can conclude by noting his love for foreign films and hosting movie nights with these friends. Although these details would be unrelated to the given topic, they provide essential background and contribute to a natural-sounding biography.\\

    \textit{\textbf{So, a good biography would be:}} "Jamal lives in San Francisco where he and his mother run a coffee shop, a local favorite among the community. Jamal has a deep passion for painting portraits, and wants to pursue a career as an artist. Having lost his father at a young age, Jamal is very close with his childhood friends, who have become like family to him. He is also a big fan of foreign films, a hobby he regularly enjoys by hosting movie nights with his friends."\\

  \midrule

  \noindent $\triangleright$ {\bf Example 6} for user profile generation with chain of thought reasoning:

    \textit{\textbf{The given topics}} are $\langle$\textbf{\it canoeing, kayaking, sunny fish, deck, beautiful garden, hummingbirds, songbirds}$\rangle$.\\

    \textit{\textbf{Let's think step by step}} about how to write a good biography for an individual using the given topics. Here, since there are some topics with similar themes, we express these through a more general category i.e, the topics 'canoeing, kayaking' with 'adventure sports'. We can begin by discussing the individual's hometown and current profession, such as being a journalist from Toronto. Next, we can mention his passion for adventure sports and explicitly integrate the given topic into the narrative, aligning it with the individual's hobbies and passions and also underscoring the interest or appeal that the individual would feel toward the given topics. Since we identified journalism as his profession, it would be fitting to introduce photography as one of his passions. Let's specify it further by categorizing it as 'wildlife photography. Thus, we can implicitly integrate the remaining given topics 'sunny fish, beautiful garden, hummingbirds' into the narrative as a wildlife photographer can be reasonably expected to enjoy taking pictures of hummingbirds and sunny fishes. Additionally, we can mention that he likes to spend his time mentoring aspiring journalists and photographers, an activity unrelated to the given topics, but which complements the established narrative and enriches his background.\\

    \textit{\textbf{So, a good biography would be:}} "Alan Fang is a journalist from Toronto who loves adventure and is passionate about sports such as rock climbing and kayaking. He is also passionate about photography and skillfully merges his journalistic instincts with wildlife photography. Alan's talent for photography has led him to explore distant natural areas, using his photos to show conservation needs through his lens. Alan also likes to spend his time mentoring aspiring journalists and photographers."\\

  \midrule

  \noindent $\triangleright$ {\bf Example 7} for user profile generation with chain of thought reasoning:

    \textit{\textbf{The given topics}} are $\langle$\textbf{\it capoeira/Zumba studio, dance floor, salsa, bachata, merengue, cha cha cha dancers, dance lesson}$\rangle$.\\

    \textit{\textbf{Let's think step by step}} about how to write a good biography for an individual using the given topics. Here, since there are multiple topics with a similar theme, we express these through a more general category i.e, the topics 'dance floor, salsa, bachata, merengue, cha cha cha dancers, dance lesson' with 'dance and dancing'. This can be more effectively integrated as a hobby. We can begin by exploring the individual's upbringing and family history, highlighting that she is from New Orleans, where she resides with her parents and brother. Next, we can focus on her passion for dancing, further enriched by the detail that her mother introduced her to the art. In doing so, we explicitly integrate the given topic into the narrative, aligning it with the individual's hobbies and passions, and emphasizing the interest or appeal that the individual would feel toward the given topics. We can also add some social activities such as she also organizes local dance workshops and performances with her trope, further integrating her engagement with the dance community. To ensure the biography feels well-rounded and natural, we can include her professional background, choosing something unrelated to the given topics. For the given the topic 'capoeira/Zumba studio', we can assume that it is implicit for a person who is interested in dancing.\\

    \textit{\textbf{So, a good biography would be:}} "Chanda is from New Orleans, where lives with her parents and brother. She splits her time between working as a nurse and a dance instructor. She also organizes local dance workshops and performances with her trope. Her mother introduced her to the classical dance form Kathak, which has significantly influenced her approach to dancing."\\

  \midrule

  \noindent $\triangleright$ {\bf Example 8} for user profile generation with chain of thought reasoning:

    \textit{\textbf{The given topics}} are $\langle$\textbf{\it new artists, songwriters night, open mic}$\rangle$.\\

    \textit{\textbf{Let's think step by step}} about how to write a good biography for an individual using the given topics. Here, again, we can see there are multiple topics 'new artists, songwriters night, open mic' with a similar theme, we can express these through a more general category i.e., music. This can be more effectively integrated as a hobby. We can start with the individual's professional background. To make the biography sound well-rounded and natural, we can choose a profession unrelated to the given topics. Next, we can delve into his personal history and attempt to tie it with his passion for music such as when he met his wife in college and bonded over their shared love of music. This detail explicitly integrates the given topic into the narrative, aligning it with the individual's hobbies and passions, and highlighting the interest or appeal that the individual would feel toward the given topics. Furthermore, we can delve deeper into his passion for music by mentioning that he loves spending hours discovering new bands and unique sounds, often sharing these finds with his bandmates. These details further enrich the biography.\\

    \textit{\textbf{So, a good biography would be:}} "Manish works as an IT manager. His passion for technology and interpersonal skills makes him very competent in his job. He met his wife in college and bonded over their shared love of music. Manish also has a small band with his wife and 2 other friends. They love performing local clubs and are typically a hit at office parties. Manish loves spending hours discovering new bands and unique sounds, often sharing these finds with his bandmates, inspiring their next session"\\

  \midrule

  \noindent $\triangleright$ {\bf Example 9} for user profile generation with chain of thought reasoning:

    \textit{\textbf{The given topics}} are $\langle$\textbf{\it panoramic view of the beach}$\rangle$.\\

    \textit{\textbf{Let's think step by step}} about how to write a good biography for an individual using the given topics. Here, we can try not to quote the given topic verbatim and instead explore related concepts that allow us to integrate the topic implicitly, aligning it with the individual's hobbies or passions.  It would be fitting to assume that someone who has a passion for marine photography would also relish spending time at the beach or appreciating its panoramic views. There exists a natural connection or affinity. Let's begin by discussing the individual's current profession and hometown, selecting a location away from coastal cities, and highlighting their preference for traveling to destinations close to the ocean, for instance, a software engineer from Denver, Colorado with a deep passion for marine photography. We are, thus, implicitly integrating the given topic into the narrative. Furthermore, we can mention his work on travel-related projects and apps that help others discover hidden beach getaways, as well as his subsequent social media presence. These details enrich his story, ensuring the biography is well-rounded and natural.\\

    \textit{\textbf{So, a good biography would be:}} "Tom is a software developer from the bustling city of Denver, Colorado. His day-to-day life revolves around coding and developing applications for tech startups. Tom harbors a deep passion for marine photography, particularly fascinated by the beauty of shores and the diverse ecosystems found beneath ocean waves. It inspires Tom to work on travel-related projects and apps that help others discover hidden beach getaways. He often shares stories of his seaside excursions on a personal blog, where he connects with fellow beach lovers and plans future journeys to coastal landscapes, which have started gaining attention on social media platforms."\\
\end{longtable}

\section{Examples of Utility Assignments}
\label{app:utility-examples}

We present examples illustrating each utility level, as described in the annotation guidelines in Section~\ref{sec:utility-dataset}. Each example consists of a user profile and a restaurant review containing an atypical aspect annotated with the corresponding utility level for that user.

\subsection{None Utility Examples}

\noindent\rule{\linewidth}{0.4pt}
\textbf{User Profile:} Elena, living in sunny San Diego, is a market research analyst by profession but finds her true passion in capturing the essence of beach life through her hobby of photography. Elena cherishes moments spent under cheerful beach umbrellas, which have often provided a colorful backdrop to her seaside portraits. Her fascination with visual aesthetics doesn't stop there; she owns an eclectic collection of cute wallets, each picked for its unique design and functionality. The highlight of Elena's year is typically the Fourth of July fireworks display over the ocean, an event that combines her love for vibrant spectacles with the serene beach atmosphere she adores.

\textbf{Review:} Came for Valentine's dinner, and I would recommend this place for special occasions. The ambiance was wonderful, our server was terrific (I wish I could remember her name, she had an accent if that helps), the food was delicious. I ordered the duck breast. It was phenomenal, prepared with sweet potato mash. I also had the prosciutto and pear salad, which was great, but I (unlike most) would have preferred less meat and more salad. My date ordered the chicken, she liked hers but said that the gnocchi and tomato underneath stole the show from the chicken. Her shrimp bisque was amazing, one of the best dishes of the night (along with the duck). The shrimp cocktail had colossal shrimp, but there were only three shrimp and I thought that was kinda lame for \$21. I would recommend asking for a table in the \textbf{piano room}, I would have liked to sit there but didn't see it until we were leaving.

\textbf{Utility} for \textbf{piano room} is \textbf{None}, per guideline \textbf{None}.\ref{item:noneone}\\
\noindent\rule{\linewidth}{0.4pt}
\textbf{User Profile:} Sophie is a tech consultant in San Francisco with a keen eye for design and an appreciation for the aesthetics of her living space. Outside her demanding job, \ul{\it Sophie indulges in redesigning her home to reflect a high-end look that's both functional and playful}. She has a particular fondness for modern plastic furniture, which she finds adds a modern touch to rooms without sacrificing comfort or durability. Living with her partner and their two young children, Sophie is always on the lookout for ways to incorporate fun elements that appeal to the whole family. One such highlight is the 'big Connect 4' game in their family room, a hit not only with the kids but also provides endless entertainment during family gatherings. Her passion for blending sleek design with practical, kid-friendly features makes her home uniquely inviting.

\textbf{Review:} I met my Meetup group there for lunch, this was my first visit here. Beautiful establishment, classy and stately. My order consisted of 2 appetizers and dessert: 1 pound of clams in a brothy-creamy sauce, grilled artichokes and the bread pudding. Clams were great, artichokes were the best I have ever had. Probably lightly steamed, then brushed with oil and grilled! Wowzer, that the chokes elevated to nirvana-like level Bread pudding was good and best described as a prelude to fall served in a cast iron skillet. Prompt and friendly service. It was enjoyable to walk the long hallway in the building leading to a garden, and take in the numerous pieces of art lining the walls. We took a glimpse at their larger dining room all set up for high tea. I plan on going there and experience it on a cold, snowy day this winter. We also visited the gift shop and found that the price point of most of their items was very reasonable. I purchase a \textbf{cute wallet}, a shawl and a necklace!!What a fun and productive experience!

\textbf{Utility} for \textbf{cute wallet} is \textbf{None} per guideline \textbf{None}.\ref{item:nonetwo}\\
\noindent\rule{\linewidth}{0.4pt}

\textbf{User Profile:} Mia is a graphic designer from Honolulu, Hawaii, where she runs her freelance business. She grew up near the beach and developed a passion for surfing early on, incorporating it into her daily routine. Before starting work each day, Mia enjoys catching the morning waves, which helps clear her mind and often inspires her designs. Surfing keeps her physically active and mentally sharp, essential for tackling her creative projects. \ul{\it Mia also participates in local initiatives to protect marine life} and enjoys guiding young designers and surfers by sharing practical tips on design and surfing fundamentals.

\textbf{Review:} Very cool place and PERFECT for the family. This is one of the best places that you can take the family for a meal. The Pop Shop is all about the kids. We brought our 4 kids in tonight and was pleasantly surprised that they had a woman making balloon puppets and \textbf{animals}. Our kids loved it. They also have a big menu. The food is great and the staff are nice.

\textbf{Utility} for \textbf{animals} is \textbf{None} per guideline \textbf{None}.\ref{item:nonethree}\\
\noindent\rule{\linewidth}{0.4pt}

\subsection{Low Utility Examples}

\noindent\rule{\linewidth}{0.4pt}
\textbf{User Profile:} Lily, a dedicated early childhood educator in Burlington, excels at weaving creativity and interactive play into her curriculum. She's known for incorporating engaging games that challenge her students' strategic thinking while nurturing their patience and focus. Outside of lesson planning, \ul{\it Lily has a profound love for crafting and visual arts. She delights in painting and creating cheerful, imaginative decorations for her classroom to spark curiosity and joy among her pupils}. On weekends, she extends her \ul{\it passion for creativity by hosting art sessions for local children}, aiming to broaden their artistic horizons from an early age.

\textbf{Review:} The \textbf{lime green counter}, shiny plastic furniture, and stained concrete floors at the Fat Straw fit right in with and outdo the futuristic style of bubble tea cafes I've been to in other cities. The wall of polaroids of customers does a nice job balancing out all the geometric shapes and synthetic materials, and keeps the Fat Straw from feeling sterile. I would expect the high-end look of the Fat Straw, along with its prime location at Edgehill Village, to mean that the prices would be higher than the usual bubble tea hole in the wall, but they're actually pretty standard (\$3.50 for a 16oz bubble tea). Along with the usual flavors, some more interesting possibilities that caught my eye were red bean, honeydew, lychee, pomegranate, ginger, elvis (banana + peanut butter), and sesame. The only drawbacks to the Fat Straw that I can see are the small size of the cafe (it would feel crowded with any more than 15 customers inside) and their cash only policy.

\textbf{Utility} for \textbf{lime green counter} is \textbf{Low} per guideline \textbf{Low}.\ref{item:lowone}\\
\noindent\rule{\linewidth}{0.4pt}

\textbf{User Profile:} Elena, living in sunny San Diego, is a market research analyst by profession but finds her true passion in capturing the essence of beach life through her hobby of photography. Elena cherishes moments spent under cheerful beach umbrellas, which have often provided a colorful backdrop to her seaside portraits. \ul{\it Her fascination with visual aesthetics} doesn't stop there; she owns an eclectic collection of cute wallets, each picked for its unique design and functionality. The highlight of Elena's year is typically the Fourth of July fireworks display over the ocean, an event that combines her love for vibrant spectacles with the serene beach atmosphere she adores.

\textbf{Review:} I met my Meetup group there for lunch, this was my first visit here. Beautiful establishment, classy and stately. My order consisted of 2 appetizers and dessert: 1 pound of clams in a brothy-creamy sauce, grilled artichokes and the bread pudding. Clams were great, artichokes were the best I have ever had. Probably lightly steamed, then brushed with oil and grilled! Wowzer, that the chokes elevated to nirvana-like level Bread pudding was good and best described as a prelude to fall served in a cast iron skillet. Prompt and friendly service. It was enjoyable to walk the long hallway in the building leading to a garden, and take in the numerous pieces of \textbf{art} lining the walls. We took a glimpse at their larger dining room all set up for high tea. I plan on going there and experience it on a cold, snowy day this winter. We also visited the gift shop and found that the price point of most of their items was very reasonable. I purchase a cute wallet, a shawl and a necklace!!What a fun and productive experience!

\textbf{Utility} for \textbf{art} is \textbf{Low} per guideline \textbf{Low}.\ref{item:lowtwo}\\
\noindent\rule{\linewidth}{0.4pt}

\subsection{Medium Utility Examples}

\noindent\rule{\linewidth}{0.4pt}
\textbf{User Profile:} Sophie is a tech consultant in San Francisco with a keen eye for design and an appreciation for the aesthetics of her living space. Outside her demanding job, Sophie indulges in redesigning her home to reflect a high-end look that's both functional and playful. She has a particular fondness for modern plastic furniture, which she finds adds a modern touch to rooms without sacrificing comfort or durability. \ul{\it Living with her partner and their two young children, Sophie is always on the lookout for ways to incorporate fun elements that appeal to the whole family}. One such highlight is the 'big Connect 4' game in their family room, a hit not only with the kids but also provides endless entertainment during family gatherings. Her passion for blending sleek design with practical, kid-friendly features makes her home uniquely inviting.

\textbf{Review:} Amazing new pizza joint. The service was outstanding - they gave my kiddo a \textbf{free comic} and were always attentive and just willing to chat!!! The gasparilla pizza was so good - it was a Cuban sandwich in pizza form! Wings were crispy and saucey, and the fire sticks for dessert were addictive. I think this is our new favorite pizza place!!!

\textbf{Utility} for \textbf{free comic} is \textbf{Medium} per guideline \textbf{Medium}.\ref{item:mediumone}\\
\noindent\rule{\linewidth}{0.4pt}

\textbf{User Profile:} Charlie is a Tampa resident who has always had an affinity for diverse hobbies and interests, reflecting his eclectic personality. After moving to the area for a job opportunity in digital marketing, he quickly fell in love with the vibrant Channelside district, often spending weekends exploring its lively atmosphere and waterfront views. His fascination with aquatic life led him to set up a large aquarium at home, where he meticulously cares for various fish species, often sourced from specialized stores in the local Asian market, a place he frequents not just for pet supplies but also to indulge in authentic cuisine and unique cultural experiences. Additionally, \ul{\it Charlie is passionate about playing ping pong; it's not only a way to stay active but also a cherished activity that allows him to connect with friends over friendly matches}.

\textbf{Review:} I have been wanting to venture over to Garrison since their opening and last night I finally made it! My friends and I made the short walk from Sparkman's Warf and found ourselves here to play some virtual golf. Little did we know how much fun we were actually going to have. I'm not much of a golfer but I did love that you can switch it to \textbf{games} like zombie dodgeball. When we ordered drinks I was even happier, my old fashioned was perfect and not overly sweet like most bars make it. The poutine was soooooo good and perfect for the for of us to split. To top everything off our waitress was very friendly. Overall really happy with our experience and look forward to coming back.Ps. If you dont want to walk from Channelside just take the trolly!

\textbf{Utility} for \textbf{games} is \textbf{Medium} per guideline \textbf{Medium}.\ref{item:mediumtwo}\\
\noindent\rule{\linewidth}{0.4pt}

\subsection{High Utility Examples}

\noindent\rule{\linewidth}{0.4pt}
\textbf{User Profile:} Lena is a community center coordinator in the bustling city of Seattle, Washington. \ul{\it She has always had a profound passion for both physical and mental wellness, which naturally led her to fall in love with yoga and meditation at an early age}. Recognizing their benefits, she incorporated these practices into her daily routine and often champions them within her community programs. Besides this, Lena has a vibrant side that lights up whenever she hits the dance floor on weekends with her friends. Dancing gives her joy and freedom to express herself away from the structured environment of her job.

\textbf{Review:} I love the atmosphere here but have to admit that the coffee lacked the robustness \& aromatic flavor I've grown to expect in New Orleans. Overjoyed by the veg options, though the vegan chili was less tasty than it sounded. That said I'll definitely return and am willing to revise my opinions. Looking forward to checking out the \textbf{yoga \& meditation} too!

\textbf{Utility} for \textbf{yoga \& meditation} is \textbf{High} per guideline \textbf{High}.\ref{item:high}\\
\noindent\rule{\linewidth}{0.4pt}

\section{In-context Learning Examples for Classifying Utility of Atypical Aspects}
\label{app:utilityclassificationcot-examples}

We used 4 in-context learning examples in the prompt for classifying the utility of an atypical aspect as None, Low, Medium, or High with respect to a user profile, as shown in Table~\ref{tab:prompt_utilityclassification}. As described in Section~\ref{sec:utility-llm}, we adopted two distinct ICL strategies: {\it Fixed} few-shot with chain-of-thought rationale and {\it Dynamic} few-shot with RAG. Here, we present the full text of 4 ICL examples used in the {\it Fixed} few-shot with chain-of-thought rationale setting for restaurants, hotels and hair salons in Table~\ref{tab:restutilityclassificationcot-examples-lt}, Table~\ref{tab:hotelutilityclassificationcot-examples-lt}, and Table~\ref{tab:hsutilityclassificationcot-examples-lt}, respectively.

\subsection{Restaurants}

\begin{longtable}{p{14.5cm}}
  \caption{In-context learning examples for classifying the utility of atypical aspects for restaurants as None, Low, Medium, or High with respect to a user profile.} 
  \label{tab:restutilityclassificationcot-examples-lt} \\
  \toprule
  \endfirsthead

  \multicolumn{1}{p{14.5cm}}{\tablename\ \thetable\ -- \textit{Continued from previous page}}\\
  \toprule
  \endhead

  \bottomrule
  \multicolumn{1}{r}{\textit{Continued on next page}}\\
  \endfoot

  \bottomrule
  \endlastfoot

  \noindent $\triangleright$ {\bf Example 1:}

      \textbf{U:} Lily, a dedicated early childhood educator in Burlington, excels at weaving creativity and interactive play into her curriculum. She's known for incorporating engaging games that challenge her students' strategic thinking while nurturing their patience and focus. Outside of lesson planning, Lily has a profound love for crafting and visual arts. She delights in painting and creating cheerful, imaginative decorations for her classroom to spark curiosity and joy among her pupils. On weekends, she extends her passion for creativity by hosting art sessions for local children, aiming to broaden their artistic horizons from an early age.\\

      \textbf{R:} The Fat Straw features a \textbf{<ata>lime green counter</ata>}, shiny plastic furniture, and stained concrete floors\\
    
       \hspace{2em}\textbf{Output}:  A' = ("lime green counter", "Low")\\
      
      \textbf{Explanation} = lime green counter: Assigned a relevance label of "Low". \textit{\textbf{Reasoning:}} \textit{\textbf{Let's think step by step}}, Lily has a passion for crafting and visual arts, which could extend to an appreciation for unique design elements like a lime green counter. However, this is a weak association as her interest is more focused on creating art and decorations. Therefore, it is labeled as Low. \textit{\textbf{Guideline applied:}} "The attractor and the atypical aspect are distinct and specific concepts with only a small degree of conceptual similarity between them."\\
  
  \midrule

  \noindent $\triangleright$ {\bf Example 2:}

      \textbf{U:} Sophie is a tech consultant in San Francisco with a keen eye for design and an appreciation for the aesthetics of her living space. Outside her demanding job, Sophie indulges in redesigning her home to reflect a high-end look that's both functional and playful. She has a particular fondness for modern plastic furniture, which she finds adds a modern touch to rooms without sacrificing comfort or durability. Living with her partner and their two young children, Sophie is always on the lookout for ways to incorporate fun elements that appeal to the whole family. One such highlight is the 'big Connect 4' game in their family room, a hit not only with the kids but also provides endless entertainment during family gatherings. Her passion for blending sleek design with practical, kid-friendly features makes her home uniquely inviting.\\

      \textbf{R:} The service at the restaurant is outstanding, with staff giving out \textbf{<ata>free comics</ata>} to kids and being attentive and willing to chat\\
    
       \hspace{2em}\textbf{Output}:  A' = ("free comics", "Medium")\\
      
      \textbf{Explanation} = free comics: Assigned a relevance label of "Medium". \textit{\textbf{Reasoning: Let's think step by step}}, Sophie has two young children and is always looking for ways to incorporate fun elements that appeal to the whole family. A free comic could be of interest to her children, making it moderately relevant to her interests. Therefore, it is labeled as Medium. \textit{\textbf{Guideline applied:}} "The attractor is a broad concept, while the atypical aspect is a more specific concept subsumed by that broad concept."\\

  \midrule

  \noindent $\triangleright$ {\bf Example 3:}

      \textbf{U:} Lena is a community center coordinator in the bustling city of Seattle, Washington. She has always had a profound passion for both physical and mental wellness, which naturally led her to fall in love with yoga and meditation at an early age. Recognizing their benefits, she incorporated these practices into her daily routine and often champions them within her community programs. Besides this, Lena has a vibrant side that lights up whenever she hits the dance floor on weekends with her friends. Dancing gives her joy and freedom to express herself away from the structured environment of her job.\\

      \textbf{R:} The restaurant also offers \textbf{<ata>yoga \& meditation</ata>}\\
    
      \hspace{2em}\textbf{Output}:  A' = ("yoga \& meditation", "High")\\
      
      \textbf{Explanation} = yoga \& meditation: Assigned a relevance label of "High". \textit{\textbf{Reasoning: Let's think step by step}}, Lena is very passionate about physical and mental wellness, particularly through yoga and meditation. Thus, the aspect of yoga \& meditation is a close match with her interests and it is labeled as High. \textit{\textbf{Guideline applied:}} "The atypical aspect closely matches an attractor in the user profile. Both the attractor and the atypical aspect must involve the same specific context.\\

  \midrule

  \noindent $\triangleright$ {\bf Example 4:}

      \textbf{U:} Elena, living in sunny San Diego, is a market research analyst by profession but finds her true passion in capturing the essence of beach life through her hobby of photography. Elena cherishes moments spent under cheerful beach umbrellas, which have often provided a colorful backdrop to her seaside portraits. Her fascination with visual aesthetics doesn't stop there; she owns an eclectic collection of cute wallets, each picked for its unique design and functionality. The highlight of Elena's year is typically the Fourth of July fireworks display over the ocean, an event that combines her love for vibrant spectacles with the serene beach atmosphere she adores.\\

      \textbf{R:} The restaurant has a \textbf{<ata>piano room</ata>} where customers can request to be seated\\
    
      \hspace{2em}\textbf{Output}:  A' = ("piano room", "None")\\
      
      \textbf{Explanation} = piano room: Assigned a relevance label of "None". \textit{\textbf{Reasoning: Let's think step by step}}, Elena's profile focuses on her passion for beach life, photography, visual aesthetics, and her collection of cute wallets. The piano room does not relate to these interests or passions, therefore it is labeled as None. \textit{\textbf{Guideline applied:}} "The atypical aspect lacks any relevance to the attractors in the user profile."\\
\end{longtable}

\subsection{Hotels}

\begin{longtable}{p{14.5cm}}
  \caption{In-context learning examples for classifying the utility of atypical aspects for hotels as None, Low, Medium, or High with respect to a user profile.} 
  \label{tab:hotelutilityclassificationcot-examples-lt} \\
  \toprule
  \endfirsthead

  \multicolumn{1}{p{14.5cm}}{\tablename\ \thetable\ -- \textit{Continued from previous page}}\\
  \toprule
  \endhead

  \bottomrule
  \multicolumn{1}{r}{\textit{Continued on next page}}\\
  \endfoot

  \bottomrule
  \endlastfoot

  \noindent $\triangleright$ {\bf Example 1:} 

      \textbf{U:} Sara is an elementary school teacher in the small, vibrant town of Maplewood, nestled in the lush valleys of Northern California. Growing up in a family that placed a strong emphasis on community engagement and healthy living, Sara naturally gravitated toward activities that reflect these values. She spends most of her Saturday mornings browsing through local produce at the farmer's market, where she also participates in various community events. One event close to her heart is the annual hula hoop contest hosted right there at the market. This passion not only allows her to connect deeper with her roots but also brings a unique element to her teaching style, as she often incorporates lessons on nutrition and physical wellness into her curriculum. Additionally, Sara uses these experiences to foster a sense of community and teamwork among her students by organizing mini contests and group activities.\\

      \textbf{R:} At night, the hotel transforms into a lively social scene with \textbf{<ata>live music</ata>} and dancing.\\
    
      \hspace{2em}\textbf{Output}:  A' = ("live music", "Low")\\
      
      \textbf{Explanation} = live music: Assigned a relevance label of "Low". \textit{\textbf{Reasoning: Let's think step by step}}, live music might have a small degree of conceptual similarity to Sara's interests in community engagement and healthy living. While music can be part of community events, it is not explicitly related to her specific interests or activities. Therefore it is labeled as Low. \textit{\textbf{Guideline applied:}} "The attractor and the atypical aspect are distinct and specific concepts with only a small degree of conceptual similarity between them."\\

  \midrule

  \noindent $\triangleright$ {\bf Example 2:} 

      \textbf{U:} Lena Thomas is a retired school teacher residing in the bustling heart of Orlando, Florida. After retirement, Lena channelled her lifelong passion for nature into creating a butterfly garden at her home. This project not only reflects her commitment to environmental conservation but also provides a sanctuary where she can immerse herself in the nature. Lena's childhood by the beach instilled in her a love for the ocean; however, residing inland, she makes it a point to regularly visit the ocean. Her trips are less about sunbathing on sandy beaches and more focused on experiencing the majesty of the sea from scenic overlooks or participating in guided marine tours.\\

      \textbf{R:} The location of the hotel is great, half a block from the beach and an easy walk to State St. and \textbf{<ata>Trolley tours</ata>} and Land Shark tour leave from less than a block away.\\
    
      \hspace{2em}\textbf{Output}:  A' = ("Trolley tours", "Medium")\\
      
      \textbf{Explanation} = Trolley tours: Assigned a relevance label of "Medium". \textit{\textbf{Reasoning: Let's think step by step}}, trolley tours may provide scenic views and information about the local area, which could interest Lena due to her appreciation for nature and scenic overlooks. Thus it is labeled with Medium annotation. \textit{\textbf{Guideline applied:}} "The attractor is a broad concept, while the atypical aspect is a more specific concept subsumed by that broad concept."\\

  \midrule

  \noindent $\triangleright$ {\bf Example 3:} 

      \textbf{U:} Lena Thomas is a retired school teacher residing in the bustling heart of Orlando, Florida. After retirement, Lena channelled her lifelong passion for nature into creating a butterfly garden at her home. This project not only reflects her commitment to environmental conservation but also provides a sanctuary where she can immerse herself in the nature. Lena's childhood by the beach instilled in her a love for the ocean; however, residing inland, she makes it a point to regularly visit the ocean. Her trips are less about sunbathing on sandy beaches and more focused on experiencing the majesty of the sea from scenic overlooks or participating in guided marine tours.\\

      \textbf{R:} The hotel has a nice pool and a veranda to check out the \textbf{<ata>butterfly garden</ata>}\\
    
      \hspace{1em}\textbf{Output}:  A' = ("butterfly garden", "High")\\
      
      \textbf{Explanation} = butterfly garden: Assigned a relevance label of "High". \textit{\textbf{Reasoning: Let's think step by step}}, Lena has a lifelong passion for nature, which she channels into creating and maintaining a butterfly garden at her home. A butterfly garden at the hotel aligns perfectly with her interests in environmental conservation and nature, making it highly relevant to her. Therefore, it is labeled as High. \textit{\textbf{Guideline applied:}} "The atypical aspect closely matches an attractor in the user profile. Both the attractor and the atypical aspect must involve the same specific context."\\

  \midrule

  \noindent $\triangleright$ {\bf Example 4:} 

      \textbf{U:} Amelia, a marine biologist based in Key West, Florida, has always been fascinated by the natural world. With her work focused on wildlife conservation, she spends much of her time researching and advocating for the protection of sea and bird species. Her home reflects her interests with simple decor inspired by the ocean and tropical forests. Outside work, Amelia enjoys volunteering at local animal rescue centers and giving talks at schools to educate children about environmental preservation. She's an avid kayaker, often spending weekends on the water where she feels closest to nature. Through social media, she shares tips on how individuals can help protect endangered habitats in their daily lives.\\

      \textbf{R:} There is a \textbf{<ata>pianist</ata>} in the lobby of the hotel\\
    
      \hspace{1em}\textbf{Output}:  A' = ("Pianist", "None")\\
      
      \textbf{Explanation} = Pianist: Assigned a relevance label of "None". \textit{\textbf{Reasoning: Let's think step by step}}, a pianist in the lobby is a form of entertainment that does not connect with Amelia's specific interests in nature, wildlife conservation, or marine biology. Therefore it is labeled as None. \textit{\textbf{Guideline applied:}} "The atypical aspect lacks any relevance to the attractors in the user profile."\\
\end{longtable}

\subsection{Hair Salons}

\begin{longtable}{p{14.5cm}}
  \caption{In-context learning examples for classifying the utility of atypical aspects for hair salons as None, Low, Medium, or High with respect to a user profile.} 
  \label{tab:hsutilityclassificationcot-examples-lt} \\
  \toprule
  \endfirsthead

  \multicolumn{1}{p{14.5cm}}{\tablename\ \thetable\ -- \textit{Continued from previous page}}\\
  \toprule
  \endhead

  \bottomrule
  \multicolumn{1}{r}{\textit{Continued on next page}}\\
  \endfoot

  \bottomrule
  \endlastfoot

  \noindent $\triangleright$ {\bf Example 1:} 

      \textbf{U:} Simone is a devoted single mom living in Orlando, Florida, with her 6-year-old daughter Mia. Their home is a testament to their shared love for Disney princesses, with each room sprinkled with themed decorations that spark imagination and joy. Simone's passion for bringing fairy tales to life goes beyond her doorstep; she actively participates in charitable activities within her community. As an avid crafter, she combines her knack for creativity with philanthropy by making blankets for local animal shelters. This activity not only allows her to help animals in need but also serves as valuable teaching moments for Mia about kindness and giving back. During weekends, Simone and Mia often volunteer at these shelters, turning each visit into a heartwarming adventure of compassion and care.\\

      \textbf{R:} The hair salon has <ata>play activities</ata> in the waiting area\\
    
      \hspace{2em}\textbf{Output}:  A' = ("play activities", "Low")\\
      
      \textbf{Explanation} = play activities: Assigned a relevance label of "Low". \textit{\textbf{Reasoning: Let's think step by step}}, Simone engages in activities with her daughter that bring joy and imagination. While play activities are not directly related to Simone's main interests, they could be somewhat appealing as they might entertain Mia. Thus, it is labeled as Low. \textit{\textbf{Guideline Applied:}} "The attractor and the atypical aspect are distinct and specific concepts with only a small degree of conceptual similarity between them."\\

  \midrule

  \noindent $\triangleright$ {\bf Example 2:} 

      \textbf{U:} Mia is a graphic designer from Honolulu, Hawaii, where she runs her freelance business. She grew up near the beach and developed a passion for surfing early on, incorporating it into her daily routine. Before starting work each day, Mia enjoys catching the morning waves, which helps clear her mind and often inspires her designs. Surfing keeps her physically active and mentally sharp, essential for tackling her creative projects. Mia also participates in local initiatives to protect marine life and enjoys guiding young designers and surfers by sharing practical tips on design and surfing fundamentals.\\

      \textbf{R:} The hair salon makes <ata>hair donations</ata> to soak up the oil spills\\
    
      \hspace{2em}\textbf{Output}:  A' = ("hair donations", "Medium")\\
      
      \textbf{Explanation} = hair donations: Assigned a relevance label of "Medium". \textit{\textbf{Reasoning: Let's think step by step}}, Mia participates in local initiatives to protect marine life and is involved in community activities. Hair donations to soak up oil spills have some relevance to her passion for environmental conservation, as it could be considered a specific activity within the broad concept of environmental conservation. Thus, it is labeled as Medium. \textit{\textbf{Guideline Applied:}} "The attractor is a broad concept, while the atypical aspect is a more specific concept subsumed by that broad concept."\\

  \midrule

  \noindent $\triangleright$ {\bf Example 3:} 

      \textbf{U:} Jessica, a devoted mother and passionate advocate for animal welfare, resides in the vibrant community of Boulder, Colorado. Working as a high school guidance counselor, she dedicates her time to nurturing young minds and actively participates in local charity events focused on supporting both people and animals in need. Jessica is always on the lookout for businesses that align with her values, especially those that offer meaningful experiences for her family while contributing positively to the community. She particularly treasures places where around the holidays they run specials actions that teach compassion and kindness to her children. Additionally, Jessica values self-care and enjoys pampering sessions such as waxing that understand the importance of 'me time' amidst a busy lifestyle.\\

      \textbf{R:} The salon runs specials around the holidays, such as offering a discount for bringing in a \textbf{<ata>bag of dog food</ata>} or blankets for local shelters.\\
    
      \hspace{1em}\textbf{Output}:  A' = ("bag of dog food", "High")\\
      
      \textbf{Explanation} = bag of dog food: Assigned a relevance label of "High". \textit{\textbf{Reasoning: Let's think step by step}}, Jessica is a passionate advocate for animal welfare and actively participates in charity events supporting animals in need. Bringing a bag of dog food to support local shelters aligns closely with her values. Thus, it is labeled as High. \textit{\textbf{Guideline Applied:}} "The atypical aspect closely matches an attractor in the user profile. Both the attractor and the atypical aspect must involve the same specific context."\\

  \midrule

  \noindent $\triangleright$ {\bf Example 4:} 

      \textbf{U:} Jessica, a devoted mother and passionate advocate for animal welfare, resides in the vibrant community of Boulder, Colorado. Working as a high school guidance counselor, she dedicates her time to nurturing young minds and actively participates in local charity events focused on supporting both people and animals in need. Jessica is always on the lookout for businesses that align with her values, especially those that offer meaningful experiences for her family while contributing positively to the community. She particularly treasures places where around the holidays they run specials actions that teach compassion and kindness to her children. Additionally, Jessica values self-care and enjoys pampering sessions such as waxing that understand the importance of 'me time' amidst a busy lifestyle.\\

      \textbf{R:} The decor of the salon is simple and pretty, featuring \textbf{<ata>Craftsman tool boxes</ata>} that serve as beauty tool boxes\\
    
      \hspace{1em}\textbf{Output}:  A' = ("Craftsman tool boxes", "None")\\
      
      \textbf{Explanation} = Craftsman tool boxes: Assigned a relevance label of "None". \textit{\textbf{Reasoning: Let's think step by step}}, while the use of Craftsman tool boxes as beauty tool boxes is a unique aspect, it does not align with Jessica's interests in family experiences, animal welfare, or self-care. Thus, it is labeled as None. \textit{\textbf{Guideline Applied:}} "The atypical aspect lacks any relevance to the attractors in the user profile."\\
\end{longtable}

\section{Design of Amazon Mechanical Turk User Interface}
\label{app:mturk}

\begin{figure}[t]
    \centering
    \includegraphics[width=\linewidth]{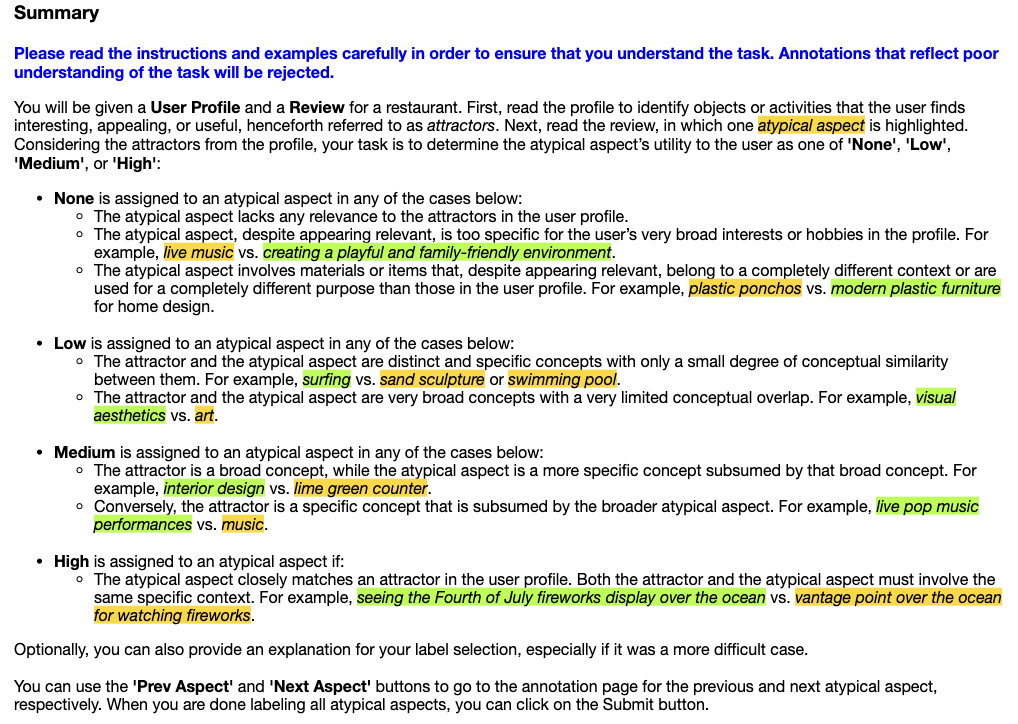}
    \caption{MTurk user interface with summary instructions as displayed on the platform.}
    \label{fig:amt_interface-example}
\end{figure}

\begin{figure}[]
    \centering
    \includegraphics[width=\linewidth]{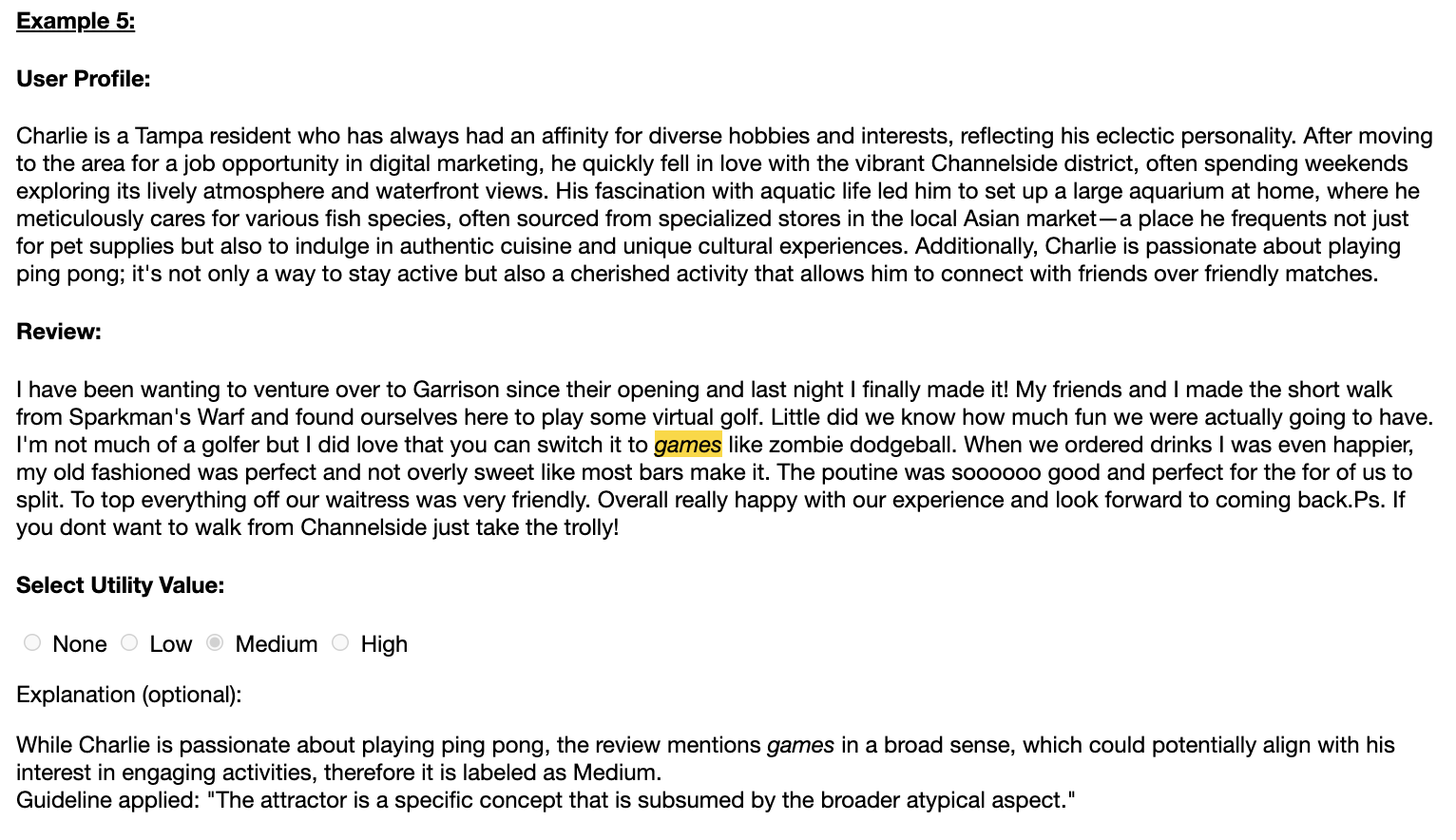}
    \caption{Example provided to MTurk workers for annotating utility value to an atypical aspect. }
    \label{fig:enter-label}
\end{figure}

\begin{figure}[]
    \centering
    \includegraphics[width=\linewidth]{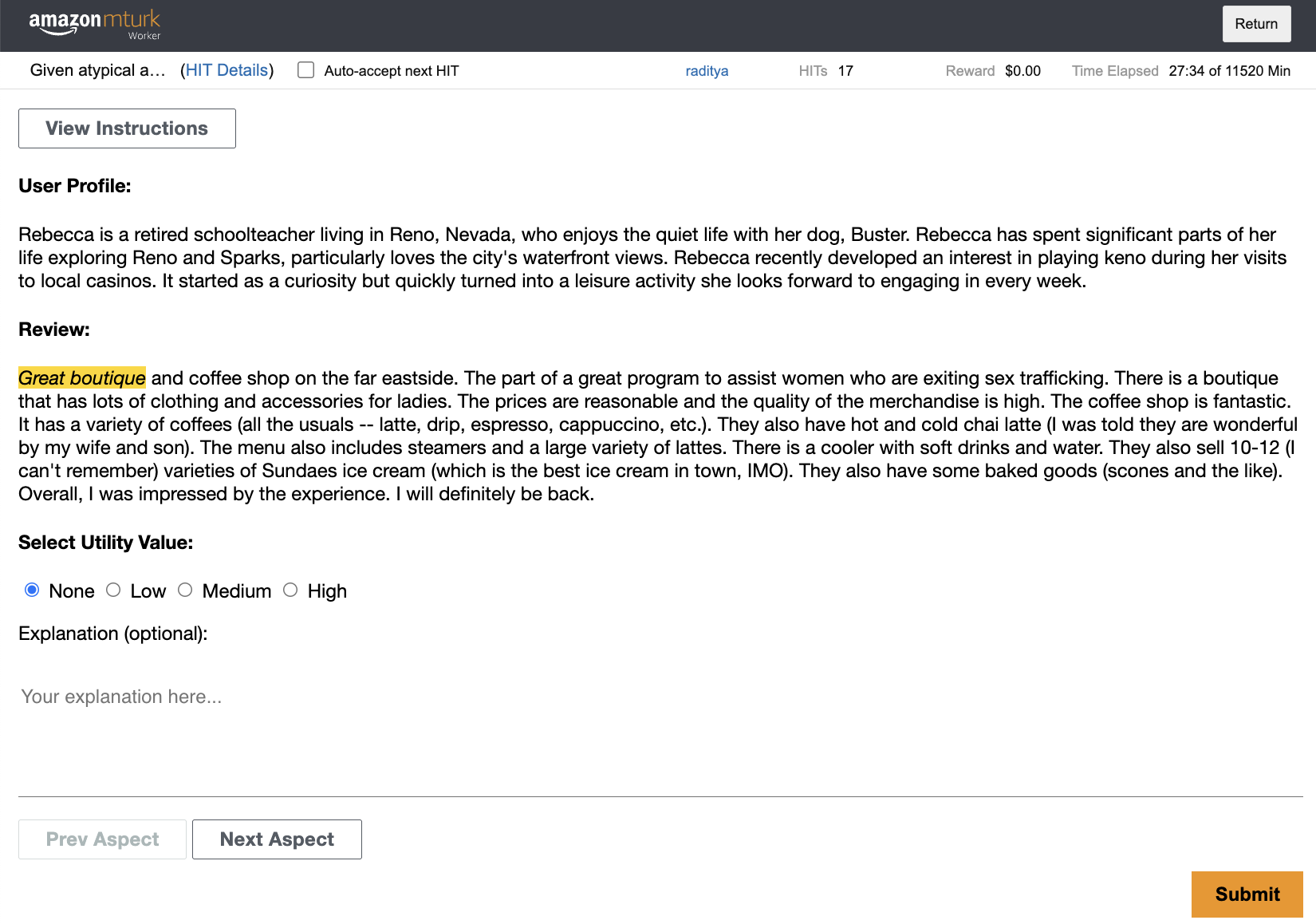}
    \caption{Sample Human Intelligence Task (HIT) provided to MTurk workers for annotating utility value to an atypical aspect.}
    \label{fig:enter-label-2}
\end{figure}

We present the design of the MTurk user interface used for crowd-sourcing ground truth annotations of utility values for atypical aspects with respect to a user profile (Section~\ref{sec:crowdsourcing}). Figure~\ref{fig:amt_interface-example} shows the summary instructions and annotation guidelines provided to MTurk workers. Figure~\ref{fig:enter-label} presents an example of annotating the utility value of an atypical aspect, while Figure~\ref{fig:enter-label-2} displays a sample Human Intelligence Task (HIT) used for this annotation process.

\end{document}